\documentclass[]{article}
\usepackage{amsthm}
\usepackage{amsmath}
\usepackage{natbib} 
\usepackage[colorlinks,citecolor=blue,urlcolor=blue,filecolor=blue,backref=page]{hyperref}
\usepackage{graphicx} 
\usepackage[utf8]{inputenc}
\usepackage{subcaption}
\usepackage{tikz}
\usetikzlibrary{bayesnet}
\usetikzlibrary{positioning}
\usetikzlibrary{decorations.text}
\usetikzlibrary{decorations.pathmorphing}
\usetikzlibrary{backgrounds}
\usepackage{bm}
\usepackage{multirow}
\usepackage{amssymb}
\usepackage{float}
\usepackage{units}
\usepackage[margin = 1.6in]{geometry}
\setcitestyle{authoryear,open={(},close={)}}

\usepackage{parskip}
\numberwithin{equation}{section}
\theoremstyle{plain}
\newtheorem{thm}{Theorem}
\newtheorem{assumption}{Assumption}
\newtheorem{result}{Prior class}
\theoremstyle{definition}
\newtheorem{exmp}{Example}
\newtheorem{definition}{Definition}

\definecolor{basecolor}{HTML}{C4C4C4}

\newcommand{\logit}{\text{logit}}

\newcommand\xqed[1]{%
  \leavevmode\unskip\penalty9999 \hbox{}\nobreak\hfill
  \quad\hbox{#1}}
\newcommand\demo{\xqed{$\triangle$}}

\begin{document}

\title{Intuitive joint priors for variance parameters}
\newcommand{\footremember}[2]{%
    \footnote{#2}
    \newcounter{#1}
    \setcounter{#1}{\value{footnote}}%
}
\newcommand{\footrecall}[1]{%
    \footnotemark[\value{#1}]%
} 
\author{%
  Geir-Arne Fuglstad\footremember{ntnu}{
  	Department of Mathematical Sciences, 
	Norwegian University of Science and Technology,
	Alfred Getz' vei 1,
	7034 Trondheim, Norway. 
	Corresponding author:
	\href{mailto:geir-arne.fuglstad@ntnu.no}{geir-arne.fuglstad@ntnu.no}
	}%
  \and Ingeborg Gullikstad Hem\footrecall{ntnu}%
  \and Alexander Knight\footrecall{ntnu}%
  \and Håvard Rue\footremember{saudi}{CEMSE Division,
	King Abdullah University of Science and Technology,
	Thuwal 23955-6900, Saudi Arabia.}%
  \and Andrea Riebler\footrecall{ntnu}%
  }

\date{August 2019}

\maketitle

\begin{abstract}

Variance parameters in additive models are typically assigned independent priors that 
do not account for model structure. We present a new framework for prior selection 
based on a hierarchical decomposition of the total variance along a tree structure
to the individual model components.
For each split in the tree, an analyst may be ignorant or 
have a sound intuition on how to attribute variance
to the branches. In the former case a Dirichlet prior is appropriate to use, while in the 
latter case a penalised complexity (PC) prior provides robust shrinkage. A bottom-up combination
of the conditional priors results in a proper joint prior.
We suggest default values for the hyperparameters and offer intuitive statements for
eliciting the hyperparameters based on expert knowledge. 
The prior framework is applicable for
\texttt{R} packages  for Bayesian inference such as \texttt{INLA} and \texttt{RStan}.

Three simulation studies show that, in terms of the application-specific measures of
interest, PC priors improve inference over Dirichlet priors when used to penalise
different levels of complexity in splits.
However, when expressing ignorance in a split, Dirichlet priors perform equally well and are 
preferred for their simplicity. 
We find that assigning current state-of-the-art default priors for each variance
parameter individually is less transparent and does not perform better than using the
proposed joint priors. We demonstrate
practical use of the new framework by analysing spatial heterogeneity in neonatal mortality
in Kenya in 2010--2014 based on complex survey data.

\end{abstract}

\graphicspath{{./img/}}

\section{Introduction}
\label{sec:introduction}


Bayesian hierachical models (BHMs) are ubiquitous in science due to their flexibility
and interpretablity \citep{gelmanbook, gelman2013bayesian, banerjee2014spatial}. In 
this paper, we consider BHMs where the latent level consists of an additive
combination of model components that are classified as fixed effects and random effects. 
This subclass covers a range of
common model classes such as generalised linear mixed models (GLMMs) and
generalised additive mixed models (GAMMs) \citep{fahrmeir2001bayesian}. In additive models, 
the total latent variance of the sum of the random effects
decomposes into the sum of the variance contributed by each random effect, and each 
random effect has a variance parameter that controls its \emph{a priori} contribution. 
We present a general framework for constructing joint priors for these variance parameters
for BHMs, and suggest robust shrinkage priors for the reduced class 
of latent Gaussian models (LGMs) where the model components are
Gaussian conditional on the model parameters \citep{rue2009,inlaReview, bakka2018, book126}.

There is no concensus on priors for variance parameters in BHMs
\citep{lambert2005, gelman2006, gelman2017}.
The default prior in the \texttt{R} package \texttt{INLA} \citep{inlaSoftwarePaper} is an inverse-gamma distribution $\text{InvGamma}(1, 5\cdot 10^{-5})$
\citep{blangiardo2015}, and the \texttt{R} package \texttt{RStan} 
\citep{carpenter2017,rstan_package} has implicit priors that are uniform on the range
of legal values for the parameters \citep{stan_manual}. WinBUGS, OpenBUGS and JAGS 
used $\text{InvGamma}(0.001, 0.001)$ distributions in their examples 
\citep{bugs_manual,jags_manual}, and the Stata manual employs
$\text{InvGamma}(0.01, 0.01)$ priors \citep{stata_manual}. Conjugacy provides
$\text{InvGamma}(\epsilon, \epsilon)$ distributions with computational advantages, but
their use may result in severe problems \citep{gelman2006} and they are generally inappropriate
for variances of random effects \citep{lunn2009}. \citet{gelman2006} proposed heavier 
tails through $\text{Half-Cauchy}(25)$ distributions on the standard deviations, and
others have investigated bounded uniform densities on the variances or the logarithms
of the variances \citep{lambert2005} and bounded uniform priors on the standard deviations
\citep{10.1093/biomet/ast023}. Recently, \citet{simpson2017} proposed a 
principle-based, robust prior termed penalised complexity (PC) prior that offers 
shrinkage towards zero variance. In the case of LGMs, the PC prior is an exponential
distribution on the standard deviation.

However, general-purpose priors may not be suitable for a given application
\citep{gelman2017prior} and independent priors for each random effect cannot 
exploit the structure of the model \citep[Section 7]{simpson2017}. For example, in 
disease mapping, prior elicitation is more meaningful for the total variance of the
random effects than their separate variances \citep{wakefield2006}, and, for 
animal models in genetic settings, the proportion of variability in a phenotypic 
trait being accounted for by genes  is important \citep{holand2013}. 
Further, the intraclass correlation (ICC) \citep{mcgraw1996} in a
random intercept model is linked to a generalised version of the coefficient of 
determination \citep{gelmanbook}, also known as $R^2$, which expresses the
proportion of the total variance explained by the model components. 
However, putting a prior on $R^2$
requires a joint prior on the two variance parameters in the random intercept model.
Additionally, in the context of regression, \citet{som2014block} discuss block 
g-priors where regression coefficients are partitioned and shrinkage is applied to 
the $R^2$ of each partition.

Consider a simple multilevel model with responses $y_{i,j,k}\vert \eta_{i,j,k} \sim \text{Poisson}(\exp(\eta_{i,j,k}))$, where
\(
	\eta_{i,j,k} = a_i + b_{i,j} + c_{i,j,k}
\)  
for experiment $k$ on individual $j$ in group $i$. We will term the group effect,
individual effect and measurement effect for A, B, and C, respectively, and write the latent model
as A+B+C for short hand. The total latent variance $t$ of A+B+C decomposes as
$t = \sigma_\mathrm{A}^2+\sigma_\mathrm{B}^2+\sigma_\mathrm{C}^2$, where 
$\sigma_\mathrm{A}^2$, $\sigma_\mathrm{B}^2$ and $\sigma_\mathrm{C}^2$ are the variances
of A, B and C, respectively. This standard parametrization facilitates independent priors
on the variances and can be used to achieve the desired  \emph{a priori} marginal properties
for the random effects. However, it is difficult to encode \emph{a priori} knowledge on joint
properties such as the size of $t$ or preference for A over B or A+B over C in a 
transparent and intuitive way. 

\begin{figure}
	\centering
	\begin{subfigure}[t]{0.3\textwidth}
		\centering
		\begin{tikzpicture} %
			\node[draw, rounded corners, minimum height = 0.6cm] (top) {A+B+C};
			\node[draw, left=of top, yshift = -.9cm, xshift = 0.9cm, rounded corners, minimum height = 0.6cm] (CD) {A};
			\node[draw, left=of top, yshift = -.9cm, xshift = 2cm, rounded corners, minimum height = 0.6cm] (SD) {B};
			\node[draw,right=of top, yshift = -.9cm, xshift = -1cm, rounded corners, minimum height = 0.6cm] (C) {C};
		    \path[->, every node/.style={midway}]
		   		(top) edge node[xshift=0.65cm, yshift = 0.1cm] {} (CD)
		   		(top) edge node[xshift= -0.65cm, yshift = 0.10cm] {} (SD)
		   		(top) edge node[xshift = 0.65cm, yshift = 0.1cm] {} (C);
		\end{tikzpicture}
		\caption{Unstructured}
		\label{fig:intro:example:unS}
	\end{subfigure}
	\begin{subfigure}[t]{0.3\textwidth}
		\centering
		\begin{tikzpicture} %
			\node[draw, rounded corners, minimum height = 0.6cm] (top) {A+B+C};
			\node[draw, right=of top, yshift = -.9cm, xshift = -1.3cm, rounded corners, minimum height = 0.6cm] (CD) {C};
			\node[draw, left=of top, yshift = -.9cm, xshift = 1.5cm, rounded corners, minimum height = 0.6cm] (SD) {A+B};
			\node[draw,right=of SD, yshift = -.9cm, xshift = -1cm, rounded corners, minimum height = 0.6cm] (C) {B};
			\node[draw, left=of SD, yshift = -.9cm, xshift = +1cm, rounded corners, minimum height = 0.6cm] (D) {A};
		    \path[->, every node/.style={midway}]
		   		(top) edge node[xshift=0.65cm, yshift = 0.1cm] {} (CD)
		   		(top) edge node[xshift= -0.65cm, yshift = 0.10cm] {} (SD)
		   		(SD) edge node[xshift = 0.65cm, yshift = 0.1cm] {} (C)
		   		(SD) edge node[xshift = -0.7cm, yshift = 0.08cm] {} (D);
		\end{tikzpicture}
		\caption{Structured}
		\label{fig:intro:example:S}
	\end{subfigure}
	\begin{subfigure}[t]{0.3\textwidth}
		\centering
		\begin{tikzpicture} %
			\node[draw, rounded corners, minimum height = 0.6cm] (top) {A+B+C};
			\node[draw, right=of top, yshift = -.9cm, xshift = -1.3cm, rounded corners, minimum height = 0.6cm, fill = basecolor] (CD) {C};
			\node[draw, left=of top, yshift = -.9cm, xshift = 1.5cm, rounded corners, minimum height = 0.6cm] (SD) {A+B};
			\node[draw,right=of SD, yshift = -.9cm, xshift = -1cm, rounded corners, minimum height = 0.6cm, fill = basecolor] (C) {B};
			\node[draw, left=of SD, yshift = -.9cm, xshift = +1cm, rounded corners, minimum height = 0.6cm] (D) {A};
		    \path[->, every node/.style={midway}]
		   		(top) edge node[xshift=0.65cm, yshift = 0.1cm] {} (CD)
		   		(top) edge node[xshift= -0.65cm, yshift = 0.10cm] {} (SD)
		   		(SD) edge node[xshift = 0.65cm, yshift = 0.1cm] {} (C)
		   		(SD) edge node[xshift = -0.7cm, yshift = 0.08cm] {} (D);
		\end{tikzpicture}
		\caption{Structured shrinkage}
		\label{fig:intro:example:Ssh}
	\end{subfigure}
	\caption{Hierarchical model decomposition. Gray boxes indicate preferred branches.}
	\label{fig:intro:example}
\end{figure}
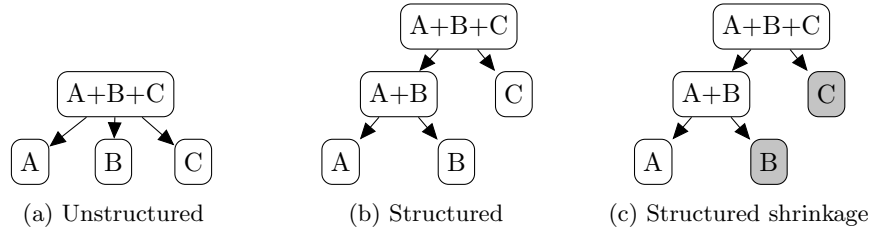

An obvious alternative is to parametrize the variance parameters as $t$ and the 
proportion of $t$ assigned to each random effect 
$(\omega_\mathrm{A}, \omega_\mathrm{B}, \omega_\mathrm{C})$, 
where $0 \leq \omega_\mathrm{A}, \omega_\mathrm{B}, \omega_\mathrm{C} \leq 1$ and
$\omega_\mathrm{A} + \omega_\mathrm{B} + \omega_\mathrm{C} = 1$. This is illustrated in Figure \ref{fig:intro:example:unS} by splitting A+B+C into the models A, B and C.
This parametrization is suitable for expressing ignorance about how the 
variance should be attributed to the random effects. A simple way to assign the 
joint prior is to set $(\omega_\mathrm{A}, \omega_\mathrm{B}, \omega_\mathrm{C}) \sim \text{Dir}(a,a,a)$,
$a>0$, where $\text{Dir}$ denotes the Dirichlet distribution \citep{balakrishnan2004primer}. This prior has no
preference for one of the random effects over the other and is invariant to the ordering
of the random effects, 
and we can 
select $a > 0$ to make the prior suitably vague. Together with the conditional prior
$\pi(t | \omega_\mathrm{A}, \omega_\mathrm{B}, \omega_\mathrm{C})$, this implicitly defines a 
proper joint prior for $(\sigma_\mathrm{A}^2, \sigma_\mathrm{B}^2, \sigma_\mathrm{C}^2)$
that is invariant to permutations in the order of the random effects, but
can incorporate prior knowledge on $t$. This has a similar flavor as the 
Dirichlet-Laplace prior by \citet{bhattacharya2015dirichlet}, which is a 
global-local shrinkage prior \citep{polson2010shrink} that induces sparsity in 
regression. However, in this paper we will focus on random effects and not fixed effects.

The simple split strategy is not always suitable and \citet{riebler2016}
demonstrated that for the BYM (Besag, York and Mollié) model, which
is a sum of a Besag random effect and an unstructured random effect,
a  PC prior that penalises the added complexity
of the structured effect relative to the unstructured effect
improves inference. For A+B+C, fewer levels of hierarchy may be preferred so
that B is preferred to A and C is preferred over A+B. This knowledge about
relative complexity of the random effects can be incorporated by splitting A+B+C
hierarchically as shown in Figure \ref{fig:intro:example:S}. Here
we first split A+B+C into A+B  and C 
through $\omega_1 = (\sigma_\mathrm{A}^2+\sigma_\mathrm{B}^2)/t$, and then split A+B 
into A and B 
through $\omega_2 = \sigma_\mathrm{A}^2/(\sigma_\mathrm{A}^2+\sigma_\mathrm{B}^2)$,
where $0 \leq \omega_1, \omega_2 \leq 1$. The joint prior 
for $(\sigma_\mathrm{A}^2, \sigma_\mathrm{B}^2, \sigma_\mathrm{C}^2)$ is then 
constructed by first selecting $\pi(\omega_2)$, then
$\pi(\omega_1\vert \omega_2)$, and finally 
$\pi(t \vert \omega_1, \omega_2)$. Priors inducing shrinkage towards
$\omega_2 = 0$ and $\omega_1 = 0$ can be chosen in the lower and upper split,
respectively. The shrinkage can be illustrated graphically as shown in
Figure \ref{fig:intro:example:Ssh}. For LGMs, PC priors offer a robust choice, but 
the framework is general and other priors can be selected by the analyst. 
For example,
if shrinkage is only required at the top level, a Dirichlet prior 
for $(\omega_2, 1-\omega_2)$ could be combined with a shrinkage prior 
for $\omega_1|\omega_2$. 

The ideas generalize to more random effects through the selection of
a hierarchical decomposition 
of the model in the form of a tree, and the selection of a conditional distribution for
the attribution of the total variance to the branches for each split.
The joint prior is calculated in a bottom-up approach using these conditional 
distributions. We suggest default values for 
the hyperparameters of the Dirichlet distribution based on the marginal prior 
distributions for the proportions of variance assigned to each branch of the split.
This ensures that the default setting for the prior is well-behaved as the number of 
branches in a split increases. Default values for the PC priors can be selected
based on moderate shrinkage of the proportion of variance. Additionally, we discuss 
how to include expert knowledge through interpretable statements on the
total variance and the distribution of variance in the tree. The joint prior can
contain a mix of expert knowledge and default values that provide a weakly 
informative prior \citep{gelman2008weakly,simpson2017}.  This means the prior
framework with joint priors is appropriate for default priors for software 
packages such as \texttt{INLA} and \texttt{RStan}.

The properties of the proposed priors are compared to the properties of default priors from
software and vague priors from literature. This is a fair comparison since
even though the new priors account for
model structure, they do not incorporate strong 
expert knowledge and are  suggested to be used in a default way in Bayesian software. 
The comparison is performed through
three simulation studies: a simple random intercept model with Gaussian responses, 
a latin square experiment with Gaussian responses, and a spatial model with 
Binomial responses. To ease the presentation of the comparisons and not 
overload the reader with results, we choose a set of targets for each simulation
study and compare the posteriors resulting from the different prior choices 
with respect to the targets. 
Additional results are 
provided in the Supplementary Materials. Furthermore, we provide
example code in the Supplementary Materials for producing results
for different priors for the latin square model in Section \ref{sec:gauss:LSE}. The
code is described 
in Section \ref{sec:supp:latinsquare:exCode} in the Supplementary Materials.

We start by introducing the general framework 
in Section \ref{sec:prelim}, then we introduce LGMs and suitable
priors for developing a new class of priors for LGMs in Section \ref{sec:priors}. The
new class of priors for LGMs is introduced in Section \ref{sec:method} and is 
applied to simulation studies with Gaussian responses in Section \ref{sec:gaussian}. In
Section \ref{sec:nonGauss} we present one simulation study with Binomial 
response and explain how the approach can be used in practice.
The paper ends with a discussion in Section \ref{sec:discussion}.

\section{Tree-based hierarchical variance decomposition}
\label{sec:prelim}


In this section we cover basic notation, and formally introduce additive models, 
hierarchical variance decomposition, and the new framework for joint priors for variances.

\subsection{Additive models}
Let $\boldsymbol{y} = (y_1, \ldots, y_n)$ be a vector of $n > 0$ observations. We model the 
expected values $\mathrm{E}(y_i) = g^{-1}(\eta_i)$, $i = 1,\ldots, n$, 
through a vector of linear predictors 
$\boldsymbol{\eta} = (\eta_1, \ldots, \eta_n)$
and a link function $g:\mathbb{R}\rightarrow\mathbb{R}$. We consider models
where the likelihood has parameters $\boldsymbol{\theta}_\mathrm{L}$ and factors as
$\pi(\boldsymbol{y}\vert \boldsymbol{\eta}, \boldsymbol{\theta}_\mathrm{L}) = \prod_{i = 1}^n \pi(y_i\vert \eta_i, \boldsymbol{\theta}_\mathrm{L})$. 
This covers models such
as GLMMs and GAMMs. We term $\boldsymbol{\eta}$ and its description as the latent part
of the model.

We assume that the linear predictor is described as
\begin{equation}
	\eta_i = \beta_0 + \boldsymbol{x}_i^\mathrm{T}\boldsymbol{\beta} + \sum_{j = 1}^N u_{j, k_j[i]}, \quad i = 1, \ldots, n,
	\label{eq:TBD:linMod}
\end{equation}
where $\beta_0$ is the intercept, $\boldsymbol{x}_i$ is the vector of covariates
associated with observation $i$, 
$\boldsymbol{\beta}$ is a vector of coefficients, and $\boldsymbol{u}_j = (u_1, \ldots, u_{m_j})$
is a random vector and $k_j[i]$ is the associated element of $\boldsymbol{u}_j$ for
observation $i$ for $j = 1,\ldots, N$. The two first terms will be called fixed effects
and the last $N$ terms will be called random effects. 
To focus on the joint prior for variance parameters, we will assume
that each random effect $\boldsymbol{u}_j$ has a single model parameter, which is a 
variance $\sigma_j^2$. In general, the random effects may have other parameters such as
correlation parameters and we discuss how to handle this in Section \ref{sec:discussion}.

We denote the vector of model parameters by
$\boldsymbol{\theta}_\mathrm{M} = (\sigma_1^2, \ldots, \sigma_N^2)$. The BHM is completed by
specifying the latent model through $\pi(\boldsymbol{u}_j | \sigma_j^2)$ for
$j = 1,\ldots, N$, and the 
prior $\pi(\beta_0, \boldsymbol{\beta},\boldsymbol{\theta}_\mathrm{L}, \boldsymbol{\theta}_\mathrm{M})$. We follow common practice so that the prior satisfies
$\pi(\beta_0, \boldsymbol{\beta},\boldsymbol{\theta}_\mathrm{L}, \boldsymbol{\theta}_\mathrm{M}) = \pi(\beta_0)\pi(\boldsymbol{\beta})\pi(\boldsymbol{\theta}_\mathrm{L})\pi(\boldsymbol{\theta}_\mathrm{M})$. The major improvement over common practice is that we will 
develop a framework for selecting intuitive joint priors for the variance parameters
that does not require that
$\pi(\boldsymbol{\theta}_\mathrm{M}) = \prod_{j=1}^N \pi(\sigma_j^2)$.

\subsection{Hierarchical variance decomposition}
The additivity in Equation \eqref{eq:TBD:linMod} causes the total latent variance $\text{Var}[\eta_i| \beta_0, \boldsymbol{\beta}, \boldsymbol{\theta}_\mathrm{M}]$ of linear predictor $i$ to decompose
as the variance contributed by each random effect $\text{Var}[u_{k_j[i]} |\beta_0, \boldsymbol{\beta}, \sigma_j^2]$, $j = 1, \ldots, N$, for $i = 1,\ldots, n$.
If random effect j is homogeneous, the variance parameter of random effect $j$
will be a marginal variance in the sense that
$\text{Var}[u_{k_j[i]} |\beta_0, \boldsymbol{\beta}, \sigma_j^2] = \sigma_j^2$
for $i = 1,\ldots, n$. If all random effects are homogeneous, 
the total latent variance of the linear predictors is homogeneous,
$t = \text{Var}[\eta_1| \beta_0, \boldsymbol{\beta},  \boldsymbol{\theta}_\mathrm{M}] = \cdots = \text{Var}[\eta_n| \beta_0, \boldsymbol{\beta},  \boldsymbol{\theta}_\mathrm{M}] = \sigma_1^2 + \ldots +\sigma_N^2$.
If random effect $j$ is heterogenous so that 
 $\text{Var}[u_{k_j[i]}| \beta_0, \boldsymbol{\beta},\sigma_j^2]$
varies for different values of $i$, the variance parameter $\sigma_j^2$ is selected
to be comparable to a marginal variance; 
see the discussion in Section \ref{sec:method:preliminary}. We term the parameter
$t = \sigma_1^2 + \ldots+ \sigma_N^2$ the total latent variance.

We describe the attribution of $t$ to the individual random effects
through a tree $\mathcal{T}$. The construction of $\mathcal{T}$ starts with a root 
node $T_0 = \{1,\ldots,N\}$ that contains all the random effects, and
in the first step we introduce $K_1 > 1$ child nodes $T_1, \ldots, T_{K_1}$ that
partition  $T_0$ into
$T_0 = T_1 \cup \cdots \cup T_{K_1}$. We continue this recursively for each child
node until all leaf nodes are singletons. This results in a tree $\mathcal{T}$ with
$S$ splits where there are $K_s$ child nodes for split $s = 1,\ldots, S$. We have
$S \leq N-1$, where 
$S = 1$ is achieved by directly splitting the root node to singletons as in Figure \ref{fig:intro:example:unS}
and the maximum value is achieved by only using dual 
splits such as in Figure \ref{fig:intro:example:S}.

For each split $s$, the parent node $P_s$ is split into $K_s$ child nodes
$C_1, \ldots, C_{K_s}$ and we will define a vector of parameters
$\boldsymbol{\omega}_s = (\omega_{s,1}, \ldots,\omega_{s,K_s})$,
$s = 1,\ldots, S$. The child nodes describe a partitioning of the 
random effects in the parent node, and we let $\boldsymbol{\omega}_s$ 
describe the proportion of the total
variance in the parent node, $\sum_{j\in P_s} \sigma_j^2$, that is assigned
to each child node through
\[
	\boldsymbol{\omega}_s = \frac{1}{\sum_{j\in P_s}\sigma_j^2}\left(\sum_{j\in C_1} \sigma_j^2, \ldots, \sum_{j \in C_{K_s}}\sigma_j^2\right), \quad s = 1, \ldots, S.
\]
We denote the $K-1$ simplex by 
\(
	\Delta^K = \{(x_1, \ldots, x_K)\vert \sum_{k = 1}^{K} x_k = 1, x_k \geq 0 \,\,\, \forall k\}
\)
so that the restrictions are $\boldsymbol{\omega}_s \in \Delta^{K_s}$ 
for $s = 1,\dots, S$. This means
that the parameters $\omega_{s,K_s}$ are superfluous for $s = 1,\ldots,S$,
but we keep them for ease of notation and interpretability.

For any split $s = 1,\ldots,S$, we term a child node and its decendants
as a branch of the split. The description of the model structure through a
tree structure defines a re-parametrization of
$(\sigma_1^2, \ldots, \sigma_N^2)$ to $(t, \boldsymbol{\omega}_1, \ldots, \boldsymbol{\omega}_S)$, where $S$ is the number of splits in the tree. The examples discussed in
the introduction can be rephrased in this terminology, and demostrate that
there is no unique selection of the tree.

\begin{exmp}[Tree structure]
Consider three random effects A, B and C with marginal variances 
 $(\sigma_\mathrm{A}^2, \sigma_\mathrm{B}^2, \sigma_\mathrm{C}^2)$.
Let the root node be $T_0 = \{\mathrm{A}, \mathrm{B},\mathrm{C}\}$.

Figure \ref{fig:intro:example:unS}, describes the case that 
the root node is partitioned into three children $T_1 = \{\mathrm{A}\}$,
 $T_2 = \{\mathrm{B}\}$ and $T_3 = \{\mathrm{C}\}$. This leads to a reparametrization
$(t, \boldsymbol{\omega})$, where $t = \sigma_\mathrm{A}^2 + \sigma_\mathrm{B}^2 + \sigma_\mathrm{C}^2$ and $\boldsymbol{\omega} = (\sigma_\mathrm{A}^2, \sigma_\mathrm{B}^2, \sigma_\mathrm{C}^2)/t$.

Figure \ref{fig:intro:example:S}
shows the case that $T_0$ is 
first partitioned into $T_1 = \{\mathrm{A},\mathrm{B}\}$ and
$T_2 = \{\mathrm{C}\}$, and then $T_1$ is partitioned 
into $T_3 = \{\mathrm{A}\}$ and $T_4 = \{\mathrm{B}\}$. This results in a reparamerization
$(t, \boldsymbol{\omega}_1, \boldsymbol{\omega}_2)$, where 
$t = \sigma_\mathrm{A}^2 + \sigma_\mathrm{B}^2 + \sigma_\mathrm{C}^2$,
$\boldsymbol{\omega}_1 = (\sigma_\mathrm{A}^2 + \sigma_\mathrm{B}^2, \sigma_\mathrm{C}^2)/t$
and $\boldsymbol{\omega}_2 = (\sigma_\mathrm{A}^2, \sigma_\mathrm{B}^2)/(\sigma_\mathrm{A}^2+ \sigma_\mathrm{B}^2)$.
\demo
\end{exmp}

\subsection{Hierachical decomposition priors}
The tree-based hierarchical variance decomposition facilitates the construction of
joint priors that include prior belief about the relative sizes of groups of random 
effects. The tree structure must be selected so that the desired comparisons can be made.
Trees such as shown in Figure \ref{fig:intro:example:unS} are useful for expressing
ignorance about the attribution of variance to the random effects, whereas trees such
as shown in Figure \ref{fig:intro:example:S} are useful for imposing shrinkage to one of 
the branches in each dual split. Generally, a tree may consist of a mixture of splits 
where the analyst wants to be informative and splits where the analyst wants to 
express ignorance.

We propose to construct a joint prior for the marginal variance parameters 
in a bottom-up approach where the prior for a given split only depends on descendant
nodes of the parent node.

\begin{assumption}[Bottom-up approach]
For a tree structure with $S$ splits,
\(
	\pi(\{\boldsymbol{\omega}_s\}_{s=1}^S) = \prod_{s = 1}^S \pi(\boldsymbol{\omega}_s \vert \{\boldsymbol{\omega}_j\}_{j\in D(s)}),
\) 
where $D(s)$ is the set of decscendant splits for split $s = 1, \ldots, S$.
\end{assumption}

This means that the joint prior for the decomposition uses a
directed acyclic graph so that parameters that belong to subsplits in different branches of 
a split are marginally independent. We combine the prior for the decomposition
of the variance with a conditional prior on the total variance of the random
effects to form what we will call
\emph{hierarchical decomposition} (HD) priors.

\begin{definition}[Hierarchical decomposition (HD) priors]
	\label{defn:HDpriors}
	Consider a BHM with an additive latent structure with $N$ random effects with
	marginal variance parameters $\sigma_1^2, \ldots, \sigma_N^2$. Assume that
	the model structure is described by a tree that recursively partitions the set of random
	effects into singletons. Then a hierarchical decomposition (HD) prior is given by
	\[
		\pi(\sigma_1^2, \ldots, \sigma_N^2) = \pi(t \vert \{\boldsymbol{\omega}_s\}_{s=1}^S)\prod_{s = 1}^S \pi(\boldsymbol{\omega}_s | \{\boldsymbol{\omega}_j\}_{j\in D(s)}),
	\]
	where $t = \sigma_1^2 + \ldots + \sigma_N^2$, $S$ is the number of splits, and
	 $D(s)$ denotes the set of descendant splits for
	the parent node in 	split $s$ and $\boldsymbol{\omega}_s$ describes the proportions
	of the total variance of a parent node assigned to its branches for $s = 1, \ldots, S$.
\end{definition}

\section{Latent Gaussian models and priors for the splits}
\label{sec:priors}

This section introduces LGMs and the priors we will use for the splits 
to build the intuitive
class of joint priors for the variance parameters for LGMs.

\subsection{Latent Gaussian models}
\label{sec:method:preliminary}

LGMs constitute a subclass of BHMs with additive latent structure where the
model components are Gaussian conditional on the model parameters. We write the
additive model in Equation \eqref{eq:TBD:linMod} in vector form,
\(
	\boldsymbol{\eta} = \boldsymbol{1}\beta_0 + \mathbf{X}\boldsymbol{\beta} + \sum_{j = 1}^N \mathbf{A}_j\boldsymbol{u}_j,
\)
where $\boldsymbol{1} = (1,\ldots, 1)$ is a column vector of length $n$, $\mathbf{X}$ is the 
$n \times p$ design matrix that contains the covariates for each observation 
as rows, and $\mathbf{A}_j$ are sparse $n \times m_j$ matrices that select the
 appropriate elements
of the random effects for $j = 1,\ldots, N$. The latent Gaussian structure is achieved by
$\beta_0 \sim \mathcal{N}(0, \sigma_\mathrm{I}^2)$,
$\boldsymbol{\beta} \sim \mathcal{N}_p(\boldsymbol{0}, \sigma_\mathrm{F}^2 \mathbf{I}_p )$, 
and $\boldsymbol{u}_j \vert \sigma_j^2 \sim \mathcal{N}_{m_j}(\boldsymbol{0}, \sigma_j^2\Sigma_j)$ for $j = 1, \ldots, N$. It is common to give
$\sigma_\mathrm{I}^2$ and $\sigma_\mathrm{F}^2$ suitably vague values, and we will assume
that $\sigma_\mathrm{I}^2$ and $\sigma_\mathrm{F}^2$ are fixed and 
focus on the variance parameters $\sigma_1^2, \ldots, \sigma_N^2$.

For non-intrinsic Gaussian random effects, such as 
independent and identically distributed  (i.i.d.)
random effects, stationary autoregressive processes and Matérn Gaussian random fields,
the covariance matrix $\Sigma$ of the random effect $\boldsymbol{u}$
is chosen to be a correlation matrix 
and the variance parameter $\sigma^2$ is the marginal variance.
However, this does not work for intrinsic Gaussian Markov random fields (GMRFs)
\citep{rue2005gaussian} 
such as the Besag model \citep{Besag1991}, the first-order
random walk and the second-order random walk \citep[Chapter 3]{rue2005gaussian}.
In this case there is no well-defined concept of a marginal variance since they are
defined through singular precision matrices that cannot be inverted to find
a covariance matrix. We follow
\citet{sorbye2014} and choose the variance parameter $\sigma^2$ to be a
representative value for the marginal variance.

\subsection{Introducing shrinkage towards branches}
\label{sec:blocks:shrink}
\subsubsection{Penalising complexity}
\label{sec:prelim:pcpriors}
The fundamental basis for introducing robust shrinkage in our proposed class of priors
are the PC priors introduced in \citet{simpson2017}, which uses a set of principles to derive 
model-component-specific prior distributions. The main idea is to regard a single model 
component as a flexible extension of a so-called base model. In the simplest case 
of an unstructured random effect, the base model would be to remove the effect 
entirely from the linear predictor by letting the variance parameter go to zero. 
The idea is to follow Occam's razor and favour a simpler, more sparse
or more intuitive model as long as the data does not indicate
otherwise.
The PC priors have been used successfully in a variety of contexts such as
BYM models \citep{riebler2016}, correlation parameters \citep{guo-etal-2017}, 
autoregressive processes \citep{sorbye-rue-2018} and
Mat\'ern Gaussian random fields \citep{fuglstad2018}.

\cite{simpson2017} proposed to compute the complexity of the alternative 
model relative to the base model using the Kullback-Leibler divergence (KLD) defined as
\begin{equation}
  \text{KLD}(\pi(\boldsymbol{u}|\xi) \mid\mid \pi(\boldsymbol{u}|\xi=0)) = \int \pi(\boldsymbol{u}|\xi) \log\left( \frac{\pi(\boldsymbol{u}|\xi)}{\pi(\boldsymbol{u}|\xi=0)} \right) \mathrm{d}\boldsymbol{u},
  \label{eq:KLD}
\end{equation}
where $\xi$ is the flexibility parameter, and $\xi = 0$ at the base model. The KLD is consequently transformed to an interpretable distance measure between two densities $f_1$ and $f_2$: $d(f_1 \mid\mid f_2) = \sqrt{2 \text{KLD}(f_1 \mid \mid f_2)}$. In contrast to defining a prior for 
$\xi$ directly, a prior is defined for $d$. See \cite{simpson2017} for detailed motivation.

We follow \citet{simpson2017} and select an exponential distribution, where information 
provided by the user is used to determine 
the rate $\lambda$. Usually this information is provided by a probability statement about the tail probability of the prior,
\begin{equation*}
  P(X(\xi) > U) = \alpha.
\end{equation*} 
Here, $X(\xi)$ is an interpretable transformation of the parameter of the flexible extension, $U$  can be thought of as a sensible upper bound, and $\alpha$ is a small probability.  A user can express their knowledge by constraining tail probabilities of $X(\xi)$ as above. Selecting $U$ near a large plausible value for $X(\xi)$ and $\alpha$ small encodes weak information about $\xi$ \citep{simpson2017}. This means
that it is \emph{a priori} unlikely that the value of $X(\xi)$ exceeds $U$. Finally, the prior can be transformed to the corresponding prior for the flexibility parameter $\xi$. An 
attractive feature of this principle-based construction is that the resulting priors are proper and have 
a natural link to Jeffreys' priors. 

\subsubsection{Shrinking a marginal variance parameter}
In the case of a single Gaussian random effect with marginal variance $\sigma^2$, the PC prior
with base model $\sigma^2 = 0$ is an exponential prior on $\sigma$. The rate parameter $\lambda$
can be set, for example, by an \emph{a priori}
statement $\mathrm{P}(\sigma > U) = 0.05$ so that the 95th percentile of the prior for $\sigma$ is $U > 0$. 
Then the prior is an exponential prior with rate
parameter $\lambda = -\log(\alpha)/U$ which we denote as $\sigma \sim \text{PC}_{\mathrm{SD}}(U, \alpha)$; see \citet{simpson2017} for details and derivation.

\subsubsection{Shrinking a weight parameter}
Consider the situation that the linear predictor only contains two random effects $A$ and $B$
with variances $\sigma_\mathrm{A}^2$ and $\sigma_\mathrm{B}^2$, respectively. The proportion
of $t = \sigma_\mathrm{A}^2+\sigma_\mathrm{B}^2$ assigned to each random effect
is described by
$\boldsymbol{\omega} = (1-\omega, \omega) = (\sigma_\mathrm{A}^2, \sigma_\mathrm{B}^2)/(\sigma_\mathrm{A}^2+\sigma_\mathrm{B}^2)$. If one \emph{a priori} prefers
the attribution $\boldsymbol{\omega} = \boldsymbol{\omega}^0 = (1-\omega_0,\omega_0)$, shrinkage can
be induced in the joint prior for the variance parameters using a PC prior where 
$\boldsymbol{\omega} = \boldsymbol{\omega}^0$ is the base model.
Here we apply the KLD from Equation
\eqref{eq:KLD} to express distance from the base model $\boldsymbol{\omega}^0$ to the
alternative model $\boldsymbol{\omega}$, and penalise deviations from
the base model according to the difference in model complexity.

\begin{thm}[PC prior for dual split]
	\label{thm:method:dualSplit}
	Let $\boldsymbol{u}_1$ and $\boldsymbol{u}_2$ be random effects of an LGM that
	enter the linear predictor 
	through $\mathbf{A}_1\boldsymbol{u}_1\sim\mathcal{N}_n(\boldsymbol{0}, \sigma_1^2\tilde{\Sigma}_1)$ and
	$\mathbf{A}_2\boldsymbol{u}_2\sim\mathcal{N}_n(\boldsymbol{0}, \sigma_2^2\tilde{\Sigma}_2)$.
	Assume that $\tilde{\Sigma}_1+\tilde{\Sigma}_2$ is non-singular\footnote{If this
	were not the case, some elements of the sum of $\mathbf{A}_1\boldsymbol{u}_1$
	and $\mathbf{A}_2\boldsymbol{u}_2$ would be exactly equal and
	we would choose a subset of maximal size so that $\tilde{\Sigma}_1+\tilde{\Sigma}_2$ was
	non-singular for comparing the effects
	of $\mathbf{A}_1\boldsymbol{u}_1$ and $\mathbf{A}_2\boldsymbol{u}_2$.}. Let $\omega = \sigma_2^2/(\sigma_1^2+\sigma_2^2)$ and 
	$\Sigma(w) = (1-\omega)\tilde{\Sigma}_1 + \omega\tilde{\Sigma}_2$. Then the distance from 
	the base model $\Sigma(\omega_0)$ to the alternative model $\Sigma(\omega)$ is given by
	$d(\omega) = \sqrt{\mathrm{tr}(\Sigma(\omega_0)^{-1}\Sigma(\omega))-n-\log\vert\Sigma(\omega_0)^{-1}\Sigma(\omega)\vert}$ for $0 \leq \omega_0 \leq 1$.

	The PC prior for $\omega$ with base model $\omega_0 = 0$ is
	\[
		\pi(\omega) =  \begin{cases} \frac{\lambda \left\vert d'\left(\omega\right)\right\vert}{1-\exp(-\lambda d(1))} \exp\left(-\lambda d\left(\omega\right)\right),  & \text{$0 < w < 1$, $\tilde{\Sigma}_1$ non-singular}, \\
		\frac{\lambda}{2\sqrt{\omega}(1-\exp(-\lambda))}\exp(-\lambda \sqrt{\omega}), & \text{$0 < \omega < 1$, $\tilde{\Sigma}_1$ singular},
		\end{cases}
	\]
	where $\lambda > 0$ is the hyperparameter. We suggest to set $\lambda$ so that the
	median is $\omega_\mathrm{m} = 0.25$. 

	For base model $0 < \omega_0 < 1$,  the PC prior whose median is equal to
	$\omega_0$ is
	\[
		\pi(\omega) = \begin{cases}
					\frac{\lambda \left\vert d'\left(\omega\right)\right\vert}{2[1-\exp(-\lambda d(0))]} \exp\left(-\lambda d\left(\omega\right)\right), &  0 < \omega < \omega_0, \\
					\frac{\lambda \left\vert d'\left(\omega\right)\right\vert}{2[1-\exp(-\lambda d(1))]} \exp\left(-\lambda d\left(\omega\right)\right), &  \omega_0 < \omega < 1,

		\end{cases}
	\]
	where $\lambda>0$ is a hyperparameter. We suggest to set $\lambda$ so that
	\[
		\mathrm{P}(\mathrm{logit}(1/4) + \mathrm{logit}(\omega_0) < \mathrm{logit}(\omega) < \mathrm{logit}(\omega_0)+\mathrm{logit}(3/4)) = 1/2.
	\]
	
	Base model equal to $\omega_0 = 1$ follows directly by reversing the roles of $\boldsymbol{u}_1$ and $\boldsymbol{u}_2$.

	\begin{proof}
		See Section \ref{sec:supp:proof1} in the Supplementary Materials.
	\end{proof}
\end{thm}

The default values in each case are specified as to place most of the prior mass in a small interval 
on the $\omega$ scale around $\omega_0$, but to also ensure large deviations from $\omega_0$
are \emph{a priori} plausible; in this sense they are weakly 
informative \citep{gelman2006,gelman2008weakly}. Sections \ref{sec:gauss:RIM} and 
\ref{sec:gauss:LSE} show that the results from the inference are stable to changes in 
these hyperparameters; which in turn shows that these $\lambda$'s provide weak information. If 
the analyst has expert knowledge this should be used instead of the default values. 
Large $\omega$ might be 0.75 for test-retest reliability in 
a psychology study \citep{cicchetti1994guidelines} but 0.4 for the genetic heritability of a trait 
\citep{shen2016heritability}. 

\subsection{Expressing \emph{a priori} ignorance about a split}
\label{sec:blocks:ignorance}
\subsubsection{Exchangeability}
In some cases the analyst does not want to express an \emph{a priori} preference
for any of the branches in a split in the tree. This can be achieved indirectly through
a series of dual splits. For example, by replacing the split in 
Figure \ref{fig:intro:example:unS} by the series of dual splits as shown in Figure
\ref{fig:intro:example:S} where the left-hand side has a base model of $2/3$ in the first
split and the left-hand side has a base model of $1/2$ for the second split. In total this
is specifying a base model of $1/3$ of the total variance to each random effect, but the
resulting prior is not invariant to permutations of A, B and C in
Figure \ref{fig:intro:example:S}. See Section \ref{sec:supp:multPC} of the Supplementary
Materials for details.
 When the goal is to express ignorance about the
decomposition of the variance, one can use a base model of equal attribution of
the total variance to each random effect and choose an exchangeable prior
for $(\sigma_\mathrm{A}^2, \sigma_\mathrm{B}^2, \sigma_\mathrm{B}^2)$. This can be done,
for example, through a Dirichlet prior.

\subsubsection{Dirichlet prior}
The Dirichlet prior of order $K \geq 2$ with parameters $a_1, \ldots, a_K > 0$ is
given by
\[
	\pi(\boldsymbol{\omega}) = \frac{1}{B(a_1, \ldots, a_K)}\prod_{k = 1}^K \omega_k^{a_k-1}, \quad \boldsymbol{\omega} = (\omega_1, \ldots, \omega_K) \in \Delta^K,
\]
where $B$ is the multivariate beta function, and $\Delta^K$ is the $K-1$ simplex.
Since there is no preference for any random effect, we consider the symmetric Dirichlet
distribution where $a_1 = \ldots = a_K = a > 0$, where $a$ is the hyperparameter that
must be selected by the analyst.
For $a = 1$ the prior is uniform, for $a < 1$ the prior has peaks at the 
vertrices of $\Delta^K$, and for $a > 1$ the mode is $\boldsymbol{\omega} = (1,\ldots, 1)/K$.
The prior is invariant to permutations of the elements of $\boldsymbol{\omega}$ for
any value of $a > 0$ and it is computationally cheap for
arbitrary dimensions $K$.

The hyperparameter $a$ can be selected by considering the marginal properties of
$\pi(\boldsymbol{\omega})$. The marginal prior
$\pi(\omega_1) \propto \omega_1^{a-1}(1-\omega_1)^{(K-1)a-1}$, $0 < \omega_1 < 1$, is 
a Beta distribution whose quantiles are dependent both on the values of $a$ and $K$. We select $a$
by requiring $\text{P}(\text{logit}(1/4)  < \text{logit}(\omega_1)-\text{logit}(\omega_0) < \text{logit}(3/4)) = 1/2$. By symmetry the same marginal properties are
satisfied for $\omega_i$, $i = 2, \ldots, K$.

\section{Hierarchical decomposition priors for LGMs}
\label{sec:method}


In this section we introduce the new class of intuitive joint priors for the variance
parameters in LGMs.

\subsection{Accounting for model structure}
In the general formulation of HD priors in Definition
\ref{defn:HDpriors}, the prior is composed of conditional priors that
for each split depends on all descendant splits. This is impractical because
computing PC priors would require new KLDs to be computed every 
time the prior is evaluated. We
take a pragmatic approach where we decide on a set of base models,
which expresses our best prior guess, and condition on these.

\begin{assumption}[Simplified conditioning]
	\label{assume:method:cond}
	For a given tree with $S$ splits and base models
	$\{\boldsymbol{\omega}_1^0, \ldots, \boldsymbol{\omega}_S^0\}$, we
	replace 
	$\pi(\boldsymbol{\omega}_s| \{\boldsymbol{\omega}_j\}_{j\in\mathrm{D}(s)})$ with 
	\(
		\pi(\boldsymbol{\omega}_s | \{\boldsymbol{\omega}_j = \boldsymbol{\omega}_j^{0}\}_{j\in\mathrm{D}(s)}),
	\)
	$s = 1, \ldots, S$.
\end{assumption}

Under this assumption a new class of  HD
priors for LGMs are constructed by combining intuition
about shrinkage and ignorance through independent priors for the splits.

\begin{result}[HD priors for LGMs]
	\label{thm:method:jointWeight}
	Assume the LGM contains $N$ random effects with variances
	$\sigma_1^2, \ldots, \sigma_N^2$ and that the hierarchical
	decomposition of the variance is described through a tree with $S$ splits.
	Under base models $\{\boldsymbol{\omega}_1^0, \ldots, \boldsymbol{\omega}_S^0\}$,
	the prior is
	\[
		\pi(\sigma_1^2, \ldots, \sigma_N^2) = 
		\pi(t\vert \{\boldsymbol{\omega}_s\}_{s=1}^S)\prod_{s = 1}^{S} \pi(\boldsymbol{\omega}_s | \{\boldsymbol{\omega}_j = \boldsymbol{\omega}_j^0\}_{j\in\mathrm{D}(s)}),
	\]
	where the total latent variance is $t = \sigma_1^2 + \ldots+ \sigma_N^2$, and
	$\boldsymbol{\omega}_i \in \Delta^{l_s}$, where $l_s$ is the number of branches
	in split $s$, $s = 1 , \ldots, S$.

	For each of the $S$ splits, the analyst can express ignorance through a Dirichlet prior
	or sequence of PC priors as described in Section \ref{sec:blocks:ignorance}, or 
	express preference to the selected base models as described in Section \ref{sec:blocks:shrink}.
	The selection of $\pi(t\vert \{\boldsymbol{\omega}_s\}_{s=1}^S)$ must
	be done in the context of the likelihood as described in Section \ref{sec:method:likelihood}.
\end{result}

This prior is computationally inexpensive
since the overall prior probability density factorises into independent
conditional distributions that consist of PC priors, which can be precomputed, 
and Dirichlet priors,
which are cheap to compute.

We demonstrate the use of HD priors through one example where the analyst wants to express
ignorance and one example where the analyst wants to penalise complexity.

\begin{exmp}[Non-nested random effects]
	\label{exmp:nonNested}
	Consider responses $y_1, \ldots, y_n$, described by the Gaussian linear model
	$y_i|\eta_i \sim \mathcal{N}(\eta_i, \sigma_\mathrm{R}^2)$ with 
	\[
		\eta_i = \mu + h_1(\text{Age}_i) + h_2(\text{Weight}_i) + h_3(\text{Income}_i), \quad i = 1, 2, \ldots, n,
	\]
	where $\mu$ is the intercept, $h_1$, $h_2$ and $h_3$ are smooth effects of the covariates expressed as
	second-order random walks \citep{rue2005gaussian}, and $\sigma_\mathrm{R}^2$ is the
	residual variance.
	Assume that one has no \emph{a priori} preference for the three
	smooth effects, and decide to encode the decomposition of the total latent variance as shown
	Figure \ref{fig:intro:example:unS}, where A, B and C represents the three smooth
	of covariates effects. 	Let $\boldsymbol{\omega}_1$ denote the proportions of variance assigned to model 
	components and let $t$ denote the total latent variance. We construct an HD prior
	by assigning a Dirichlet prior to $\boldsymbol{\omega}_1$, and handle $t | \boldsymbol{\omega}_1$ as discussed in Section
	\ref{sec:method:likelihood}. 
	\demo
\end{exmp}

\begin{exmp}[Shrinkage in multilevel models]
	\label{exmp:Nested1}
	The latent part of the multilevel model in Section \ref{sec:introduction} can
	be written in vector form as 
	$\boldsymbol{\eta} = \mathbf{A}_\mathrm{A} \boldsymbol{u}_\mathrm{A} + \mathbf{A}_\mathrm{B} \boldsymbol{u}_\mathrm{B} + \mathbf{A}_\mathrm{C} \boldsymbol{u}_\mathrm{C}$, where
	$\mathbf{A}_\mathrm{A}$, $\mathbf{A}_\mathrm{B}$ and $\mathbf{A}_\mathrm{C}$ are sparse
	matrices selecting the appropriate group, individual and measurement effects, respectively. 
	Assume we use an LGM, then 
	$\boldsymbol{u}_1 \sim \mathcal{N}_{G}(\boldsymbol{0}, \sigma_\mathrm{A}^2 \mathbf{I}_{G})$,
	$\boldsymbol{u}_2 \sim \mathcal{N}_{GP}(\boldsymbol{0}, \sigma_\mathrm{B}^2 \mathbf{I}_{GP})$ and
	$\boldsymbol{u}_3 \sim \mathcal{N}_{GPK}(\boldsymbol{0}, \sigma_\mathrm{C}^2 \mathbf{I}_{GPK})$, where
	$G$ is the number of groups, $P$ is the number of individuals per group, and $K$ is the
	number of measurements per individual.

	If we prefer shrinkage towards fewer levels in the multilevel model
	as shown in Figure \ref{fig:intro:example:Ssh}, we decompose the total latent variance 
	$t = \sigma_\mathrm{A}^2+\sigma_\mathrm{B}^2+\sigma_\mathrm{C}^2$ through two splits. For the split at the root node,
	we decompose $t$ according to the proportions $\boldsymbol{\omega}_1 = (\sigma_\mathrm{A}^2+\sigma_\mathrm{B}^2, \sigma_\mathrm{C}^2)/t$. Then in the 
	second split we decompose $\sigma_\mathrm{A}^2+\sigma_\mathrm{B}^2$ according to 
	the proportions $\boldsymbol{\omega}_2 = (\sigma_\mathrm{A}^2, \sigma_\mathrm{B}^2)/(\sigma_\mathrm{A}^2+\sigma_\mathrm{B}^2)$. 

	We use an HD prior where we apply base models $\boldsymbol{\omega}_1^0 = (0,1)$,
	which prefers C over A+B, and $\boldsymbol{\omega}_2^0 = (0,1)$, which prefers
	B over A. Due to the desire for shrinkage we apply PC priors and use 
	Theorem \ref{thm:method:dualSplit} with base model $\boldsymbol{\omega}_2^0$ to compute
	$\pi(\boldsymbol{\omega}_2)$. We define
	$\tilde{\boldsymbol{u}}_1 = \mathbf{A}_\mathrm{A} \boldsymbol{u}_\mathrm{A} +  \mathbf{A}_\mathrm{B} \boldsymbol{u}_\mathrm{B}$ 
	and $\tilde{\boldsymbol{u}}_2 = \mathbf{A}_\mathrm{C} \boldsymbol{u}_\mathrm{C}$. 
	Then if we condition on $\boldsymbol{\omega}_2$, the top
	split in Figure \ref{fig:intro:example:Ssh} compares 
	$\tilde{\boldsymbol{u}}_1|\boldsymbol{\omega}_2 \sim \mathcal{N}_n(\boldsymbol{0}, (\sigma_\mathrm{A}^2+\sigma_\mathrm{B}^2)(\omega_{2,1}\mathbf{A}_\mathrm{A}\mathbf{A}_\mathrm{A}^\mathrm{T}+\omega_{2,2}\mathbf{A}_\mathrm{B}\mathbf{A}_\mathrm{B}^\mathrm{T}))$
	and $\tilde{\boldsymbol{u}}_2 \sim \mathcal{N}_n(\boldsymbol{0}, \sigma_3^2\mathbf{A}_3\mathbf{A}_3^\mathrm{T})$, and the conditional prior $\pi(\boldsymbol{\omega}_1\vert \boldsymbol{\omega}_2 = \boldsymbol{\omega}_2^0)$ can be computed
	using Theorem \ref{thm:method:dualSplit}  with base model $\boldsymbol{\omega}_1^0$ conditional on $\boldsymbol{\omega}_2 = \boldsymbol{\omega}_2^0$. The joint prior is then
	$\pi(\boldsymbol{\omega}_1, \boldsymbol{\omega}_2) = \pi(\boldsymbol{\omega}_1|\boldsymbol{\omega}_2= \boldsymbol{\omega}_2^0) \pi(\boldsymbol{\omega}_2)$,
	and an appropriate prior is chosen for $\pi(t|\boldsymbol{\omega}_1, \boldsymbol{\omega}_2)$
	as described in Section \ref{sec:method:likelihood}.
	\demo
\end{exmp}

\subsection{Accounting for the likelihood}
\label{sec:method:likelihood}
Meaningful priors for the total latent variance $t$ 
depend on the likelihood and prior 
beliefs about the responses in the
specific application \citep{gelman2017prior}. We provide tools for expressing
scale-invariance for the variances of the random effects and 
the measurement error when the responses
are Gaussian, or shrinkage for the total latent variance of the random effects.

Under a Gaussian likelihood, the selection of the unit of measurement by the analyst affects
the sizes of the variances. However, when the 
residual variance $\sigma_\mathrm{R}^2$ is expected to be 
well-identified, we can define the prior on $t$ relative 
 to $\sigma_\mathrm{R}^2$ and shrink $t$ by preferring
to describe the total variance $V=t + \sigma_\mathrm{R}^2$ in the model by $ \sigma_\mathrm{R}^2$. This can be complemented by a scale-independent
Jeffreys' prior on $V$ to achieve a scale-invariant joint prior for the variance parameters.

\begin{result}[HD priors with Gaussian likelihoods]
	\label{thm:method:fullnormal}
	Assume an HD prior from Prior class \ref{thm:method:jointWeight} 
	is desired for an LGM with Gaussian responses with residual
	variance $\sigma_\mathrm{R}^2$. First select the prior on
	the decomposition of the total latent variance $t$. Then augment the tree 
	by an extra node on the top with  variance $V= t + \sigma_\mathrm{R}^2$. 
	The new top node has one branch with residual variance and the other branch
	is the subtree describing the latent model. 
	Let $\boldsymbol{\omega}_\mathrm{R} = (1-\sigma_\mathrm{R}^2/V, \sigma_\mathrm{R}^2/V)$ and
	assume shrinkage through a PC prior $\pi(\boldsymbol{\omega}_\mathrm{R} | \{\boldsymbol{\omega}_s = \boldsymbol{\omega}_s^0\}_{s = 1}^S)$ 
	with base
	model $\boldsymbol{\omega}_\mathrm{R}^0 = (0,1)$.

	If $V$ is assigned a scale-invariant prior, the full joint prior is
	\[
		\pi(V, \boldsymbol{\omega}_\mathrm{R}, \{\boldsymbol{\omega}_s\}_{s = 1}^S) \propto \pi(\boldsymbol{\omega}_\mathrm{R} | \{\boldsymbol{\omega}_s = \boldsymbol{\omega}_s^0\}_{s = 1}^S) \pi( \{\boldsymbol{\omega}_s\}_{s = 1}^S)/V, \quad V > 0, \boldsymbol{\omega}_\mathrm{R} \in \Delta^2,
	\]
	 and $\boldsymbol{\omega}_s \in \Delta^{l_s}$, where $l_s$ is the number of branches
	in split $s$, for $s = 1 , \ldots, S$.
	\begin{proof}
		The scale-invariant prior is $\pi(V | \boldsymbol{\omega}_\mathrm{R}, \{\boldsymbol{\omega}_s\}_{s = 1}^S) \propto 1/V$, and $\pi(\boldsymbol{\omega}_\mathrm{R}, \{\boldsymbol{\omega}_s\}_{s = 1}^S) = \pi(\boldsymbol{\omega}_\mathrm{R}| \{\boldsymbol{\omega}_s\}_{s = 1}^S)\pi(\{\boldsymbol{\omega}_s\}_{s = 1}^S)$
	\end{proof}
\end{result}

If the likelihood is binomial with a logit link function, 
a scale for the random effects is induced through
their effects on the odds-ratio. Similarily, for a Poisson likelihood with a log link
function, 
there is a scale for the random effects through their effects on the relative risk. 
In these cases, scale-invariance is not meaningful and we can induce shrinkage on the total variance of the random effects by using the 
PC prior for variance from \cite{simpson2017}.

\begin{result}[HD priors with shrinkage on latent variance]
	\label{thm:method:fullBP}
	Assume an HD prior from Prior class
	\ref{thm:method:jointWeight} is desired for an LGM where shrinkage on the total
	latent variance is appropriate. 
	First select the prior on the decomposition 
	of the total latent variance $t$. Then $t$ can be shrunk towards $0$ by
	a PC prior $\pi(t \vert \{\boldsymbol{\omega}_s\}_{s = 1}^S)$
	with base model $t_0 = 0$. This results in
	\[
		\pi(t, \{\boldsymbol{\omega}_s\}_{s = 1}^S) = \frac{\lambda}{2\sqrt{t}}\exp(-\lambda\sqrt{t})\pi( \{\boldsymbol{\omega}_s\}_{s = 1}^S), 
	\]
	$t > 0$, and $\boldsymbol{\omega}_i \in \Delta^{l_s}$, where $l_s$ is the number of branches
	in split $s$, for $s = 1 , \ldots, S$, and $\lambda > 0$ is a hyperparameter.
	\begin{proof}
		The conditional PC prior for $t$ with base model $t_0 = 0$ 
		is given by
		$\pi(t | \{\boldsymbol{\omega}_s\}_{s = 1}^S) = \lambda\exp(-\lambda\sqrt{t})/(2\sqrt{t})$, $t>0$ \citep{simpson2017}.
	\end{proof}
\end{result}

We illustrate how the hyperparameter can be selected 
by considering the prior on the total latent
variance in the case of a Binomial likelihood.

\begin{exmp}[Shrinking latent variance]
	\label{exmp:Nested4}
	Let $\logit(p) = \mu + x$, where
	$x \sim \mathcal{N}(0, t)$, for a $t >0 $, and 
	$\mu$ is considered fixed. The latent variance $t$
	is difficult to interpret directly due to the non-linear link function, 
	but we can interpret it through the effect on the odds-ratio,
	\(
		p/(1-p) = \exp(\mu) \exp(x).
	\)
	The hyperparameter $\lambda$ in Prior class \ref{thm:method:fullBP} can, for example,
	be set so that the relative change in the odds-ratio, $\exp(x)$, is between $1/2$ and $2$ with 
	probability $90\%$, $\mathrm{P}(1/2 < \exp(x) < 2)$ = 0.90.
	\demo 
\end{exmp}

\section{Case studies: Gaussian responses}
\label{sec:gaussian}


In this section we investigate the performance of HD priors compared to a set of commonly used standard priors for two simulation studies with Gaussian responses.

\subsection{Random intercept model}
\label{sec:gauss:RIM}

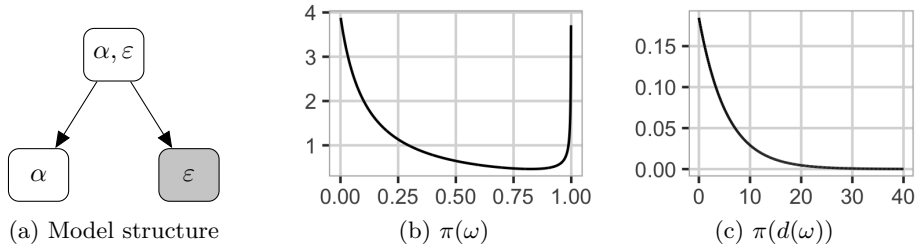
\begin{figure}[t!]
\centering
	\begin{subfigure}[b]{0.3\textwidth}
		\centering
		\begin{tikzpicture}[
			  rounded corners,
			  minimum height = 0.7cm, 
			  minimum width = 0.8cm,
			]
				\node[draw] (linpred) [fill = white] {$\alpha, \varepsilon$}; 
				\node[draw] (group1) [below of = linpred, xshift = -1cm, yshift = -0.6cm] {$\alpha$};
				\node[draw] (noise1) [below of = linpred, xshift = 1cm, yshift = -0.6cm, fill = basecolor] {$\varepsilon$};
				\path[->, every node/.style = {midway, auto = false, anchor = mid, yshift = 0.1cm}]
					(linpred) edge[] node[xshift = 0.15cm] {} (noise1)
					(linpred) edge[] node[xshift = -0.15cm] {} (group1);
		\end{tikzpicture}
		\caption{Model structure}
		\label{fig:gaussian:randint:graph}
	\end{subfigure}
	~
	\begin{subfigure}[b]{0.3\textwidth}
		\centering
		\includegraphics[width = \textwidth]{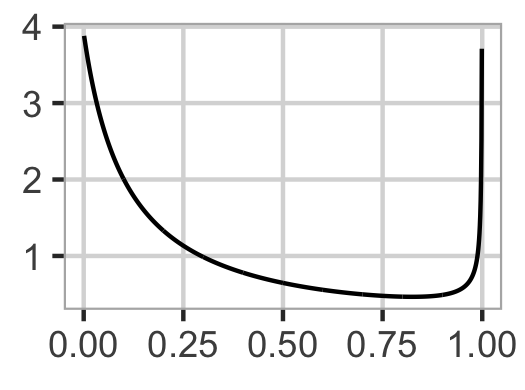}
		\vspace{-20pt}
		\caption{$\pi(\omega)$}
		\label{fig:gaussian:randint:priorw}
	\end{subfigure}
	~
	\begin{subfigure}[b]{0.3\textwidth}
		\centering
		\includegraphics[width = \textwidth]{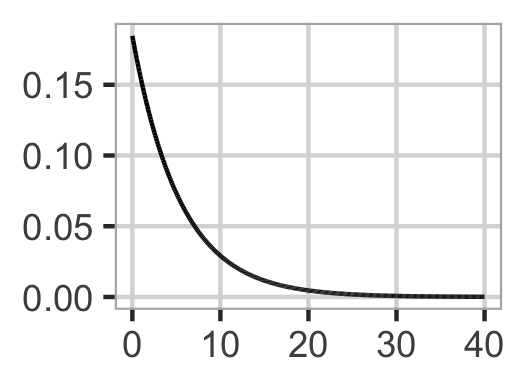}
		\vspace{-20pt}
		\caption{$\pi(d(\omega))$}
		\label{fig:gaussian:randint:prior_dist}
	\end{subfigure}
\caption{Model structure and prior for $\omega$ in the random intercept model with 10
individuals in each group and prior median $\omega_{\mathrm{m}} = 0.25$. The prior is
independent of the number of groups. \subref{fig:gaussian:randint:graph}) Tree structure, \subref{fig:gaussian:randint:priorw}) prior for $\omega$, and 
\subref{fig:gaussian:randint:prior_dist}) prior for distance $d(\omega)$.}
\label{fig:gaussian:randint:priorinfo}
\end{figure}

The \textit{random intercept model} is given by
$y_{i,j} = \alpha_i + \varepsilon_{i,j}$ for $j = 1, \dots, n_i$,
$i = 1, \dots, n_\mathrm{g}$,
where $n_i$ is the size of group $i$, and $n_\mathrm{g}$ is the number of groups. The 
random intercepts are i.i.d. Gaussian with variance $\sigma_\alpha^2$ and the residual
effects are i.i.d. Gaussian with variance $\sigma_\mathrm{R}^2$.
The total latent variance is $t = \sigma_\alpha^2$ and the total variance is
$V = \sigma_\mathrm{R}^2+\sigma_\alpha^2$. We introduce the
proportion of the total variance explained by the latent model 
$\omega = \sigma_\alpha^2/V$, and decompose $V$ as $\sigma_\alpha^2 = \omega V$
and $\sigma_\mathrm{R}^2 = (1-\omega)V$. We desire shrinkage towards the base model
$\omega^0 = 0$ and use an HD prior based on the tree structure in Figure 
\ref{fig:gaussian:randint:graph}, where the prior on $\omega$ is calculated using
Theorem \ref{thm:method:dualSplit} and we use the scale-invariant prior from
Prior class \ref{thm:method:fullnormal}. The specification of the hyperparameter of
the HD prior is done through the  median $\omega_{\mathrm{m}}$ of $\pi(\omega)$. 
The resulting prior for $\omega$ is shown in Figure \ref{fig:gaussian:randint:priorw} for
$\omega_\mathrm{m} = 0.25$ and the 
corresponding prior for the distance $d(\omega)$ discussed 
in Section \ref{sec:blocks:shrink} is shown in
\ref{fig:gaussian:randint:prior_dist}. Further details can be found in
Section \ref{sec:supp:randintback} of the Supplementary Materials.

The intraclass correlation (ICC) for the random intercept model
is given by 
$\sigma_{\alpha}^2/(\sigma_{{\mathrm{R}}}^2 + \sigma_{\alpha}^2)$,
which equals the weight parameter $\omega$. 
Thus the shrinkage 
of the ICC is 
completely 
controlled in the construction of the prior and 
expert knowledge about the ICC can be incorporated directly.
Further, $\omega$ can be linked to a generalised version of the coefficient of determination, $R^2$, suggested by \cite{gelmanbook}; see Section \ref{sec:supp:rsquared} in the Supplementary Materials for details. 

We use the \texttt{R}-package \texttt{RStan} \citep{rstan_package} to perform the 
inference for the simulation study. We use HD priors from 
Prior class \ref{thm:method:fullnormal} with shrinkage from PC priors on $\omega$ 
with hyperparameters $\omega_{\mathrm{m}} = 0.25$
(P-HD-25), $\omega_{\mathrm{m}} = 0.5$ (P-HD-50) and $\omega_{\mathrm{m}} = 0.75$ (P-HD-75),
and an HD prior from Prior class \ref{thm:method:fullnormal} where the PC prior is replaced by 
a Dirichlet prior on $(\omega, 1-\omega)$ (P-HD-D) with default hyperparameter. Additional priors are
Jeffreys' prior on the residual variance combined with different priors on
the random intercepts variance or standard deviation: the default INLA prior $\text{InvGamma}(1, 5 \times 10^{-5})$ (P-INLA), $\text{Half-Cauchy}(25)$ (P-HC), and 
$\text{PC}_{\mathrm{SD}}(3, 0.05)$  (P-PC). This gives seven joint priors.
Each scenario in the simulation study consists of $500$ datasets which are simulated from the
random intercept model 
for
$n_\mathrm{g} \in \{5, 10, 50\}$, and 
$10$, $50$, or varying number of individuals in each group. 
We select true values $\omega \in \{0.1, 0.25, 0.5, 0.75, 0.9\}$ and select true total variance
$V = 1$ in every scenario. 

We evaluate the performance of the different priors with respect to 
posterior inference for total variance $V$ and ICC $\omega$.
We use the bias of $\log(V)$ and $\logit(\omega)$, calculated using
the estimated median minus the true value, and the 80\% empirical coverage, 
found by counting the number of times the true value is contained in the 80\% 
equal-tailed credible interval.
We use the same settings for the call to the \texttt{stan} function for all priors and scenarios in the simulation study.
\texttt{RStan} reports a \textit{divergent transition} 
for each iteration of the MCMC sampler that runs
into numerical instabilities 
\citep{carpenter2017}.
In Figure \ref{fig:supp:gaussian:acc_randint} in the Supplementary Materials we
report the proportion of datasets that resulted in at most
0.1\% divergent transitions for each prior and scenario. This is used as a measure of 
stability of the inference scheme for each prior, and the dataset and prior combinations causing unstable inference are removed from the study.


\begin{figure}[t]
\centering
	\includegraphics[width = 1\textwidth]{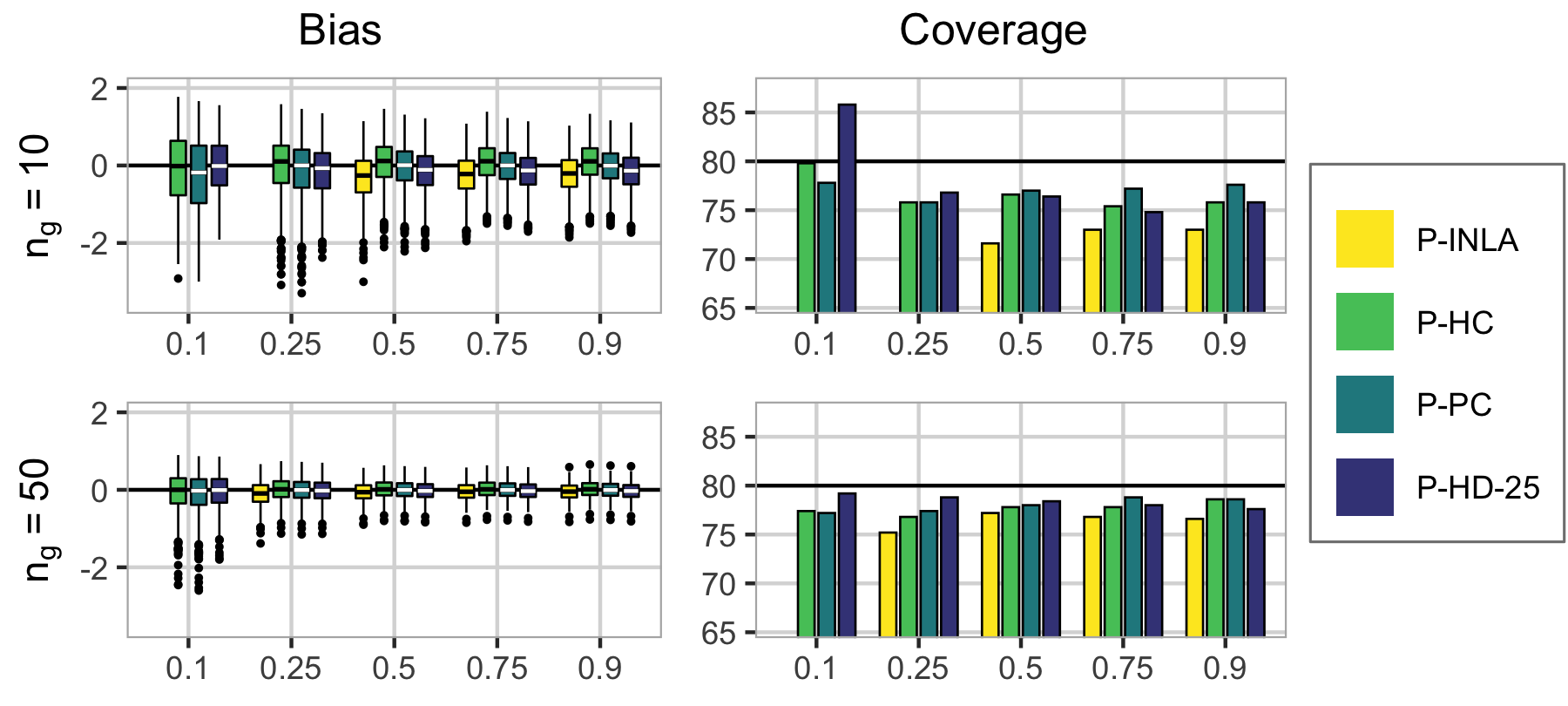}
	\caption{
	Results for $\text{logit}(\omega)$ for the random intercept simulation study.
	True value of $\omega$ shown on 
	the $x$-axis, the number of groups is shown on left-hand side, 
	and the group size is 10. 
	Results for P-INLA are only shown when it leads to stable inference.}
	\label{fig:gaussian:randint:randint_results}
\end{figure}


The results in Figure \ref{fig:gaussian:randint:randint_results} are for 
$n_g \in \{10, 50\}$ and group size $10$, and 
show that P-HD-25 performs at least as good in terms of bias and coverage of $\logit(\omega)$ as P-INLA, P-HC and P-PC. 
The magnitude of the bias decreases and the coverage approaches 80\% for all four priors when the number of groups
increases, which is expected as the amount of information about the parameters in the datasets increases. 
Figures \ref{fig:supp:gaussian:randint_ng1050_np10}--\ref{fig:supp:gaussian:randint_ng5_np50}
in the Supplementary Materials show that the HD priors perform at least as
good in terms of bias and coverage for $\logit(\omega)$ 
as P-INLA, P-HC and P-PC also for the other
combinations of the number of groups and group sizes, and that the same conclusions as
for $\logit(\omega)$ also holds for $\log(V)$.

Furthermore, Figures \ref{fig:supp:gaussian:randint_ng1050_np10}--\ref{fig:supp:gaussian:randint_ng5_np50} 
show that the behaviour of the four HD priors
is stable with respect to the choice
of $\omega_\mathrm{m}$  when group size is 10, and that P-HD-D performs worse than P-HD-25, P-HD-50 and P-HD-75 for all values of the true weight except 0.5. 
For 10 groups with two observations per group, the risk of overfitting is high
because low information about the parameters may lead to overestimating the weight
parameter and estimating spurious signals in the group effect.
In this setting, P-HD-25 leads to overfitting for true
weight equal to $0.1$, but underfitting
for true weight equal to $0.25$, $0.5$, $0.75$ and $0.9$. P-HD-50, P-HD-75 and P-HD-D
result in overfitting for true weight equal to $0.1$ and $0.25$, but underfitting
for true weight equal to $0.5$, $0.75$ and $0.9$. 
See Section \ref{sec:supp:simSmallGroup} in the Supplementary Materials for additional details.

Figure \ref{fig:supp:gaussian:acc_randint} shows that P-INLA is the only prior that is heavily affected by divergent transitions during the inference for scenarios with 10 or 50 groups. 
Part of the problem with 
P-INLA is that it results in a bi-modal posterior for $\sigma_{\alpha}^2$; see 
Figure \ref{fig:supp:gaussian:bimodal}. 
The new HD priors are preferred for the random intercept model
 due to their intuitive definition,
where the structure of the shrinkage is directly available in Figure 
\ref{fig:gaussian:randint:graph}, and interpretability of the parametrization
which aids prior elicitation.

\subsection{Latin square experiment}
\label{sec:gauss:LSE}

Consider an experiment where a latin square design \citep{hinkelmann1994} is used
to control for two nuisance sources of noise. For example, a field split into rows and
columns where different levels of strength of a new fertilizer is applied to each plot. We assume 
there are nine possible levels of the treatment so that a $9\times 9$ grid of
plots is necessary for a full latin square design. We focus on
random effects and exclude fixed effects from the model, and assume that
the responses can be modelled by 
\begin{equation}
	y_{i,j} = \alpha_{i} + \beta_{j} + \gamma_{k[i,j]} + \varepsilon_{i,j}, \quad i,j = 1, \ldots, 9,
	\label{eq:latin:model}
\end{equation}
where $\boldsymbol{\alpha} = (\alpha_1, \ldots, \alpha_9) \sim \mathcal{N}(\boldsymbol{0}, \sigma_\mathrm{r}^2\mathbf{I}_9)$ is an i.i.d. effect of row, $\boldsymbol{\beta} = (\beta_1, \ldots, \beta_9) \sim \mathcal{N}_{9}(\boldsymbol{0}, \sigma_\mathrm{c}^2\mathbf{I}_9)$ is an i.i.d. effect
of column, $\boldsymbol{\gamma} = (\gamma_1, \ldots, \gamma_9)$ is
the effect of the treatment,
$k[i,j]$ denotes the treatment assigned to row $i$ and column $j$, and
$\boldsymbol{\varepsilon} = (\varepsilon_{1,1}, \ldots, \varepsilon_{9,9}) \sim \mathcal{N}_{81}(\boldsymbol{0}, \sigma_\mathrm{R}^2\mathbf{I}_{81})$ is the residual noise. 

We believe that the effect of the treatment is ordered, and that the treatment
effect consists of a smooth signal of interest $\boldsymbol{\gamma}^{(1)} = (\gamma_1^{(1)}, \ldots, \gamma_9^{(1)})$ and random noise $\boldsymbol{\gamma}^{(2)} = (\gamma_1^{(2)}, \ldots, \gamma_9^{(2)})$ we have to control for. The signal is given a second-order random walk model described
by $\mathcal{N}_9(\boldsymbol{0}, \sigma_{\mathrm{RW2}}^2 \mathbf{Q}_{\mathrm{RW2}}^{-1})$, where $\sigma_\mathrm{RW2}^2$ is the variance and $\mathbf{Q}_{\mathrm{RW2}}^{-1}$ is a slight abuse of notation to describe
the intrinsic second-order random walk defined by the precision matrix $\mathbf{Q}_{\mathrm{RW2}}$, and the noise is
$\boldsymbol{\gamma}^{(2)}\sim \mathcal{N}_9(\boldsymbol{0}, \sigma_{\mathrm{t}}^2\mathbf{I}_9)$. We use 
the constraints $\sum_{i = 1}^9 \gamma_i^{(1)} = 0$ 
and
$\sum_{i = 1}^9 i\gamma_i^{(1)} = 0$ to remove the implicit intercept and linear effect, respectively.

We set the true standard deviations equal, 
$\sigma_\mathrm{r} = \sigma_\mathrm{c} = \sigma_\mathrm{t} = \sigma_\mathrm{R} = 0.1$, and
let the true effect of treatment be given by $x_i = C\left((i-5)^2-20/3\right)$, $i = 1, \ldots, 9$. We 
entertain 
three scenarios: $C=0$ for no effect of treatment (S1), $C = 0.05$ for  medium effect of
treatment (S2) and $C = 0.2$ for strong effect of treatment (S3). 
More details on the true treatment effect is included in Section \ref{sec:supp:latin} in the Supplementary materials, see especially Figure \ref{fig:supp:latin:trueTreat}.
We simulate $500$ datasets for each scenario and analyse them with four choices of priors.

\begin{figure}[]
    \centering
    \begin{subfigure}[t]{0.45\textwidth}
	    \begin{tikzpicture} %
		    \node[draw, rounded corners, minimum height = 0.7cm, minimum width = 0.8cm, fill = white] (top) {$\boldsymbol{\alpha}, \boldsymbol{\beta}, \boldsymbol{\gamma}^{(1)}, \boldsymbol{\gamma}^{(2)}, \boldsymbol{\varepsilon}$} ; %
		    \node[draw, rounded corners, left=of top, yshift=-1.35cm, xshift = 2.5cm, minimum height = 0.7cm, minimum width = 0.8cm] (latent) {$\boldsymbol{\alpha}, \boldsymbol{\beta}, \boldsymbol{\gamma}^{(1)}, \boldsymbol{\gamma}^{(2)}$} ; %
		    \node[draw, rounded corners, right=of top, yshift=-1.35cm, xshift = -1.5cm, minimum height = 0.7cm, minimum width = 0.8cm, fill = basecolor] (residual) {$\boldsymbol{\varepsilon}$} ; %
		    \node[draw, rounded corners, left=of latent, yshift=-1.35cm, xshift = 1.0cm, minimum height = 0.7cm, minimum width = 0.8cm, fill = basecolor] (gamma) {$\boldsymbol{\gamma}^{(1)}, \boldsymbol{\gamma}^{(2)}$};
		    \node[draw, rounded corners, right=of latent, yshift = -1.35cm, xshift = -2.46cm, minimum height = 0.7cm, minimum width = 0.8cm, fill = basecolor] (alpha) {$\boldsymbol{\alpha}$};
		    \node[draw, rounded corners, right=of latent, yshift = -1.35cm, xshift = -1.25cm, minimum height = 0.7cm, minimum width = 0.8cm, fill = basecolor] (beta) {$\boldsymbol{\beta}$};
		    \node[draw, rounded corners, left=of gamma, yshift = -1.35cm, xshift = 1.5cm, minimum height = 0.7cm, minimum width = 0.8cm] (gamma1) {$\boldsymbol{\gamma}^{(1)}$};
		    \node[draw, rounded corners, right=of gamma, yshift = -1.35cm, xshift = -1.5cm, minimum height = 0.7cm, minimum width = 0.8cm, fill = basecolor] (gamma2) {$\boldsymbol{\gamma}^{(2)}$};
		    
		    \edge {top} {latent} ; %
		    \edge {top} {residual} ; %
		    \edge {latent} {alpha};
		    \edge {latent} {beta};
		    \edge {latent} {gamma};
		    \edge {gamma} {gamma1};
		    \edge {gamma} {gamma2};
		    \path[->, every node/.style={midway}]
		   		(top) edge[] node[xshift=-0.3cm, yshift = 0.1cm] {} (latent)
		   		(top) edge[] node[xshift= 0.3cm, yshift = 0.1cm] {} (residual)
		   		(latent) edge[] node[xshift = -0.4cm, yshift = 0.05cm] {1/3} (gamma)
		   		(latent) edge[] node[xshift = -0.4cm, yshift = 0.05cm] {1/3} (alpha)
		   		(latent) edge[] node[xshift = 0.4cm, yshift = 0.05cm] {1/3} (beta)
		   		(gamma) edge[] node[xshift = -0.3cm, yshift = 0.1cm] {} (gamma1)
		   		(gamma) edge[] node[xshift = 0.3cm, yshift = 0.1cm] {} (gamma2);
	    \end{tikzpicture}
	    \caption{Original structure}
	    \label{fig:gauss:latinSquare:orig}
	\end{subfigure}
	\begin{subfigure}[t]{0.45\textwidth}
	    \begin{tikzpicture} %
		    \node[draw, rounded corners, minimum height = 0.7cm, fill = white] (top) {$\boldsymbol{\alpha}, \boldsymbol{\beta}, \boldsymbol{\gamma}^{(1)}, \boldsymbol{\gamma}^{(2)}, \boldsymbol{\varepsilon}$} ; %
		    \node[draw, rounded corners, left=of top, yshift=-1.35cm, xshift = 2.5cm, minimum height = 0.7cm] (latent) {$\boldsymbol{\alpha}, \boldsymbol{\beta}, \boldsymbol{\gamma}^{(1)}, \boldsymbol{\gamma}^{(2)}$} ; %
		    \node[draw, rounded corners, right=of top, yshift=-1.35cm, xshift = -1.5cm, minimum height = 0.7cm, minimum width = 0.8cm, fill = basecolor] (residual) {$\boldsymbol{\varepsilon}$} ; %
		    \node[draw, rounded corners, left=of latent, yshift=-1.35cm, xshift = 1.4cm, minimum height = 0.7cm, minimum width = 0.8cm, fill = basecolor] (gamma) {$\boldsymbol{\gamma}^{(1)}, \boldsymbol{\gamma}^{(2)}$};
		    \node[draw, rounded corners, right=of latent, yshift=-1.35cm, xshift = -1.4cm, minimum height = 0.7cm, minimum width = 0.8cm, fill = basecolor] (AB) {$\boldsymbol{\alpha}, \boldsymbol{\beta}$};
		    \node[draw, rounded corners, left=of AB, yshift = -1.35cm, xshift = 1.0cm, minimum height = 0.7cm, minimum width = 0.8cm, fill = basecolor] (alpha) {$\boldsymbol{\alpha}$};
		    \node[draw, rounded corners, right=of AB, yshift = -1.35cm, xshift = -1.0cm, minimum height = 0.7cm, minimum width = 0.8cm, fill = basecolor] (beta) {$\boldsymbol{\beta}$};
		    \node[draw, rounded corners, left=of gamma, yshift = -1.35cm, xshift = 1.5cm, minimum height = 0.7cm, minimum width = 0.8cm] (gamma1) {$\boldsymbol{\gamma}^{(1)}$};
		    \node[draw, rounded corners, right=of gamma, yshift = -1.35cm, xshift = -1.5cm, minimum height = 0.7cm, minimum width = 0.8cm, fill = basecolor] (gamma2) {$\boldsymbol{\gamma}^{(2)}$};
		    
		   	\path[->, every node/.style={midway}]
		   		(top) edge[] node[xshift=-0.3cm, yshift = 0.1cm] {} (latent)
		   		(top) edge[] node[xshift= 0.3cm, yshift = 0.1cm] {} (residual)
		   		(latent) edge[] node[xshift = -0.5cm, yshift = 0.05cm] {1/3} (gamma)
		   		(latent) edge[] node[xshift =  0.5cm, yshift = 0.05cm] {2/3} (AB)
		   		(gamma) edge[] node[xshift = -0.3cm, yshift = 0.1cm] {} (gamma1)
		   		(gamma) edge[] node[xshift = 0.3cm, yshift = 0.1cm] {} (gamma2)
		   		(AB) edge[] node[xshift = -0.4cm, yshift = 0.1cm] {1/2} (alpha)
		   		(AB) edge[] node[xshift = 0.4cm,  yshift = 0.1cm] {1/2} (beta);
	    \end{tikzpicture}
	    \caption{Dual-split structure}
	    \label{fig:gauss:latinSquare:new}
	\end{subfigure}
  \caption{Model structure for the latin square simulation study.
  		   Gray nodes indicate base models. $(1/3, 1/3, 1/3)$, $(1/3, 2/3)$,
  		   and $(1/2, 1/2)$ indicates that the base model for the split is a combination
  		   of the branches. 
		   \subref{fig:gauss:latinSquare:orig})
  		   Original, and
  		   \subref{fig:gauss:latinSquare:new}) alternative structure.}
  \label{fig:gauss:latinSquare}
\end{figure}
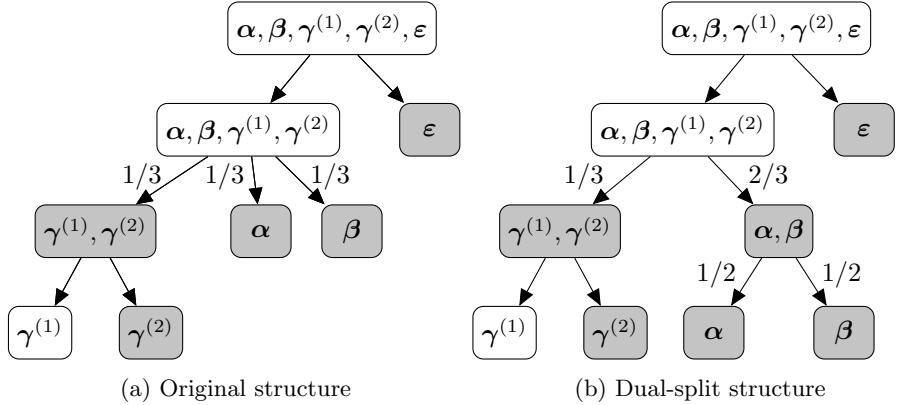

The three default priors used are Jeffreys' prior for
$\sigma_\mathrm{R}^2$ combined with $\text{InvGamma}(1, 5\times 10^{-5})$  for $\sigma_\mathrm{r}^2$,
$\sigma_\mathrm{c}^2$, $\sigma_\mathrm{t}^2$ and $\sigma_\mathrm{RW2}^2$ (P-INLA), or
$\text{Half-Cauchy}(25)$ (P-HC) or $\text{PC}_{\mathrm{SD}}(3, 0.05)$ (P-PC) for $\sigma_\mathrm{r}$,
$\sigma_\mathrm{c}$, $\sigma_\mathrm{t}$ and $\sigma_\mathrm{RW2}$. 
We select an HD prior from Prior class \ref{thm:method:fullnormal}
using the
model structure in Figure \ref{fig:gauss:latinSquare:orig}, where the triple split has a 
Dirichlet prior and the two other splits have PC priors (P-HD-D3).
We also decompose the
triple split 
into the two dual splits as shown in Figure 
\ref{fig:gauss:latinSquare:new}, and use a PC prior on all four splits according
to the shrinkage structure in the figure (P-HD-25). In all cases we use default
values for the hyperparameters. See Section \ref{sec:supp:multPC} in the
Supplementary Materials for more details on changing a triple split to two dual splits. 
Figures \ref{fig:supp:latin:orders:weight}, 
\ref{fig:supp:latin:orders:weightdiri}, \ref{fig:supp:gaussian:latin_order} and 
\ref{fig:supp:gaussian:latin_diri} in the Supplementary Materials
show that the implementation of the triple split has 
little influence on the targets of the analysis.

\begin{figure}[h!]
\centering
	\includegraphics[width = 1\textwidth]{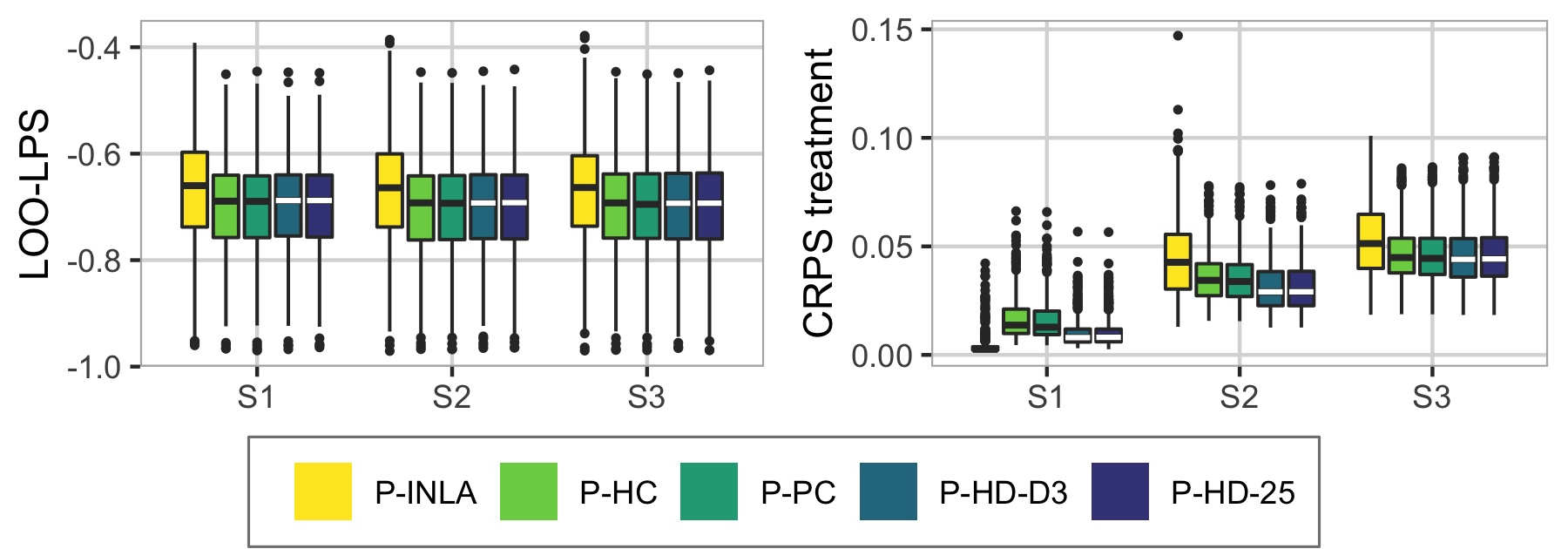}
	\caption{Results from the latin square experiment simulation study. 
	}
	\label{fig:gauss:latinSquare:results}
\end{figure}

The targets of the analysis are the posterior distribution of the structured treatment effect 
$\boldsymbol{\gamma}^{(1)}$ and the
model fit. The former will be assessed by the continuous rank probability score (CRPS) \citep{gneiting2007} and the latter by the 
leave-one-out log predictive score (LOO-LPS) $-\frac{1}{81} \sum_{i=1}^{81} \log \pi(y_i | \boldsymbol{y}_{-i})$. 
The CRPS is a proper scoring rule and given by 
\(
	\frac{1}{9}\sum_{i = 1}^9 \int_{-\infty}^{\infty} \left(F_i(x)-\mathbb{I}(x\geq x_i)\right)^2\mathrm{d}x,
\)
where $F_i$ is the cumulative distribution function for the 
posterior of $\gamma_i^{(1)}$, $x_i$ is the true effect of treatment $i$, and
$\mathbb{I}$ is the Heaviside function, and is estimated using the procedure of \cite{crps_package}. 
We report the proportion of datasets leading to no more than 0.1\% divergent transitions 
for each prior and scenario,
and use this as a measure on stability of the inference.
These numbers can be seen in Figure \ref{fig:supp:gaussian:acc_latin} in the Supplementary Materials, and show that 
all priors 
lead to similar stability.
The datasets leading to more than 0.1\% divergent transitions for one or more priors are removed 
from the study.

The main results from the simulation study are displayed in Figure 
\ref{fig:gauss:latinSquare:results}. 
Low LOO-LPS indicates good model fit and low CRPS indicates good predictive power
for the treatment effect. 
P-INLA gives a poorer model fit than the other priors, and with
respect to predictive power, the HD priors P-HD-D3 and P-HD-25
perform best for S2 and S3. The high predictive power of
P-INLA for S1 is due to the fact that P-INLA has a peak at low variance and
produces a posterior for the treatment effect with mean closer to zero and
lower variance. Overall, the HD prior performs well across all scenarios.
The results are stable to changes in the construction of the HD prior and the choice
of hyperparameters; see Section \ref{sec:supp:latin:results} in the Supplementary Materials for details.
The HD priors are preferable to the other priors because of their intuitive parametrization
and the interpretability of the \emph{a priori} assumptions placed on the joint prior
of the variance parameters.  Further, P-HD-D3 is
preferred to P-HD-25 since they perform similar and P-HD-D3 is more intuitive.


\section{Case studies: Binomial responses}
\label{sec:nonGauss}


In this section we study neonatal mortality counts arising from
complex surveys
through a simulation study, and 
show how to practically apply the HD priors.

\subsection{Background}
\label{sec:nonGauss:background}
Neonatal mortality
is an important indicator of health and well-being in a country and is included
in Goal 3.2 of the Sustainable Development Goals (SDGs) \citep{GA2015}, and mapping child
mortality is an important area of current research 
\citep{golding2017mapping,wakefield2018,li2019changes}. We
define neonatal mortality
as the rate of deaths within the first month of life per live
birth. An
important source of data for neonatal mortality is 
the nationally-representative household surveys performed by Demographic and Health Surveys (DHS).
The survey performed by DHS in 2014 in Kenya targets its 47 counties, which
is the relevant administrative level for health policies \citep{DHS2014}. 
The target of the simulation study in Section \ref{sec:nonGauss:simStudy}
and the analysis in Section \ref{sec:nonGauss:neonatal} is the spatial
heterogeneity in neonatal mortality in Kenya in the time period 2010 to
the time of the survey.

%
From the survey we can extract the number of 
live births, 
$b_{i,j,k}$, and the number of 
neonatal deaths,
$y_{i,j,k}$, in household $k$ in cluster $j$ in county $i$.
We also have an indicator $x_{i,j}$ specifying whether the cluster is rural ($0$) or
urban ($1$) and each household has an inclusion probability
$\pi_{i,j,k}$ 
of being
included in the
survey sample. 
%
%
%
See the Section \ref{sec:supp:additional} in the Supplementary
Materials for more background.

\subsection{Simulation study}
\label{sec:nonGauss:simStudy}

\begin{figure}
	\centering
	\begin{subfigure}[b]{0.45\textwidth}
		\centering
		\includegraphics[width = 1\textwidth]{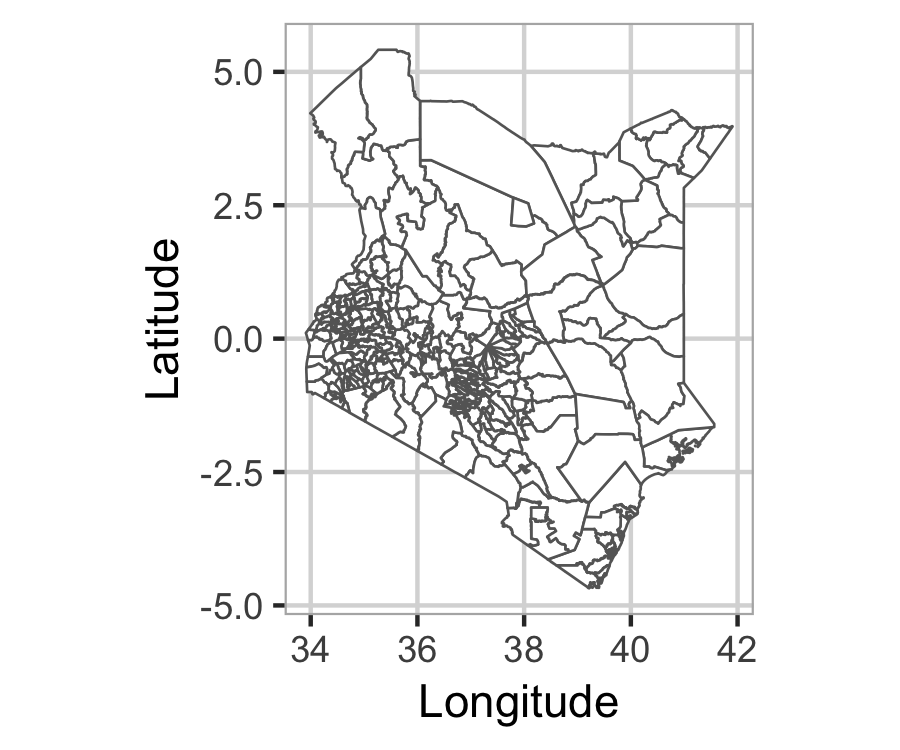}
		\caption{The 290 constituencies of Kenya.}
		\label{fig:nonGauss:simStudy:kenya_290}
	\end{subfigure}
	\begin{subfigure}[b]{0.45\textwidth}
		\centering
	    	\begin{tikzpicture}[
			  rounded corners,
			  minimum height = 0.7cm, 
			  minimum width = 0.8cm,
			]
	    		\node[draw] (linpred) [xshift = -1cm, yshift = -0.4cm, fill = white] {$\bm{u}, \bm{v}, \bm{\nu}$};
	    		\node[draw] (cluster) [below of = linpred, xshift = 1cm, yshift = -0.4cm] {$\bm{\nu}$};
	    		\node[draw] (BYM) [below of = linpred, xshift = -1cm, yshift = -0.4cm, fill = basecolor] {$\bm{u}, \bm{v}$};
	    		\node[draw] (besag) [below of = BYM, xshift = -1cm, yshift = -0.4cm] {$\bm{u}$};
	    		\node[draw] (iid) [below of = BYM, xshift = 1cm, yshift = -0.4cm, fill = basecolor] {$\bm{v}$};
	    		\path[->, every node/.style = {midway, auto = false, anchor = mid, yshift = 0.1cm}]
	    			(linpred) edge[] node[xshift = -0.1cm] {} (BYM)
	    			(linpred) edge[] node[xshift = 0.1cm] {} (cluster)
	    			(BYM) edge[] node[xshift = -0.1cm] {} (besag)
	    			(BYM) edge[] node[xshift = 0.1cm] {} (iid);
		\end{tikzpicture}
	    \vspace{15pt}
	    \caption{Model structure. Gray nodes indicate base models.}
	    \label{fig:nonGauss:simStudy:tree}
	\end{subfigure}
	\caption{Map and model structure for the Kenya neonatal simulation study.}
	\label{fig:nonGauss:simStudy:kenya}
\end{figure}
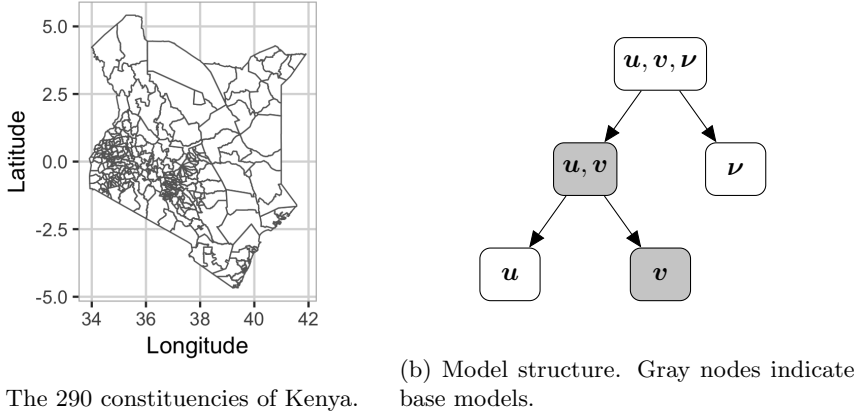

In this section we use the
$n = 290$ constituencies shown in Figure \ref{fig:nonGauss:simStudy:kenya_290}\footnote{
Preliminary investigations revealed that 47 counties provided too little
information to learn about model structure in the data. We instead use the 290 
constituencies of Kenya for the simulations study. 
}. 
We assume that $m_i = 6$ clusters are visited in constituency $i$, $i = 1, \ldots, n$, and
consider births $b_{i,j}$ and neonatal deaths $y_{i,j}$ 
in cluster $j$ in constituency $i$.
We assume that there are
$b_{i,j} = 25$ live births in each cluster and the outcomes are simulated according to the model
\(
	y_{i,j}\vert p_{i,j} \sim\text{Binomial}(b_{i,j}, p_{i,j})
\)
for 
\[
	\logit(p_{i,j}) = \eta_{i,j} = \mu + u_i + v_i + \nu_{i,j},\quad j = 1,\ldots, m_i, \ i = 1,\ldots, n,
\]
where $\mu$ is a joint intercept, $\boldsymbol{u} = (u_1, \ldots, u_{n})$ has a Besag
distribution with variance $\sigma_{\mathrm{B}}^2$ and a sum-to-zero constraint, 
$\boldsymbol{v} = (v_1, \ldots, v_{n}) \sim \mathcal{N}_n(\boldsymbol{0}, \sigma_{\mathrm{IID}}^2\mathbf{I}_{n})$, and $\boldsymbol{\nu} = (\nu_{1,1},\ldots, \nu_{n, m_{n}}) \sim \mathcal{N}_M(\boldsymbol{0}, \sigma_\mathrm{C}^2\mathbf{I}_M)$ with
$M = m_1+\ldots+m_{n} = 6 \cdot 290 = 1740$.

We use the structure for the prior shown in Figure \ref{fig:nonGauss:simStudy:tree}
to make an HD prior from Prior class \ref{thm:method:fullBP} with PC priors on all splits according to the base models indicated in
the figure (P-HD-25) and an HD prior from Prior class \ref{thm:method:fullBP} where a Dirichlet prior distributes variance to
the three model components (P-HD-D). In all cases, the splits have default hyperparameter values
and we select the hyperparameter
in the PC prior on total variance, $t = \sigma_{\mathrm{B}}^2 + \sigma_{\mathrm{IID}}^2 + \sigma_{\mathrm{C}}^2$, so that $\text{P}(t > 3) = 0.05$.
Further, we use $\text{InvGamma}(1, 5 \times 10^{-5})$ for $\sigma_{\mathrm{B}}^2$, $\sigma_{\mathrm{IID}}^2$ and $\sigma_{\mathrm{C}}^2$ 
(P-INLA), 
$\text{Half-Cauchy}(25)$
for $\sigma_{\mathrm{B}}$, $\sigma_{\mathrm{IID}}$ and $\sigma_{\mathrm{C}}$ (P-HC),
and the joint prior proposed in \cite{riebler2016} (P-PC), where $\sigma_\mathrm{B}^2$
and $\sigma_{\mathrm{IID}}^2$ has a PC prior of the type introduced in this paper with
$\mathrm{P}(\sigma_\mathrm{B}^2/(\sigma_\mathrm{B}^2+\sigma_{\mathrm{IID}}^2) < 0.5) = 2/3$ and $\sigma_{\mathrm{C}}^2$ 
is given an independent PC prior 
$\sigma_{\mathrm{C}} \sim \text{PC}_{\mathrm{SD}}(3, 0.05)$.

Based on the 
final report from the survey \citep{DHS2014} the estimated national level of neonatal mortality
is $0.022$ for 2010--2014, and 
we set $\mu = \logit(0.022)$. Further, we choose
$\sigma_{\mathrm{C}}^2 = 0.1$ and 
create five scenarios by combining this with
$\sigma_{\mathrm{IID}}^2 = \sigma_{\mathrm{B}}^2 = 0$ (S1), $\sigma_{\mathrm{IID}}^2 = 0.4$ 
and $\sigma_{\mathrm{B}}^2 = 0$ (S2), $\sigma_{\mathrm{IID}}^2 = \sigma_{\mathrm{B}}^2 = 0.2$ 
(S3), $\sigma_{\mathrm{IID}}^2 = 0.04$ and $\sigma_{\mathrm{B}}^2 = 0.36$ (S4), and 
$\sigma_{\mathrm{IID}}^2 = 0$ and $\sigma_{\mathrm{B}}^2 = 0.4$ (S5). 
We simulate 500 datasets 
for each scenario. 
The main targets of the simulation study are the structured part of the spatial heterogeneity through the
posterior of $\boldsymbol{u}$, the degree of structure in the spatial heterogeneity
through $\omega^{(2)} = \sigma_\mathrm{B}^2(\sigma_\mathrm{B}^2+\sigma_{\mathrm{IID}}^2)^{-1}$, and how well the underlying neonatal mortality is estimated through 
the posterior of the intercept $\mu$. 
The performance is assessed through the CRPS (see Section \ref{sec:gauss:LSE}) of $\bm{u}$,
the bias of the posterior median of $\omega^{(2)}$, and  the bias of the posterior
median and the coverage of the 80\% equal-tailed credible interval for $\mu$.
We use the proportion of datasets leading to at most 
0.1\% divergent transitions as a measure of stability in the inference, these numbers
can be seen in Figure \ref{fig:supp:nonGauss:acc_kenya} in the Supplementary Materials, and show that P-INLA 
leads to more unstable inference than the others.

\begin{figure}
	\centering
	\includegraphics[width = 1\textwidth]{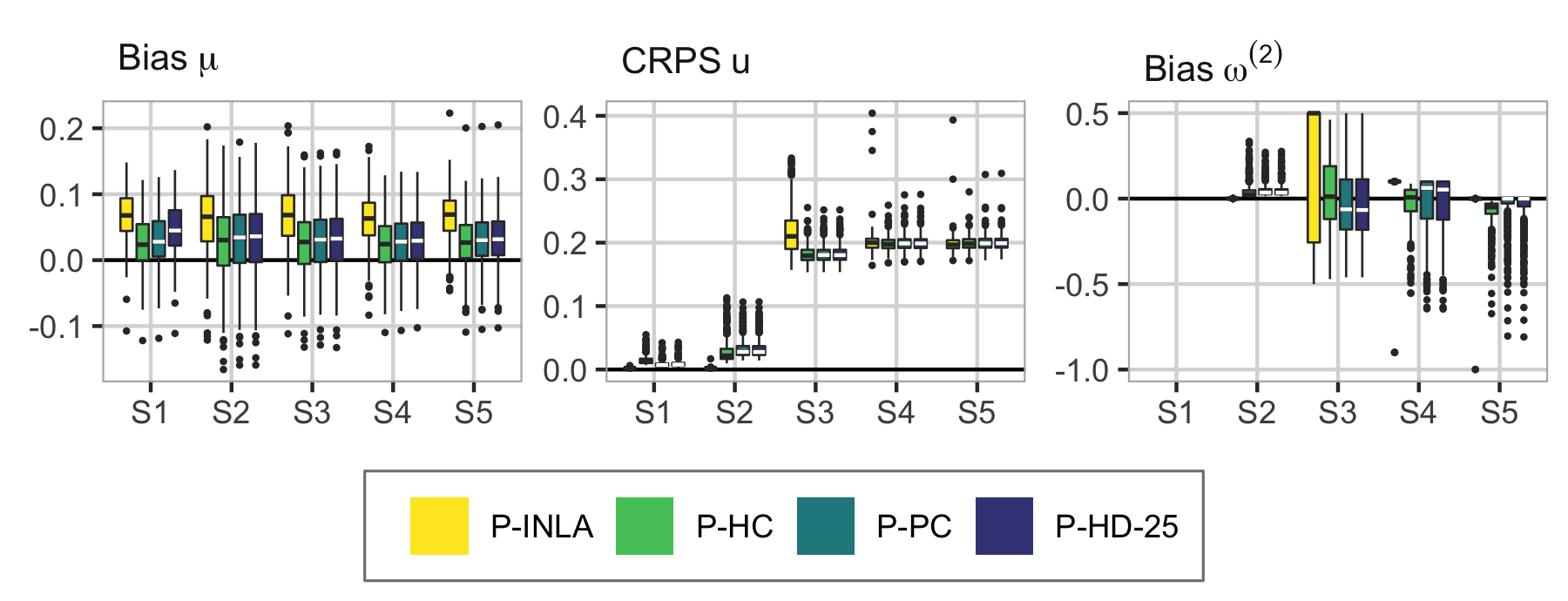}
	\caption{Main results from the Kenya neonatal mortality simulation study. Left to right: bias of the intercept $\mu$, CRPS of $\bm{u}$ and bias of $\omega^{(2)}$. Scenario shown on the x-axes.}
	\label{fig:nonGauss:simStudy:results}	
\end{figure}

Figure \ref{fig:nonGauss:simStudy:results} shows the main results from the simulation study. 
We drop datasets that cause more than 0.1\%
divergent transitions for at least one of the priors from each scenario.
All 
priors have a tendency to overestimate the intercept, with P-INLA doing worse than the others, 
%
P-INLA gives close to
exact estimates when 
the true value of 
$\omega^{(2)}$ is $0$ (in S2) and $1$ (in S5), but performs worse than
the other priors for S3 and S4.
Figure \ref{fig:supp:nonGauss:kenya_more_res} in 
the Supplementary Materials shows that
P-HD-25 performs better than P-HD-D except in S3 where 
the Dirichlet prior is closest to the truth, and that
$\omega^{(1)}$ 
tends to be underestimated under all the 
priors.
P-HD-25 is preferred because overall it performs at least as good as the other priors
P-HC and P-PC,
and P-HD-25 is an intuitive and well-behaved prior that takes
the hierarchical structure of the model into
account.

\subsection{Neonatal mortality in Kenya}
\label{sec:nonGauss:neonatal}

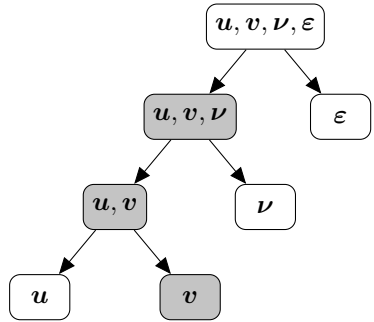
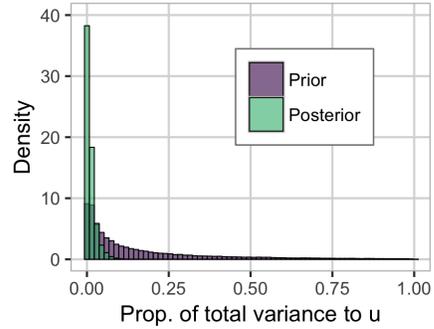
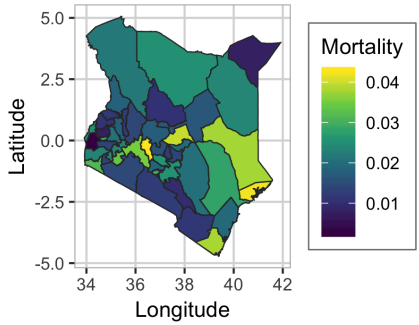
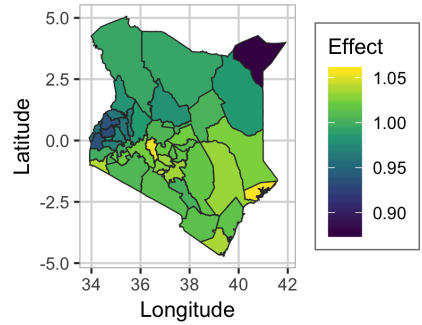
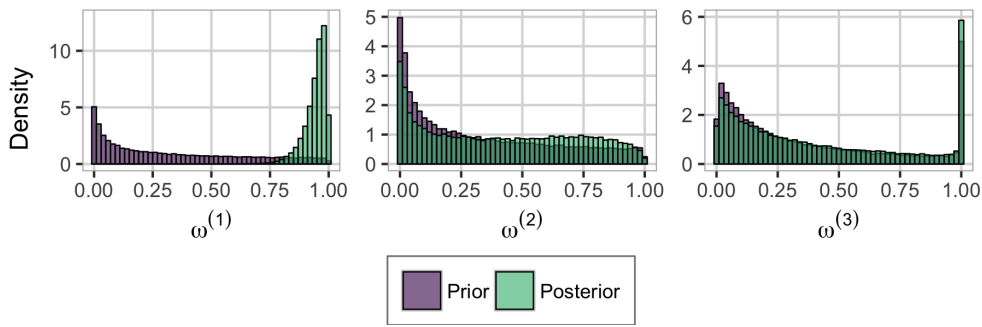
\begin{figure}
	\centering
	\begin{subfigure}[b]{0.48\textwidth}
		\centering
			\begin{tikzpicture}[
			  rounded corners,
			  minimum height = 0.6cm, 
			  minimum width = 0.8cm,
			]
	    		\node[draw] (linpred) [fill = white] {$\bm{u}, \bm{v}, \bm{\nu}, \bm{\varepsilon}$};
			\node[draw] (household) [below of = linpred, xshift = 1.0cm, yshift = -0.2cm] {$\bm{\varepsilon}$};
	    		\node[draw] (clustcount) [below of = linpred, xshift = -1.0cm, yshift = -0.2cm, fill = basecolor] {$\bm{u}, \bm{v}, \bm{\nu}$};
	    		\node[draw] (cluster) [below of = clustcount, xshift = 1.0cm, yshift = -0.2cm] {$\bm{\nu}$};
	    		\node[draw] (BYM) [below of = clustcount, xshift = -1.0cm, yshift = -0.2cm, fill = basecolor] {$\bm{u}, \bm{v}$};
	    		\node[draw] (besag) [below of = BYM, xshift = -1.0cm, yshift = -0.2cm] {$\bm{u}$};
	    		\node[draw] (iid) [below of = BYM, xshift = 1.0cm, yshift = -0.2cm, fill = basecolor] {$\bm{v}$};
	    		\path[->, every node/.style = {midway, auto = false, anchor = mid, yshift = 0.05cm}]
	    			(linpred) edge[] node[xshift = -0.2cm] {} (clustcount)
	    			(linpred) edge[] node[xshift = 0.2cm] {} (household)
	    			(clustcount) edge[] node[xshift = -0.2cm] {} (BYM)
	    			(clustcount) edge[] node[xshift = 0.2cm] {} (cluster)
	    			(BYM) edge[] node[xshift = -0.2cm] {} (besag)
	    			(BYM) edge[] node[xshift = 0.2cm] {} (iid);
		\end{tikzpicture}
		\vspace*{10pt}
		\caption{Model structure.}
		\label{fig:nonGauss:neonatal:tree}
	\end{subfigure}
	\begin{subfigure}[b]{0.48\textwidth}
		\centering
		\includegraphics[width = .90\textwidth]{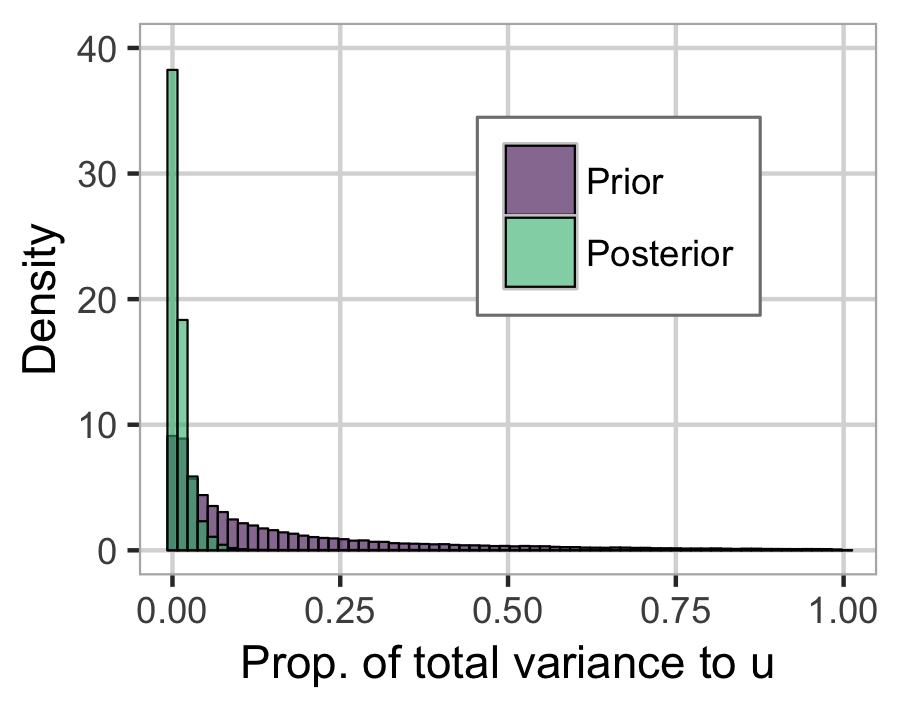}
		\caption{Variance of $\bm{u}$ relative to total variance.}
		\label{fig:nonGauss:neonatal:totweight}
	\end{subfigure}
	\begin{subfigure}{0.48\textwidth}
		\centering
		\includegraphics[height = 4.7cm]{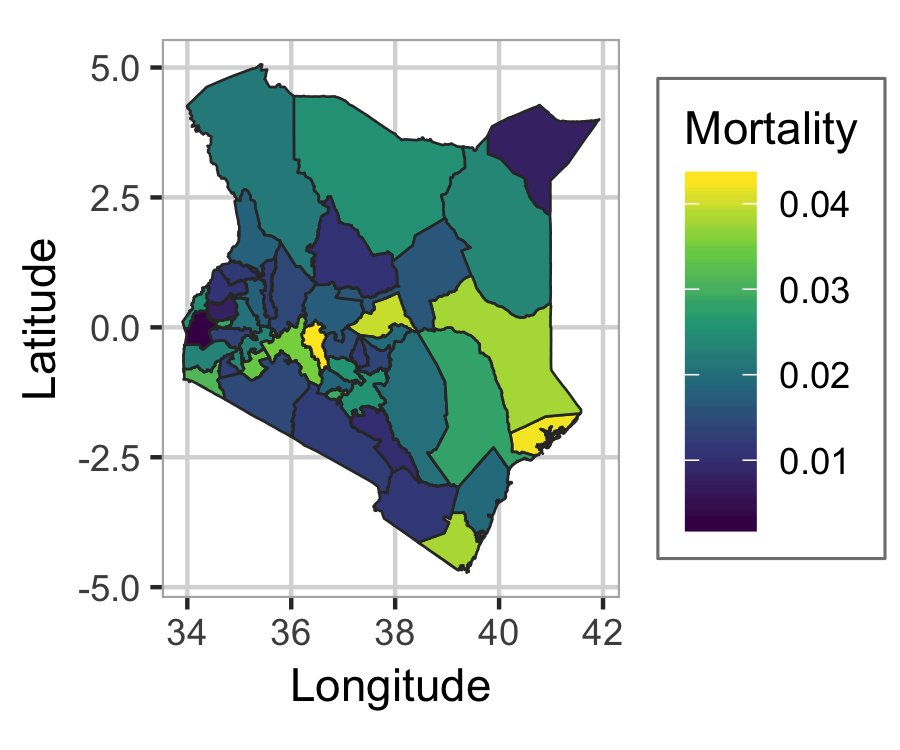}
		\caption{Weighted average of neonatal mortality.}
		\label{fig:nonGauss:neonatal:kenya_47_weighted}
	\end{subfigure}
	\begin{subfigure}{0.48\textwidth}
		\centering
		\includegraphics[height = 4.7cm]{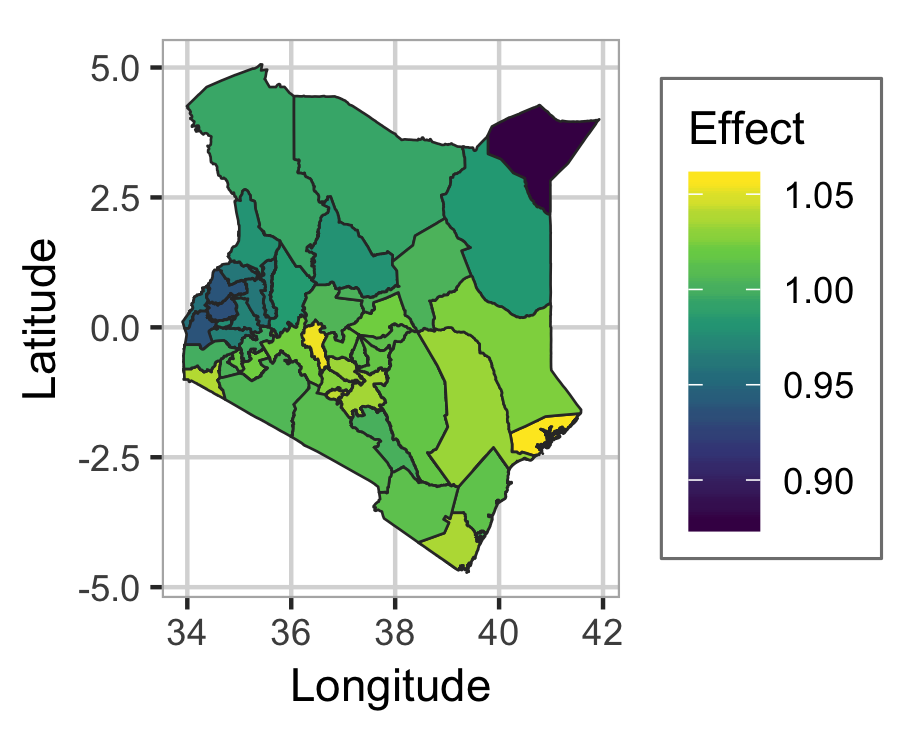}
		\caption{Posterior median of $e^{\bm{u}}$.}
		\label{fig:nonGauss:neonatal:kenya_47_posterior}
	\end{subfigure}
	\begin{subfigure}{1\textwidth}
		\centering
		\includegraphics[width = 0.97\textwidth]{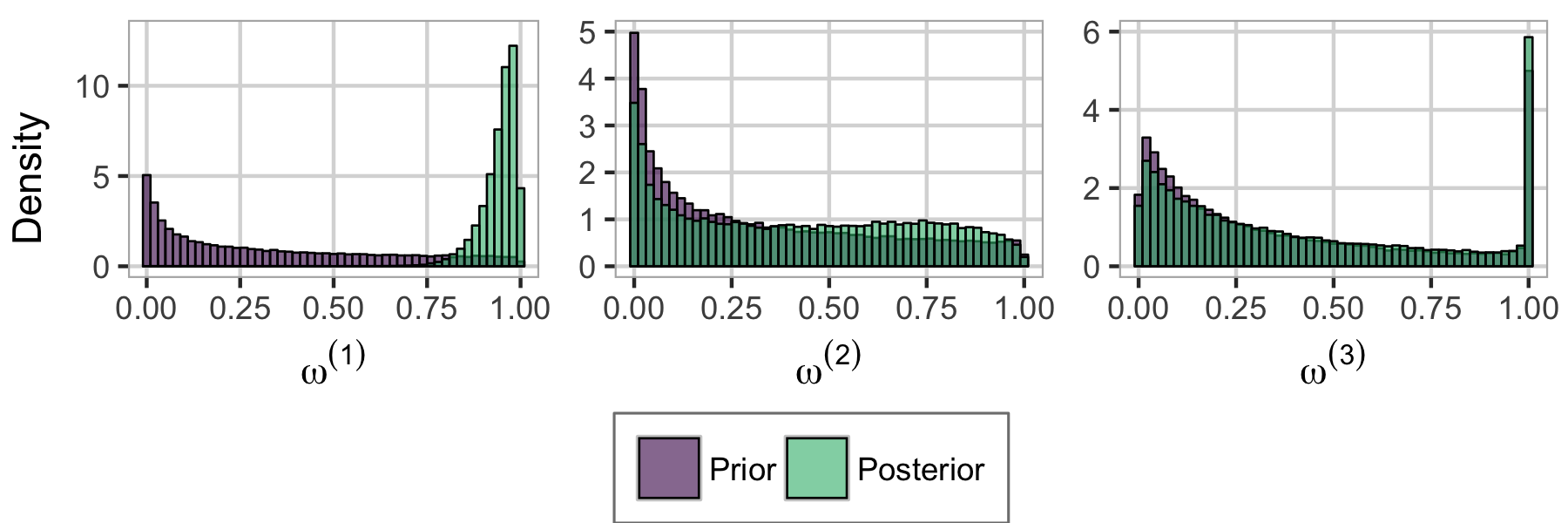}
		\caption{The priors and posteriors for the 
			proportion of household variance to total variance of the random effects $\omega^{(1)}$,
			the proportion of cluster variance to cluster- and
		         household-level variance $\omega^{(2)}$, and the
		         proportion of structured spatial variance to total between-county variance $\omega^{(3)}$.}
		\label{fig:nonGauss:neonatal:posteriors}
	\end{subfigure}
	\caption{Description of model structure, map of observed mortality, and 
		     results for neonatal mortality in Kenya.
	}
	\label{fig:nonGauss:neonatal:kenya}
\end{figure}

This section follows the notation introduced in Section \ref{sec:nonGauss:background}.
The survey consists of 13183 households with one or more live births, 
distributed over 
1593 clusters that are distributed over $n = 47$ counties. 
In total 
there are 376 deaths among 17664 children. Figure \ref{fig:nonGauss:neonatal:kenya_47_weighted} 
shows the counties and the weighted neonatal mortality by the inverse inclusion probabilities, 
and it is unclear if there is a structured spatial pattern.
The neonatal mortality is assumed to follow a survival model with constant hazard through the first month of
life, and we use 
a latent Gaussian model with a binomial likelihood,
\(	
	y_{i,j,k} | b_{i,j,k}, p_{i,j,k} \sim \text{Binomial}(b_{i,j,k}, p_{i,j,k}),
\)
a logit link function, and a linear latent Gaussian model
\begin{equation}
	\eta_{i,j,k} = \text{logit}(p_{i,j,k}) = \mu + x_{i,j}\beta + u_i + v_i + \nu_{i,j} + \varepsilon_{i,j,k},
	\label{eq:kenya_fullmodel}
\end{equation}
where $\mu$ is an overall intercept, $\beta$ is the effect of urban, $\boldsymbol{u}$ is a Besag model with
variance $\sigma_{11}^2$, $\boldsymbol{v}$ is a Gaussian i.i.d. effect of county
with variance $\sigma_{12}^2$, $\boldsymbol{\nu}$ is a Gaussian i.i.d. effect of
cluster with variance $\sigma_2^2$, and $\boldsymbol{\varepsilon}$ is a Gaussian
i.i.d. effect of household with variance $\sigma_3^2$. In this model, $\boldsymbol{u}$ and $\boldsymbol{v}$
provide structured and unstructured, respectively, between-county variation,
$\boldsymbol{\nu}$ provides between-cluster variation, and $\boldsymbol{\varepsilon}$
provides within-cluster variation. The Besag effect has a sum-to-zero constraint to make
the overall intercept identifiable. The random effects of cluster and household are
necessary to account for the dependence induced between sampled households due to the 
clustering in the sampling design. 
We assume that there is no difference between the effect of urbanicity between different
counties.

The model has four variance parameters that must be assigned a joint prior. The first
step is to choose the tree structure. For simplicity's sake,
the alternatives to the full model \eqref{eq:kenya_fullmodel}
we would entertain are first $\eta_{i,j,k} = \mu + x_{i,j}\beta + v_i$, then we would add $u_i$, so $\nu_{i,j}$, and at last $\varepsilon_{i,j,k}$.
%
%
We prefer coarser unstructured effects over finer unstructured effects since we would
like to explain the data at a coarser level if possible, and we prefer the unstructured
spatial effect over the structured spatial effect since we want to reduce the risk of
estimating spurious spatial signals. 
This gives the nested tree structure in Figure \ref{fig:nonGauss:neonatal:tree} where
the household effect, cluster effect and Besag effect are sequentially split off from the total 
latent variance.
We construct an HD prior based on the tree structure with PC priors with default hyperparameter
values for the splits, and induce
shrinkage on the total latent variance as in Prior class
\ref{thm:method:fullBP} with a PC prior
where $\mathrm{P}(\text{Total variance} > 11.296) = 0.05$. 
This corresponds
to \emph{a priori} equal-tailed 90\% credible interval of $(0.1, 10)$ for the effect of the random
effects on the odds-ratio, $\exp(u_i + v_i + \nu_{i,j} + \varepsilon_{i,j,k})$.
This
allows for 
high variation in the data and is used because the data is observed at the 
household level.
The splits in Figure \ref{fig:nonGauss:neonatal:tree} are given PC priors with
default hyperparameters and bases models as indicated in the figure.


The model is parameterized by total standard deviation $\sigma_\mathrm{T}$, and proportion
of household variance to total variance of the random effects $\omega^{(1)}$, proportion of
cluster variance to the sum of cluster and county variance $\omega^{(2)}$, and the proportion of
structured spatial variance to county variance $\omega^{(3)}$. The priors and posteriors of the proportions
$\omega^{(1)}$, $\omega^{(2)}$ and $\omega^{(3)}$ are shown in Figure \ref{fig:nonGauss:neonatal:posteriors}. The
total standard deviation has a posterior median of $1.47$, and the prior and posterior 
can be seen in Figure \ref{fig:supp:nonGauss:totstd} in the Supplementary Materials.
The results
show that the data only weakly informs about the proportion of structured to unstructured
spatial effects, which indicates that the data provide no strong evidence in favor of or against a structured
spatial effect. Also the posterior of $\omega^{(2)}$ is similar to the prior, but there
is a strong signal in the posterior of $\omega^{(1)}$ that there is 
non-negligible
household-level
dependence. A plausible explanation for the weak signals in $\omega^{(2)}$ and $\omega^{(3)}$ is that there
is substantial noise coming from high variance in the household-level random effect 
and weak information from the Binomial likelihood due to few successes and few numbers of trials.

As shown in Figure \ref{fig:nonGauss:neonatal:totweight} the
proportion of the total latent variance attributed to the structured spatial effect is low and
the posterior median is
0.56\%. The estimated spatial
effect in Figure \ref{fig:nonGauss:neonatal:kenya_47_posterior} only explains a small part
of the variation seen in the observed data in Figure \ref{fig:nonGauss:neonatal:kenya_47_weighted}.
One should be careful to draw conclusions about spatial variation based on Figure 
\ref{fig:nonGauss:neonatal:kenya_47_posterior} because
the data is only weakly informative about the split between the structured and 
the unstructured spatial random effects $\omega^{(3)}$, and there is only weak evidence
for the spatial effect being 
different from 0 as shown in Figure \ref{fig:supp:nonGauss:prob0}
in the Supplementary Materials.
The fact that the comparisons of priors and posteriors for $\omega^{(2)}$ and
$\omega^{(3)}$ directly informs about the weak signal in the data is an advantage of the parametrization
through proportions of variance, and a strong argument for setting priors on
$\omega^{(2)}$ and $\omega^{(3)}$ rather than independent priors on the variance of each effect since the resulting
posteriors for $\omega^{(2)}$ and $\omega^{(3)}$ are strongly dependent on the resulting
implicit priors for $\omega^{(2)}$ and $\omega^{(3)}$.

One could argue for other splits in the tree in Figure 
\ref{fig:nonGauss:neonatal:tree} such as preferring 
finer level effects to coarser level effects because one does not want to estimate spurious cluster-level or county-level
effects, but the key point of this application is that it is easy to set up the prior based on \emph{a priori} assumptions 
and the assumptions are available to other scientists at a glance. 
With the traditional approach of independent priors, the resulting prior on the total variance of
the random effects and the distribution of this total variance to the different random effects is obfuscated. Furthermore,
if expert knowledge indicates that stronger relative shrinkage of the variances than the default setting is needed, the medians of the conditional
priors for $\omega^{(1)}$, $\omega^{(2)}$ and $\omega^{(3)}$ can be reduced.

\section{Discussion}
\label{sec:discussion}



Independent priors for the variance parameters in a BHM result
in an implicit prior on the total variance of the
random effects, $t$, and the attribution of $t$ to the random effects.
Additive models are typically built in a modular fashion,
but these implict priors are not consistent with respect to
adding or removing random effects. In the case of Gaussian responses,
both the prior for $t$ and the prior for $t$ relative to the size of the
residual variance change. The proposed HD priors overcomes these
shortcomings, and respect the defined model structure and 
are consistent for $t$ and the attribution
of $t$ to the different random effects for different selections
of random effects.

The HD priors admit a visual representation
through trees that allow transparent communication of the assumptions
made in constructing the priors and facilitate discussion around the assumptions. 
The tree clearly specifices
where shrinkage has been applied, and in some cases lead
to more intuitive parametrization that is more suitable for elicitation of priors.
For the random intercept model, the tree-based hierarchical variance decomposition
leads to a parameterisation in terms of $t$ and the ICC. 
A prior on these parameters is more interpretable than
separate priors on the group variance and individual variance, which obfuscates the joint
effect of the priors. 
The increased interpretability of joint priors compared to
independent priors 
addresses concerns raised about transparency for
point processes where prior sensitivity is a major concern 
\citep{sorbye2018prior}.

The mix of robust PC priors for shrinkage and simple Dirichlet priors for expressing
ignorance, allows principled priors that respect the relative complexity
of the random effects when shrinkage is necessary, and intuitive exchangeability
when no random effects are preferred or no model structure is apparent. 
The simulation studies show that this approach
performs better than a completely unstructured approach with a Dirichlet prior
attributing $t$ to the different random effects, but that Dirichlet priors
perform well for subgroups of the random effects where there is no nested structure 
or difference in complexity.


HD priors with default settings for the hyperparameters performs well, but
there are corner cases like no treatment
effect in the latin square experiment and no structured spatial effect for the binomial
data, which are best handled by the default \texttt{INLA} prior. However, this prior has a 
peak in the prior distribution for low variances and generally performs surprisingly bad.
The HD priors perform comparable to component-wise PC priors and separate half-cauchy
priors for the marginal variances. The main benefit of the HD priors over other
default priors is their combination of intuitive graphical representation with robust 
inference that behaves well across a range of different scenarios.

The calculation of PC priors is more complex in the context of correlation parameters,
but multivariate PC priors have been developed for more complex random effects
such as autoregressive processes \citep{sorbye2017penalised} and spatial Mat\'ern models
\citep{fuglstad2018}. These can be integrated into the HD prior framework by
first defining priors on the correlation parameters,
and then 
constructing the joint prior for the variance parameters
with the correlation parameters fixed to reasonable values. 
This follows the pragmatic mindset of Assumption \ref{assume:method:cond}
of producing priors that are computationally feasible, intuitive and practically useful.

A key focus for future work is 
to exploit sparsity in the precision matrices of the random effects. This is important
when shrinkage is desired through PC priors because many models such
as random walks, Besag models, and Gaussian random fields \citep{lindgren2011explicit}
have dense covariance matrices, but can be expressed through sparse precision matrices. 
Assume that the total variance is split between random effects with sparse precision 
matrices $\mathbf{Q}_1$ and $\mathbf{Q}_2$, where $\mathbf{Q}_1$ corresponds to the base 
model. Let $0 < \omega < 1$, then the KLD used in Theorem \ref{thm:method:dualSplit}
consists of the trace of
$\mathbf{Q}_1[(1-\omega) \mathbf{Q}_1^{-1}+\omega \mathbf{Q}_2^{-1}]$, which can
be computed quickly through the techniques in \citet[Section 12.1.7.10]{handbook2010}, 
and the determinant
$\det[\mathbf{Q}_1[(1-\omega) \mathbf{Q}_1^{-1}+\omega \mathbf{Q}_2^{-1}] = \det [ (1-\omega) \mathbf{Q}_2+\omega \mathbf{Q}_1] (\det [\mathbf{Q}_2])^{-1}$, 
which can be computed quickly through Cholesky factorizations.

We aim to further broaden the advantages of the HD priors in the future by
constructing a joint prior for the variance parameters and the fixed effects.
However, this will require re-thinking of the concept of total latent variance as
it is the values of the coefficients of the fixed effects and not their variance 
that determines the amount of variance they explain. Instead of starting
with the concept of marginal variances, it is natural to begin with
the classical concept of explained variance and use ideas from block-wise 
g-priors \citep{som2014block} to distribute variance inside a group of
covariates. In a multilevel model this would connect
the attribution of explained variance to different levels
to generalised coefficients of determinations. Additionally, towards
non-parametric regression by including a combination of a
linear effect of a covariate and a smooth effect of a covariate, and explicitly putting a 
prior on the degree
of non-linearity \citep[Section 7]{simpson2017}. However, there are still
open questions and this addition is outside the scope of this paper.

The choice of tree structure for HD priors should be
guided by the application at hand, for example, by considering the relative complexity
of the random effects. When expert knowledge is available, 
the default values for the hyperparameters should
be replaced by values elicited based on expert knowledge.
We believe that the advantages of the HD priors over independent priors
mean that they should be used as the default option in software for Bayesian analysis.
However, it is necessary to make the selection and computation of HD prior for a specific problem easier for analysts. We plan to address this 
by providing a separate \texttt{R} package, which is compatible with \texttt{INLA}, that provides a graphical user interface for selecting
the tree structure and selecting priors for the splits, and has the option to
pre-compute priors for use in \texttt{RStan}. This will allow analysts
to experiment with different \emph{a priori} assumptions and produce graphical figures
that summarize their assumptions and can be communicated to fellow scientists. This will
encourage
transparancy and clarity in \emph{a priori} assumptions in the scientific community.

\renewcommand{\thefigure}{S\arabic{section}.\arabic{figure}}
\renewcommand{\thesection}{S\arabic{section}}
\graphicspath{{./supplementary/img/}}

\clearpage
\newpage
\section*{Appendix A: Supplementary materials}
\setcounter{section}{0}
\setcounter{thm}{0}

\section{Proofs}
\subsection{Theorem 3.1}

\begin{thm}[Prior for the case $N = 2$]
	\label{thm:method:dualSplit} 
	Let $\boldsymbol{u}_1$ and $\boldsymbol{u}_2$ be random effects of an LGM that
	enter the linear predictor 
	through $\mathbf{A}_1\boldsymbol{u}_1\sim\mathcal{N}_n(\boldsymbol{0}, \sigma_1^2\tilde{\Sigma}_1)$ and
	$\mathbf{A}_2\boldsymbol{u}_2\sim\mathcal{N}_n(\boldsymbol{0}, \sigma_2^2\tilde{\Sigma}_2)$.
	Assume that $\tilde{\Sigma}_1+\tilde{\Sigma}_2$ is non-singular\footnote{If this
	were not the case, some elements of the sum of $\mathbf{A}_1\boldsymbol{u}_1$
	and $\mathbf{A}_2\boldsymbol{u}_2$ would be exactly equal and
	we would choose a subset of maximal size so that $\tilde{\Sigma}_1+\tilde{\Sigma}_2$ was
	non-singular for comparing the effects
	of $\mathbf{A}_1\boldsymbol{u}_1$ and $\mathbf{A}_2\boldsymbol{u}_2$.}. Let $\omega = \sigma_2^2/(\sigma_1^2+\sigma_2^2)$ and 
	$\Sigma(w) = (1-\omega)\tilde{\Sigma}_1 + \omega\tilde{\Sigma}_2$. Then the distance from 
	the base model $\Sigma(\omega_0)$ to the alternative model $\Sigma(\omega)$ is given by
	$d(\omega) = \sqrt{\mathrm{tr}(\Sigma(\omega_0)^{-1}\Sigma(\omega))-n-\log\vert\Sigma(\omega_0)^{-1}\Sigma(\omega)\vert}$ for $0 \leq \omega_0 \leq 1$.

	The PC prior for $\omega$ with base model $\omega_0 = 0$ is
	\[
		\pi(\omega) =  \begin{cases} \frac{\lambda \left\vert d'\left(\omega\right)\right\vert}{1-\exp(-\lambda d(1))} \exp\left(-\lambda d\left(\omega\right)\right),  & \text{$0 < w < 1$, $\tilde{\Sigma}_1$ non-singular}, \\
		\frac{\lambda}{2\sqrt{\omega}(1-\exp(-\lambda))}\exp(-\lambda \sqrt{\omega}), & \text{$0 < \omega < 1$, $\tilde{\Sigma}_1$ singular},
		\end{cases}
	\]
	where $\lambda > 0$ is the hyperparameter. We suggest to set $\lambda$ so that the
	median is $\omega_\mathrm{m} = 0.25$. 

	For base model $0 < \omega_0 < 1$,  the PC prior whose median is equal to
	$\omega_0$ is
	\[
		\pi(\omega) = \begin{cases}
					\frac{\lambda \left\vert d'\left(\omega\right)\right\vert}{2[1-\exp(-\lambda d(0))]} \exp\left(-\lambda d\left(\omega\right)\right), &  0 < \omega < \omega_0, \\
					\frac{\lambda \left\vert d'\left(\omega\right)\right\vert}{2[1-\exp(-\lambda d(1))]} \exp\left(-\lambda d\left(\omega\right)\right), &  \omega_0 < \omega < 1,

		\end{cases}
	\]
	where $\lambda>0$ is a hyperparameter. We suggest to set $\lambda$ so that
	\[
		\mathrm{P}(\mathrm{logit}(1/4) + \mathrm{logit}(\omega_0) < \mathrm{logit}(\omega) < \mathrm{logit}(\omega_0)+\mathrm{logit}(3/4)) = 1/2.
	\]
	
	Base model equal to $\omega_0 = 1$ follows directly by reversing the roles of $\boldsymbol{u}_1$ and $\boldsymbol{u}_2$.
	
\end{thm}
\label{sec:supp:proof1}
\setcounter{figure}{0}

\noindent\textbf{Proof:}\\
First, note that since $\tilde{\Sigma}_1$ and $\tilde{\Sigma}_2$ are positive semi-definite
and $\tilde{\Sigma}_1 + \tilde{\Sigma}_2$ is non-singular, 
$\Sigma(\omega) = (1-\omega)\tilde{\Sigma}_1 + \omega\tilde{\Sigma}_2$ is positive definite
for $0 < \omega < 1$. This follows from the fact that 
$\tilde{\Sigma}_1 + \tilde{\Sigma}_2$ is non-singular means that
$\boldsymbol{v}^\mathrm{T}(\tilde{\Sigma}_1 + \tilde{\Sigma}_2)\boldsymbol{v} \neq 0$ for 
$\boldsymbol{v}\in\mathbb{R}^n$ and $\boldsymbol{v} \neq \boldsymbol{0}$, where $n$ is the dimension
of $\tilde{\Sigma}_1$, which implies
that either $\boldsymbol{v}^\mathrm{T}\tilde{\Sigma}_1\boldsymbol{v} > 0$ or
$\boldsymbol{v}^\mathrm{T}\tilde{\Sigma}_2\boldsymbol{v} > 0 $ 
for each $\boldsymbol{v}\neq \boldsymbol{0}$ so that 
$\boldsymbol{v}^\mathrm{T}[(1-\omega)\tilde{\Sigma}_1 + \omega\tilde{\Sigma}_2]\boldsymbol{v} > 0$
for $\boldsymbol{v}\in\mathbb{R}^n$ and $\boldsymbol{v}\neq \boldsymbol{0}$.

The proof of the theorem is split into three cases. 

\subsubsection{Case 1: $\omega_0 = 0$ and $\tilde{\Sigma}_1$ is non-singular}
The Kullback-Leibler divergence (KLD) from $\mathcal{N}_n(\boldsymbol{0}, \Sigma(\omega))$ 
to $\mathcal{N}_n(\boldsymbol{0}, \tilde{\Sigma}_1)$  is given by
\(
	\text{KLD}(\omega) = 0.5(\text{tr}(\tilde{\Sigma}_1^{-1}\Sigma(\omega)) - n - \log(\vert\tilde{\Sigma}_1^{-1}\Sigma(\omega)\vert))
\),
where tr denotes the trace of the matrix, and $\mathrm{KLD}(\omega)$ is finite
for $0 \leq \omega < 1$ since the KLD between two non-singular multivariate Gaussian
distributions is finite. Thus a distance can be defined through
\begin{equation}
	d(\omega) = \sqrt{\text{tr}(\tilde{\Sigma}_1^{-1}\Sigma(\omega)) - n - \log(\vert\tilde{\Sigma}_1^{-1}\Sigma(\omega)\vert)}, \quad 0 \leq \omega < 1,
	\label{eqn:proofs:3.1:dist}
\end{equation}
and we follow \citet{simpson2017} and use an exponential distribution 
on the
distance so that $\pi(d) = \lambda \exp(-\lambda d)(1-\exp(-\lambda d(1)))^{-1}$, $0 < d < d(1)$,
where $\lambda > 0$, and the possibly truncated density is normalized by
 $(1-\exp(-\lambda d(1)))$. A change of variables gives
\begin{equation}
	\pi(\omega) = \frac{\lambda \vert d'(\omega)\vert}{1-\exp(-\lambda d(1))} \exp(-\lambda d(\omega)), \quad 0 < \omega < 1.
	\label{eqn:proofs:3.1}
\end{equation}
\qed

\subsubsection{Case 2: $\omega_0 = 0$ and $\tilde{\Sigma}_1$ is singular}
If $\tilde{\Sigma}_1$ is singular and $\Sigma(\omega)$, $0 < \omega < 1$, is non-singular, the
distance $d(\omega)$ given in Equation \eqref{eqn:proofs:3.1:dist} is infinite for all $0 <\omega <1$ and the direct approach for constructing the
prior is not possible. We change the notation to $d(\omega; \omega_0)$ to make the 
dependence on the base
model explicit. For any base model $\omega_0 > 0$, $d(\omega;\omega_0)$ is finite for
$\omega_0 \leq \omega < 1$, and the prior can be constructed as for Case 1. The distance $d(\omega;\omega_0)$ is 
scaled by $\lambda$ in Equation \eqref{eqn:proofs:3.1}
and we seek an expression $\lambda(\omega_0)$ so that $\lambda(\omega_0) d(\omega;\omega_0)$ remains finite
for all $\omega_0 \leq \omega < 1$ when $\omega_0 \rightarrow 0^+$.

Since $\tilde{\Sigma}_1+\tilde{\Sigma}_2$ is positive definite, there exist an $n \times n$ matrix
$\mathbf{P}$ so that
\[
	\mathbf{P}(\tilde{\Sigma}_1+\tilde{\Sigma}_2)\mathbf{P}^\mathrm{T} = \mathbf{I}.
\]
This corresponds to a linear transformation of the
Gaussian distributions that
results in covariance matrices
 $\mathbf{S}_1 = \mathbf{P}\tilde{\Sigma}_1\mathbf{P}^\mathrm{T}$ and $\mathbf{S}_2 = \mathbf{P}\tilde{\Sigma}_2\mathbf{P}^\mathrm{T}$.
The KLD is
invariant to a linear transformation of the variables and the distance in Equation
\eqref{eqn:proofs:3.1:dist} can be calculated by
\[
	d(\omega;\omega_0)^2 = \text{tr}(\mathbf{S}(\omega_0)^{-1}\mathbf{S}(\omega)) - n - \log(\vert \mathbf{S}(\omega_0)^{-1}\mathbf{S}(\omega)\vert),
\]
where 
\[
	\mathbf{S}(\omega) = (1-\omega)\mathbf{S}_1 + \omega \mathbf{S}_2 = \omega(\mathbf{S}_1+\mathbf{S}_2) + (1-2\omega)\mathbf{S}_2 =  \omega\mathbf{I} + (1-2\omega)\mathbf{S}_1,
\]
since $\mathbf{S}_1+\mathbf{S}_2 = \mathbf{I}$.

$\mathbf{S}_1$ is symmetric and 
can be diagonalized so that $\mathbf{S}_1 = \sum_{i = 1}^n \lambda_i \boldsymbol{v}_i\boldsymbol{v}_i^\mathrm{T}$. This gives
\[
	\mathbf{S}(\omega) = \sum_{i = 1}^n [(1-2\omega) \lambda_i + \omega] \boldsymbol{v}_i\boldsymbol{v}_i^\mathrm{T}
	\] 
so that
\[
	\mathbf{S}(\omega_0)^{-1}\mathbf{S}(\omega) = \sum_{i = 1}^n \frac{[(1-2\omega)\lambda_i + \omega]}{[(1-2\omega_0)\lambda_i + \omega_0]} \boldsymbol{v}_i\boldsymbol{v}_i^\mathrm{T}.
\]
Thus the distance is given by
\[
	d(\omega; \omega_0)^2 = \sum_{i = 1}^n \frac{[(1-2\omega)\lambda_i + \omega]}{[(1-2\omega_0)\lambda_i + \omega_0]} - n - \sum_{i = 1}^n \log\left(\frac{[(1-2\omega)\lambda_i + \omega]}{[(1-2\omega_0)\lambda_i + \omega_0]}\right).
\]

Let $l$ be the rank deficency of $\tilde{\Sigma}_1$ and assume that the eigenvalues of $\mathbf{S}_1$ are sorted
from largest to smallest, then $\lambda_i > 0$ for $i = 1, \ldots, n-l$ and $\lambda_i = 0$ 
for $i = n-l+1, \ldots, n$, and the distance can be written as
\begin{align*}
	d(\omega; \omega_0)^2 = l \left(\frac{w}{\omega_0} - \log\left(\frac{\omega}{\omega_0}\right)\right) &+  \sum_{i = 1}^{n-l} \frac{[(1-2\omega)\lambda_i + \omega]}{[(1-2\omega_0)\lambda_i + \omega_0]} \\
	 &- n  - \sum_{i = 1}^{n-l} \log\left(\frac{[(1-2\omega)\lambda_i + \omega]}{[(1-2\omega_0)\lambda_i + \omega_0]}\right).
\end{align*}
The first term blows up as $\omega_0$ tends to zero, whereas the latter terms converges to a finite value. We introduce the scaled distance
\[
	\tilde{d}(\omega; \omega_0)^2 = \omega_0 d(\omega; \omega_0)^2 = l \left(\omega - \omega_0\log\left(\frac{\omega}{\omega_0}\right)\right) +  \omega_0 C(\omega_0),
\]
where $C(\omega_0) = \mathcal{O}(1)$ as $\omega_0 \rightarrow 0^+$,
and define
\(
	\tilde{d}(\omega; 0) = \lim_{\omega_0\rightarrow 0^+} \sqrt{\omega_0}d(\omega; \omega_0) = \sqrt{l w}.
\)

Thus by letting $\lambda(\omega_0) = \sqrt{\omega_0/l} \tilde{\lambda}$, 
we find the density
\begin{equation}
	\pi(\omega) = \frac{\tilde{\lambda}}{2\sqrt{\omega}(1-\exp(-\tilde{\lambda}))} \exp(-\tilde{\lambda} \sqrt{\omega}), \quad 0 < \omega < 1,
	\label{eqn:proofs:dens2}
\end{equation}
as $\omega_0\rightarrow 0^+$.

\qed

\subsubsection{Case 3: $0 < \omega_0 < 1$}
This case proceeds like Case 1 for $0 \leq \omega < \omega_0$ and for $\omega_0 < \omega < 1$.
On each side of $\omega_0$ we get a similar expression
as in Equation \eqref{eqn:proofs:3.1}. If we want to place the median at $\omega_0$ we
must place $1/2$ probability on each side of $\omega_0$ by introducing factors of
$1/2$ in the expressions. The density becomes 
\[
		\pi(\omega) = \begin{cases}
					\frac{\lambda \left\vert d'\left(\omega\right)\right\vert}{2(1-\exp(-\lambda d(0)))} \exp\left(-\lambda d\left(\omega\right)\right), &  0 < \omega < \omega_0, \\
					\frac{\lambda \left\vert d'\left(\omega\right)\right\vert}{2(1-\exp(-\lambda d(1)))} \exp\left(-\lambda d\left(\omega\right)\right), &  \omega_0 < \omega < 1,

		\end{cases}
\]
where $(1-\exp(-\lambda d(0)))$ makes sure the density in $0 < \omega < \omega_0$ integrates to 
$1/2$ and $(1-\exp(-\lambda d(1)))$ makes sure the density in $\omega_0 < \omega < 1$ integrates
to $1/2$.
\qed

\newpage
\section{Multivariate PC priors for ignorance}
\label{sec:supp:multPC}
The PC prior framework can be applied directly to dual splits since
distance can be defined as a function of a single parameter. However,
the PC prior framework does not translate
to a general approach for distances that are functions of multiple parameters without
further assumptions \citep[Section~6]{simpson2017}.
Consider a split with $K>2$ branches, and denote the proportion of
variances assigned to each branch as $\boldsymbol{\omega} = (\omega_1, \ldots, \omega_K)$.
Assume 
that the base model for the split is equal apportion of variance 
into the branches. Then the following procedure can be applied to replace
the split with a sequence of dual splits.

\begin{assumption}[Turn a multi-split into dual splits]
	\label{assume:method:multisplit}
	Consider a split in the tree structure that has $K > 2$ branches 
	and assume that the variance in each branch is
	  $\tilde{\sigma}_i^2$, for $i = 1, \ldots, K$. 
	We sequentially split out random effect 1, 2, and so on, through $K-1$ dual splits. The proportion of variance
	assigned to random effect $i$ of the total variance $\sum_{j=i}^K\tilde{\sigma}_j^2$ is
	$\omega^{(i)} = \tilde{\sigma}_i^2/\sum_{j=i}^K \tilde{\sigma}_j^2$ for $i = 1, \ldots, K-1$.
	The base models are $\omega_0^{(i)} = 1/(K+1-i)$, and ensures that
	conditioning on the base models results in a proportion of $1/K$ of the total variance to each child node.
\end{assumption}

The priors for each dual split can be precomputed before inference. 
The prior depends on the ordering of the $K-1$ dual splits, but when the hyperparameters are set according to the suggested values for dual splits in the main article,
 we do not expect the ordering of the child nodes within each multisplit to greatly affect inference because
 the conditional priors are weakly informative in the sense that they put most mass around the base models, but also ensure that large deviations from the base model are plausible. The base models are chosen so that the variance
 is split equally between the child nodes.

\newpage 
\section{Gaussian responses: Random intercept model}
\setcounter{figure}{0}

In this section we include additional background, theory and results for the random intercept model simulation study from Section 5.1 in the main article.

\subsection{Additional background}
\label{sec:supp:randintback}

The \textit{random intercept model} is given by
\begin{equation}
	y_{i,j} = \alpha_i + \varepsilon_{i,j}, \quad  j = 1, \dots, n_i,\, i = 1, \dots, n_\mathrm{g}, \, N = \sum_{i = 1}^{n_\mathrm{g}} n_i, 
	\label{gaussian:randint:randint}
\end{equation}
where $y_{i,j}$ is the $j$-th observation in group $i$, $\bm{\alpha} = (\alpha_1 \dots, \alpha_{n_\mathrm{g}})^\mathrm{T} \sim \mathcal{N}_{n_\mathrm{g}}(\bm{0}, \sigma_{\alpha}^2 \mathbf{I}_{n_\mathrm{g}})$ is a vector 
with the random intercepts (group effect), 
and $\bm{\varepsilon} = (\varepsilon_{1,1}, \varepsilon_{1, 2}, \dots, \varepsilon_{n_g, n_{n_\mathrm{g}}})^\mathrm{T} \sim \mathcal{N}_N(\bm{0}, \sigma_{\mathrm{R}}^2 \mathbf{I}_N)$ is the residual noise (individual effect). 
We denote the $N$-dimensional vector of observations $\bm{y} = (y_{1,1}, y_{1, 2}, \dots, y_{n_\mathrm{g}, n_{n_\mathrm{g}}})^\mathrm{T}$
and let $\mathbf{A}$ be a block matrix of size $N \times n_\mathrm{g}$
connecting the correct entries of $\bm{\alpha}$ to each observation in
$\bm{y}$.
Reparameterizing the model with total variance $V =
\sigma_\mathrm{R}^2+\sigma_\alpha^2$ and $\omega =
\sigma_\alpha^2/V$, the model can be written in vector form as
\begin{equation}
	\bm{y} = \sqrt{V}\left( \sqrt{\omega} \mathbf{A}\bm{\alpha} + \sqrt{1-\omega}\bm{\varepsilon}\right), \ (\bm{\alpha}, \bm{\varepsilon}) \sim \mathcal{N}_{n_\mathrm{g} + N}(\bm{0}, \mathbf{I}_{n_\mathrm{g} + N}).
	\label{gaussian:randint:randint}
\end{equation}
We use the \texttt{R} package \texttt{RStan} \citep{rstan_package} to perform the inference for all the three 
simulation studies in the paper. More 
specifically, we use the function \texttt{stan} from this package, where we use the following settings
for the random intercept model simulation study: 
burn-in of length 25 000, total sample length of 125 000 (i.e., 100 000 samples after burn-in), one chain,
we thin the chain to every fifth sample, initialize all parameters to zero, and we set
the value \texttt{adapt\_delta} to 0.95. \texttt{adapt\_delta} is the average proposal acceptance 
probability Stan aims for during the adaption (burn-in) period, and a larger value will give a smaller step
size \citep{stanwarnings}. For all other inputs we use the default values. 
We ran the simulation study on a computing cluster, where the full study runs in between a day and a week, depending on the available memory on the cluster.

\texttt{RStan} reports a \textit{divergent transition} 
for each iteration of the MCMC sampler that runs
into numerical instabilities 
\citep{carpenter2017}.
The divergent transitions are typically caused by an inappropriately large step size in the sampler or a poorly parameterized model,
and may indicate that the results are biased since the sampler had trouble exploring the posterior \citep{stanwarnings}.
It is difficult to completely avoid divergent transitions across all datasets, but to avoid reporting biased results,
we removed dataset and prior combinations that resulted in 0.1\% or more divergent transitions 
during the inference for $n_\mathrm{g} = 10$ or $50$. For $n_\mathrm{g} = 5$ we remove the 
dataset from the study if at least one prior results in too many divergent transitions.
We report the proportion of datasets that resulted in at most
0.1\% divergent transitions for each prior and scenario and use this as a measure of 
stability of the inference scheme for each prior.

\subsection{Connection to $R^2$}
\label{sec:supp:rsquared}

The coefficient of determination, commonly known as $R^2$, is a measure on how much of the data 
variance is explained by a given linear regression model \citep{gelmanbook}. In frequentistic 
statistics, the $R^2$ is used to assess model fit by comparing the variance in the residuals to 
the variance in the data. \cite{gelmanbook} generalise the $R^2$ to also make sense for 
multilevel models, such as the random intercept model. In this approach the $R^2$ is computed at each 
level of the model, which means we can assess the model fit at each level. In the case of the 
random intercept model, we have two levels in the model. The classical $R^2$ can be written as 
\begin{equation}
	R^2 = 1 - \frac{\sum_{i=1}^N (y_i - \hat{y}_i)^2}{\sum_{i=1}^N (y_i - \bar{y})^2}
\end{equation}
where $y_i$, $i = 1, \dots, N$, are observations, $\bar{y} = N^{-1}\sum_{i=1}^N y_i$, and 
$\hat{y}_i$ are the fitted values. Originally, the $R^2$ compares the model fit of any given 
linear regression model with covariates to a regression model with only an intercept. 
\cite{gelmanbook} define the generalised $R^2$ at each level $k$ in the model to be a comparison 
of the errors $\varepsilon_i^{(k)}$ at level $k$ and the total linear predictor $\eta_i^{(k)}$ at the 
same level of the model. The total linear predictor $\eta_i^{(k)}$ is the covariates and predictors at level $k$ in addition to the errors at the level, which means that $\eta_i^{(k)} \geq \varepsilon_i^{(k)}$ 
for all $k$.
We write the generalised $R^2$ as
\begin{equation}
	R_{\text{gen}}^{2,(k)} = 1 - \frac{\text{E}\left(\frac{1}{n_k}\sum_{i}\left(\varepsilon_i^{(k)} - \bar{\varepsilon}_i^{(k)}\right)^2\right)}{\text{E}\left(\frac{1}{n_k}\sum_{i}\left(\eta_i^{(k)} - \bar{\eta}_i^{(k)} \right)^2\right)}
\end{equation}
%
where $n_k$ is the number of
observations/groups at level $k$. The random intercept model has two levels, so $k \in \{1, 2\}$. 
In the main article 
we have standardised the data and omitted the intercept from the random intercept model we use, and we have no
covariates. This means that $\varepsilon_i^{(1)} = \varepsilon_i$, 
$\eta_i^{(1)} = y_i$, $\varepsilon_i^{(2)} = \alpha_i$ and $\eta_i^{(2)} = \alpha_i$, and we have that
\begin{align}
	\text{E}\left(\frac{1}{n_{\mathrm{g}}}\sum_{i}(\alpha_i - \bar{\alpha}_i)^2\right) &\stackrel{n_{\mathrm{g}} \to \infty}{=} \text{E}\left(\text{Var}\left(\bm{\alpha}\right)\right) = \sigma_{\alpha}^2, \\
	\text{E}\left(\frac{1}{N}\sum_{i}(\varepsilon_i - \bar{\varepsilon}_i)^2\right) &\stackrel{N \to \infty}{=} \text{E}\left(\text{Var}\left(\bm{\varepsilon}\right)\right) = \sigma_{\mathrm{R}}^2, \\
	\text{E}\left(\frac{1}{N}\sum_{i}(y_i - \bar{y}_i)^2\right) &\stackrel{N \to \infty}{=} \text{E}\left(\text{Var}\left(\bm{y}\right)\right) = \sigma_{\alpha}^2 + \sigma_{\mathrm{R}}^2.
\end{align}
The generalised $R^2$ at the group level ($k = 2$) for our model is zero (in the limit $n_{\mathrm{g}} \to \infty$), 
which makes sense as there is nothing more in the linear predictor than the errors at the lowest level when we have
no covariates in the model.
For the data level, the generalised $R^2$ is given by 
$1 - \sigma_{\mathrm{R}}^2/(\sigma_{\alpha}^2 + \sigma_{\mathrm{R}}^2) = 
\sigma_{\alpha}^2/(\sigma_{\alpha}^2 + \sigma_{\mathrm{R}}^2)$, 
which is the weight $\omega$ in the parametrization presented in this paper. Thus this weight 
is the asymptotic $R_{\text{gen}}^{2,(1)}$, which is also equal to the intra-class correlation.

\subsection{Results}
\label{sec:supp:randintres}

We present all the results from the random intercept model simulation study. 
The priors used in the study are the HD prior
with median $\omega_{\mathrm{m}} = 0.25$ (P-HD-25), $\omega_{\mathrm{m}} = 0.5$ (P-HD-50) and $\omega_{\mathrm{m}} = 0.75$ (P-HD-75), the HD prior with a symmetric Dirichlet prior on the weight (P-HD-D),
and the three commonly used priors P-INLA (Jeffreys' prior on residual variance and 
$\text{InvGamma}(1, 5 \times 10^{-5})$ on group variance), P-HC (Jeffreys' prior on residual variance and 
$\text{Half-Cauchy}(25)$ on group variance) and P-PC (Jeffreys' prior on residual variance and 
$\text{PC}_{\mathrm{SD}}(3, 0.05)$ on group variance). 
The different scenarios we have used are the true weight $\omega \in \{0.1, 0.25, 0.5, 0.75, 0.9\}$, 
$n_{\mathrm{g}} \in \{5, 10, 50\}$, $n_i = 10 \ \forall i$, 
and $n_i = 50 \ \forall i$, and 10 groups with varying group size where the 
group size is sampled from a $\text{Poisson}(10)$-distribution, and samples equal to 0 
or 1 is set to 10 so no group is of size smaller than 2.
As performance measures we use
the bias (estimated median minus true value) and 80\% coverage (found by counting the 
number of times the true value lies in the 80\% credible interval) of $\log(V)$ and $\logit(\omega)$,
and  
the number of datasets that leads to more than 0.1\% divergent transitions during the inference as a measure of stability.
All the box-plots show the median, the first and third quartile, 1.5 times the inter-quartile range (distance between first and third quartile), and outliers, if any. 

\begin{figure}[h!]
\centering
	\includegraphics[width = 1\textwidth]{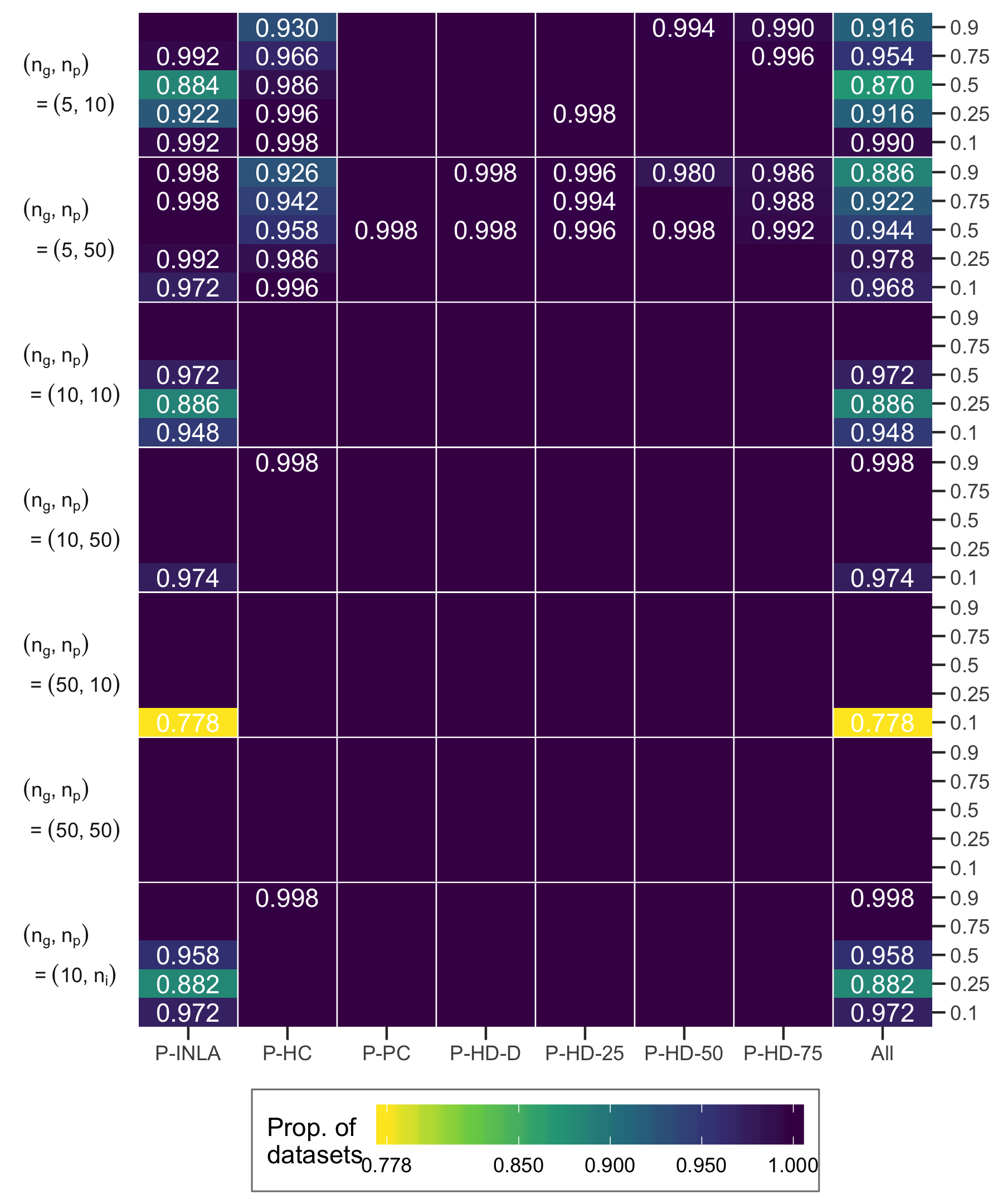}
	\caption{The proportion of datasets for each scenario and prior leading to at most 0.1\% divergent transitions during the inference in the random intercept model simulation study. We say that the stability is 1.0 if all datasets for a given prior and scenario lead to no more than 0.1\% divergent transitions. No number means that the stability is 1.0. The rightmost column, denoted ``All'', shows how many datasets must be removed from the study so all priors lead to at most 0.1\% divergent transitions for the remaining datasets.}
	\label{fig:supp:gaussian:acc_randint}
\end{figure}

From Figure \ref{fig:supp:gaussian:acc_randint} we see that P-INLA is less stable
than the other priors, except for datasets with five groups where also P-HC leads to inference with 
too many divergent transitions.
If a dataset leads to more than 0.1\% divergent transitions for a given prior, we remove the 
dataset from the study for this prior. 
For the scenarios with $n_{\mathrm{g}} = 5$, 
P-INLA and P-HC are more affected by
divergent transitions than the other priors.
In this case we remove the dataset from the 
study for all priors. This means that the results for P-INLA is based on fewer simulations than the 
other priors for $n_{\mathrm{g}} = 10$ or $50$. 

\begin{figure}[h]
\centering
	\includegraphics[width = 0.66\textwidth]{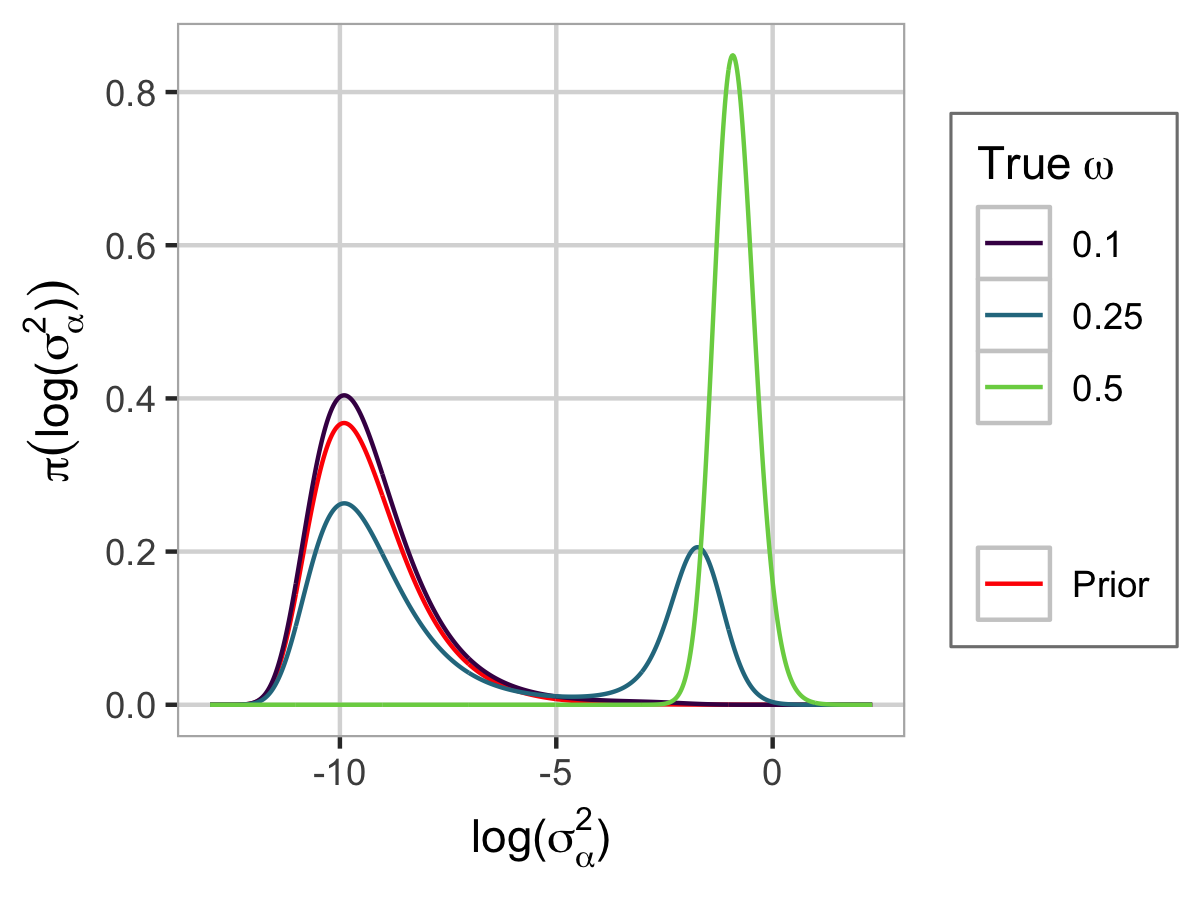}
	\caption{The posterior distribution of the logarithm of the group variance $\sigma_{\alpha}^2$ when using Jeffreys' prior on the residual variance and $\text{InvGamma}(1, 5 \times 10^{-5})$ on the group variance (P-INLA). The prior on the group variance is included in the plot. We have $n_{\mathrm{g}} = 10$ and $n_i = 10 \ \forall i$.}
	\label{fig:supp:gaussian:bimodal}
\end{figure}

Figure \ref{fig:supp:gaussian:bimodal} shows the posterior distribution of the logarithm of the group 
variance ($\log(\sigma_{\alpha}^2)$) when the priors of $\sigma_{\mathrm{R}}^2$ and 
$\sigma_{\alpha}^2$ are Jeffreys' and $\text{InvGamma}(1, 5 \times 10^{-5})$ (i.e. the 
INLA default prior), respectively.
This is the true posterior, calculated using numerical integration, with a dataset 
where the maximum likelihood (ML) estimates of the group and residual variances are 
exactly equal to $\omega$ and $1-\omega$, respectively. We vary
the value of $\omega$, and have 10 groups with 10 persons in each. When the true $\omega = 0.1$, 
and most of the variance in the model is residual variance, the posterior is highly 
influenced by the prior and we have close to no mass at the ML estimate (which is 0.1). 
When $\omega = 0.25$, the posterior is bimodal, and when $\omega = 0.5$ almost all the 
mass is at the ML estimate. This explains the bad results from P-INLA for datasets with true $\omega \leq 0.5$.

\begin{figure}[h]
\centering
	\includegraphics[width = 1\textwidth]{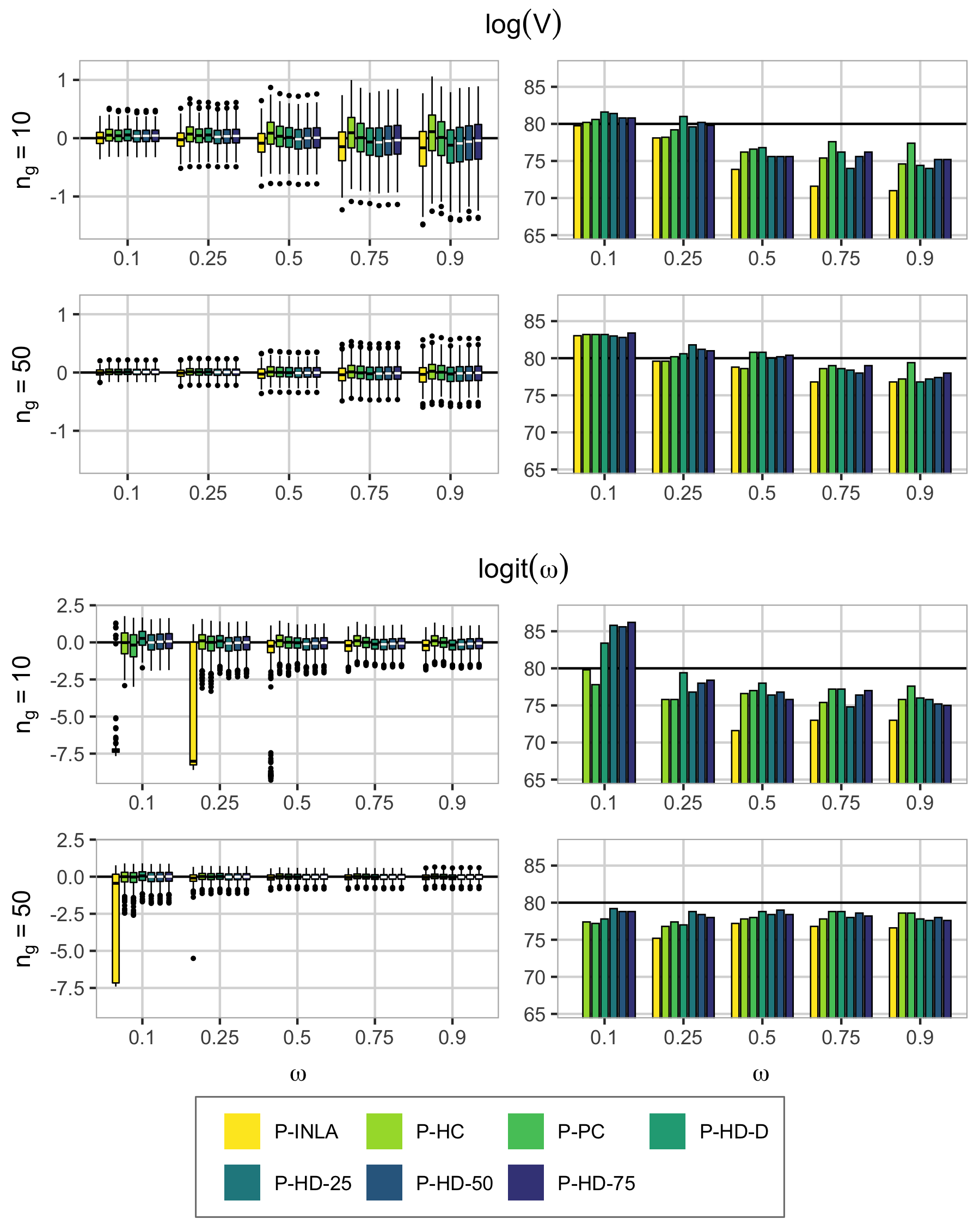}
	\caption{The true value of $\omega$ is on the x-axis in all graphs, the two upper rows contain the posterior diagnostics for the log total variance, and the two lower rows for logit weight. Bias in the left column, coverage in the right. The number of groups is indicated at the beginning of each row, either 10 or 50, and the group size $n_i = 10 \ \forall i$. 
	The order of the priors is the same in the legend and for each scenario. The coverage for P-INLA is sometimes below the 65\% and
	 not shown in the figure.}
	\label{fig:supp:gaussian:randint_ng1050_np10}
\end{figure}

\begin{figure}[h]
\centering
	\includegraphics[width = 1\textwidth]{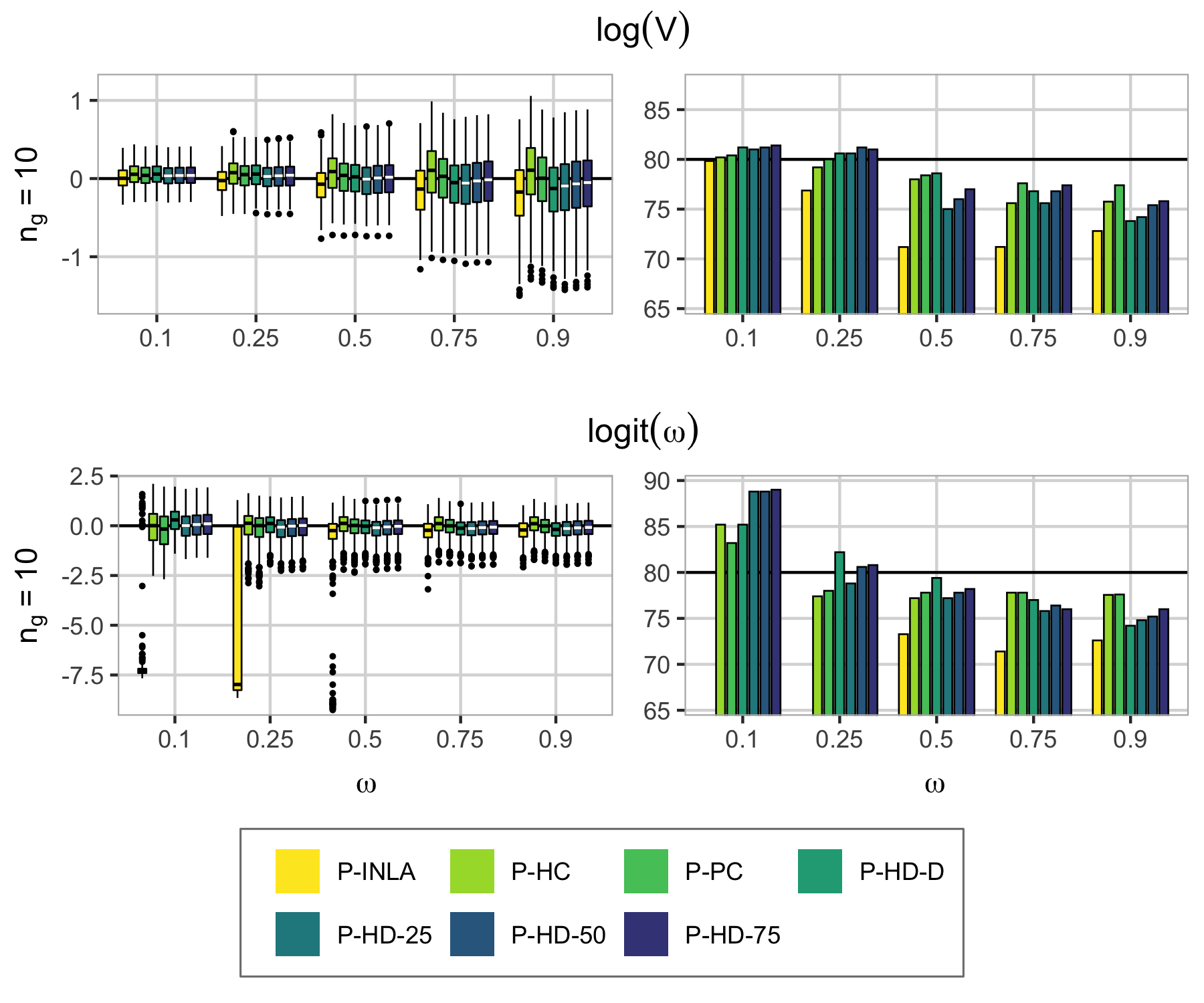}
	\caption{The true value of $\omega$ is on the x-axis in all graphs, the upper row contains the posterior diagnostics for the log total variance, and the lower row for logit weight. Bias in the left column, coverage in the right. The number of groups is 10 and the group size $n_i$ varies.
	The order of the priors is the same in the legend and for each scenario. The coverage for P-INLA is sometimes below the 65\% and
	 not shown in the figure.}
	\label{fig:supp:gaussian:randint_varies}
\end{figure}

\begin{figure}[h]
\centering
	\includegraphics[width = 1\textwidth]{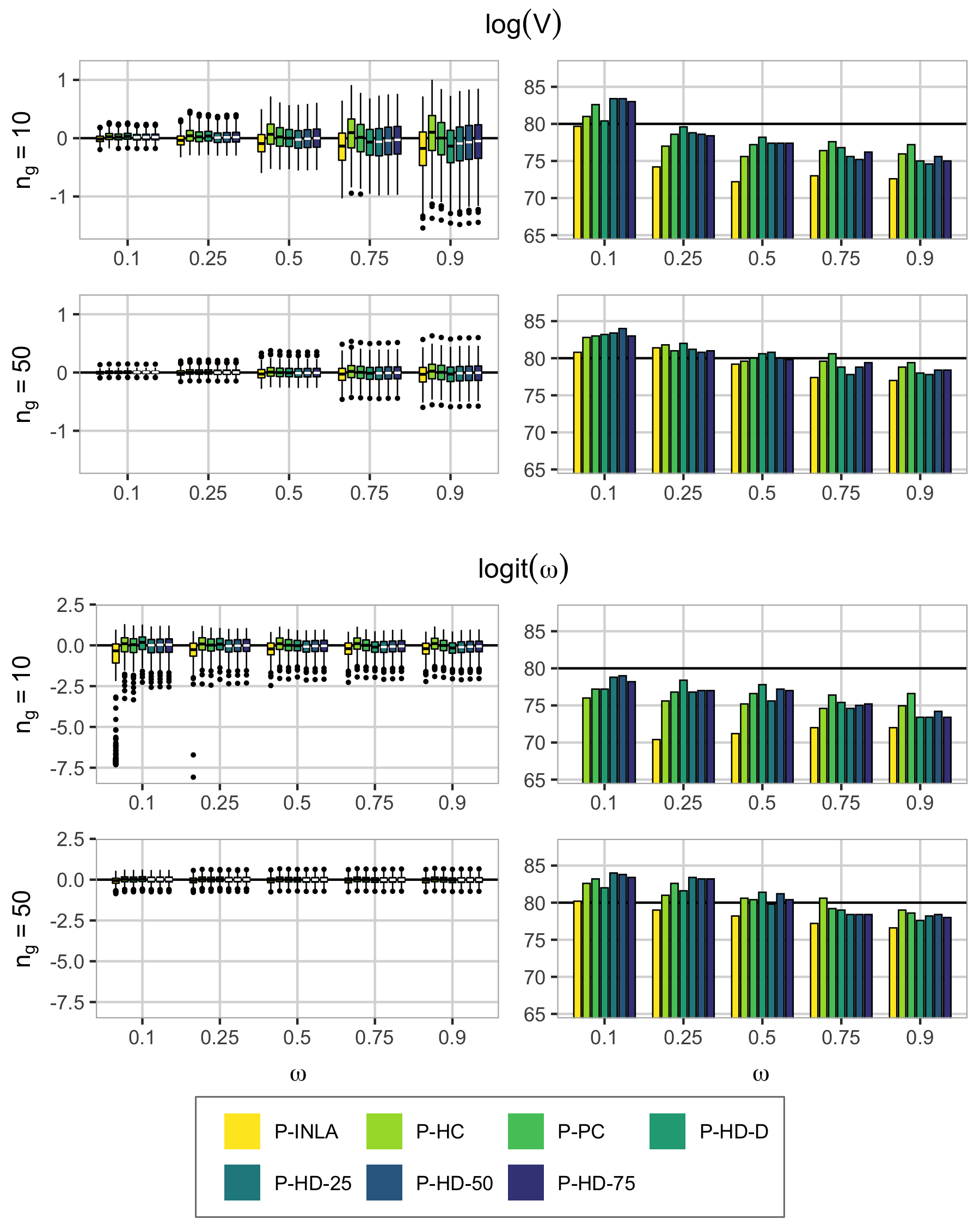}
	\caption{The true value of $\omega$ is on the x-axis in all graphs, the two upper rows contain the posterior diagnostics for the log total variance, and the two lower rows for logit weight. Bias in the left column, coverage in the right.  The number of groups is indicated at the beginning of each row, either 10 or 50, and the group size $n_i = 50 \ \forall i$.
	The order of the priors is the same in the legend and for each scenario. The coverage for P-INLA is sometimes below the 65\% and
	 not shown in the figure.}
	\label{fig:supp:gaussian:randint_ng1050_np50}
\end{figure}

\begin{figure}[h]
\centering
	\includegraphics[width = 1\textwidth]{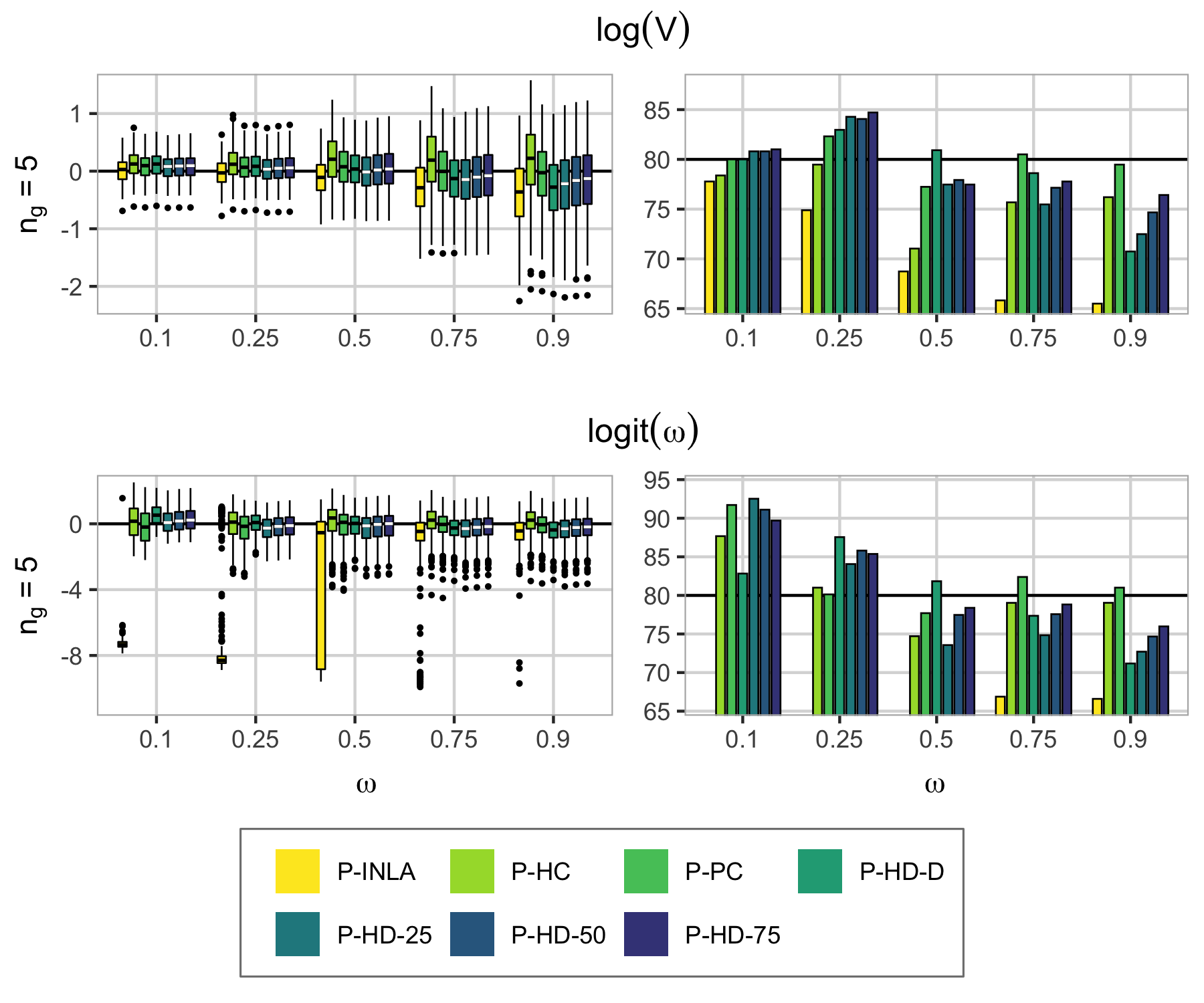}
	\caption{The true value of $\omega$ is on the x-axis in all graphs, the upper row contains the posterior diagnostics for the log total variance, and the lower row for logit weight. Bias in the left column, coverage in the right.  The number of groups $n_{\mathrm{g}} = 5$, and the group size $n_i = 10 \ \forall i$.
	The order of the priors is the same in the legend and for each scenario. The coverage for P-INLA is sometimes below the 65\% and
	 not shown in the figure.}
	\label{fig:supp:gaussian:randint_ng5_np10}
\end{figure}

\begin{figure}[h]
\centering
	\includegraphics[width = 1\textwidth]{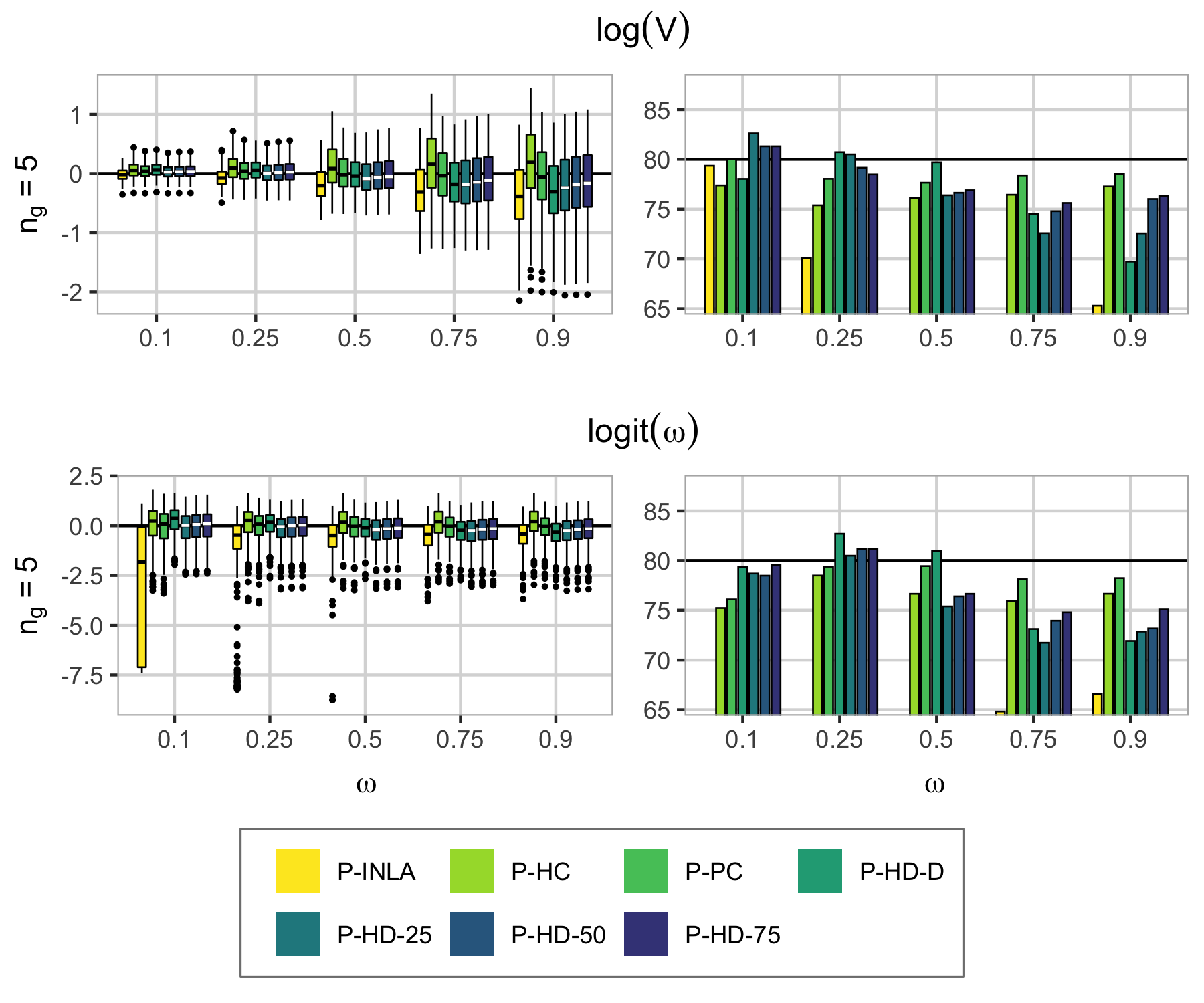}
	\caption{The true value of $\omega$ is on the x-axis in all graphs, the upper row contains the posterior diagnostics for the log total variance, and the lower row for logit weight. Bias in the left column, coverage in the right.  The number of groups $n_{\mathrm{g}} = 5$, and the group size $n_i = 50 \ \forall i$.
	The order of the priors is the same in the legend and for each scenario. The coverage for P-INLA is sometimes below the 65\% and
	 not shown in the figure.}
	\label{fig:supp:gaussian:randint_ng5_np50}
\end{figure}

Figures \ref{fig:supp:gaussian:randint_ng1050_np10}-\ref{fig:supp:gaussian:randint_ng5_np50} 
show all the bias and coverage results from the random intercept model
simulation study. 
Note that the coverage of $\omega$ is only shown for 
values larger than 65\%. The order of the priors is the same in the legend and for each scenario in all plots, so P-INLA is the leftmost, so comes P-HC and so on. 
%
For a given number of groups and group size, the magnitude of the bias for $\log(V)$ increases and for $\logit(\omega)$ 
decreases when the true value of $\omega$ increases. This is expected as a larger value of $\omega$ means that
the group variance is larger relative to the residual variance and the dataset provides more information about the
$\omega$ than would be the case when group variance is small relative to residual variance. On the other hand, 
a larger $\omega$ means the group variance dominates the total variance $V$ more and there is less information
about the group effect, which only has 5, 10 or 50 replicates, than the residual effect, which has 10 or 50 replicates for each group. This means less
information about the $V$.

In the following we list the main results from each figure.
It is clear from Figure \ref{fig:supp:gaussian:randint_ng1050_np10} that the choice of $\omega_{\mathrm{m}}$ does 
not have a large impact on the results. 
For
an HD prior with 
a Dirichlet prior on the weight $\omega$ (P-HD-D), the results are similar for the 
scenario with equal group and residual variance (true $\omega = 0.5$), and worse for the other scenarios. This is true
for all dataset sizes.
Figure \ref{fig:supp:gaussian:randint_varies} shows that also for varying 
group sizes the HD prior with a PC prior on $\omega$ behaves as well as or better than the other priors in terms of bias and coverage, and 
again the value of $\omega_{\mathrm{m}}$ does not influence the results noticeably. 
Figure \ref{fig:supp:gaussian:randint_ng1050_np50} shows that larger groups improves the 
results in terms of low bias and accurate coverage, especially for P-INLA, but not as much as larger number 
of groups improves the results.
In Figures 
\ref{fig:supp:gaussian:randint_ng5_np10} and \ref{fig:supp:gaussian:randint_ng5_np50} we include results for fewer 
groups, $n_{\mathrm{g}} = 5$, and 10 and 50 persons in each group, respectively. It is difficult to estimate the group 
variance with a low number of groups, and the results show that P-INLA is performing badly in terms of both bias and 
coverage for $V$ and $\omega$. For a given scenario with the HD prior, the bias and the coverage both increases for 
increasing values of $\omega_{\mathrm{m}}$.
P-HC leads to the least stable inference for $n_{\mathrm{g}} = 5$, and the other 
five priors give about equally stable inference. Note that for a given scenario we have removed the same datasets 
from the results for all priors, and the results may be slightly biased because of this.

\subsection{Simulation study for small group sizes}
\label{sec:supp:simSmallGroup}
We explore the properties of the HD prior when applied to problems 
with small datasets with only few observations in 
each group. Here the amount of information about the parameters is low and
the risk of overfitting is high. We define overfitting as overestimating
the value of $\omega$, and thus estimating spurious signals
in the group effect; and define underfitting as underestimating the value of $\omega$. Specifically, we use a small simulation study with two observations
per group,
and group size $n_{\mathrm{g}} \in \{10, 50, 100\}$. 
We include an additional prior denoted P-HD-10 not included in the main article, 
which is the HD prior with PC prior on weight
with median $\omega_{\mathrm{m}} = 0.1$. P-HD-10 is added
 to explore the option of higher shrinkage
in small data settings. The remaining HD priors are introduced in the main article.

\begin{figure}[h]
\centering
	\includegraphics[width = 1\textwidth]{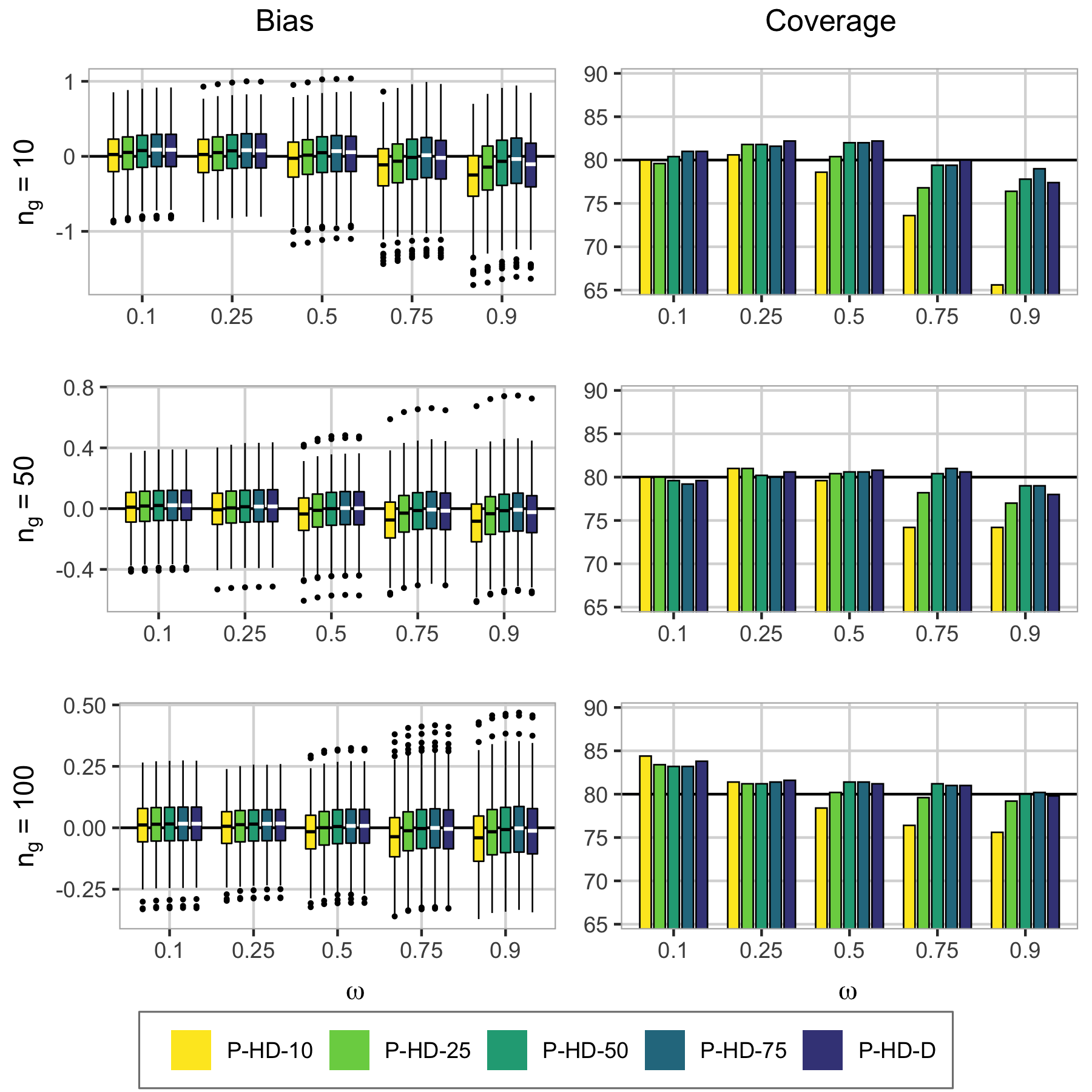}
	\caption{Results for $\log(V)$. 
	The true value of $\omega$ is on the x-axis in all graphs, bias is shown in the left column, coverage in the right. The number of groups is indicated at the beginning of each row, and there are two persons
	in each group.
	The order of the priors is the same in the legend and for each scenario. }
	\label{fig:supp:gaussian:small_groups_V}
\end{figure}

\begin{figure}[h]
\centering
	\includegraphics[width = 1\textwidth]{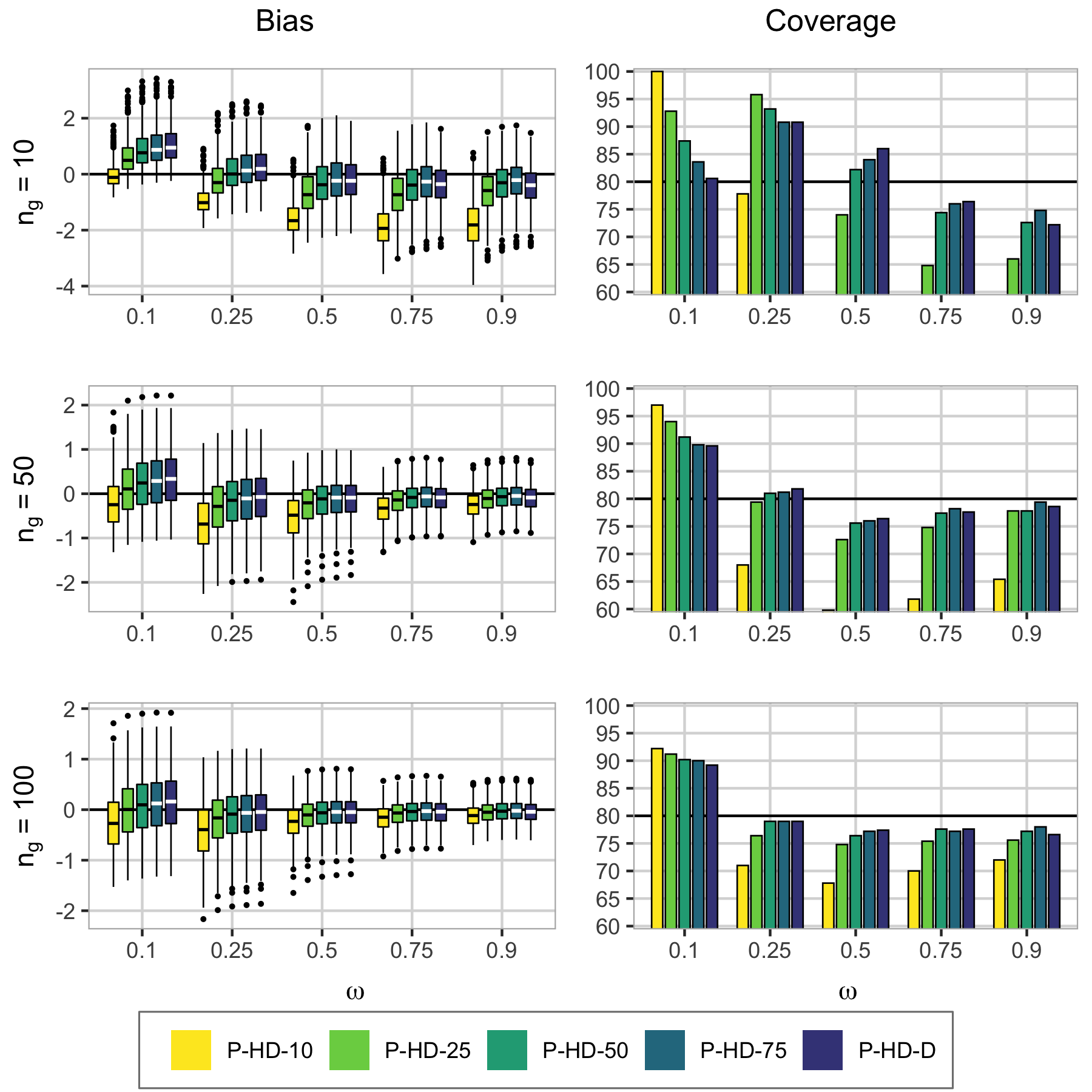}
	\caption{Results for $\logit(\omega)$. 
	The true value of $\omega$ is on the x-axis in all graphs, bias is shown in the left column, coverage in the right. The number of groups is indicated at the beginning of each row, and there are two persons
	in each group.
	The order of the priors is the same in the legend and for each scenario. The coverage for P-HD-10 is sometimes below the 65\% and
	 not shown in the figure.}
	\label{fig:supp:gaussian:small_groups_w}
\end{figure}

From Figure \ref{fig:supp:gaussian:small_groups_V} one can see that the inference
for total variance $V$ is stable in terms of bias and coverage. This indicates that
the Jefferey's prior on $V$ works well also in low information settings. From Figure 
\ref{fig:supp:gaussian:small_groups_w}, one can see that the inference for
the weight $\omega$ depends on the chosen prior. 
Using the recommended P-HD-25, we are slightly overfitting for
the scenario where the true weight is $0.1$, and we are slightly underfitting in the
other scenarios.
Using stronger shrinkage through P-HD-10 avoids overfitting for true weight equal to
$0.1$, but results in a stronger bias for higher values of the true weight, and
the resulting coverage varies from 100\% to 0\% in the scenarios.
 P-HD-50, P-HD-75 and P-HD-D result in overfitting also for
true weight equal to $0.25$ for $n_g = 10$. The results indicate that the recommended prior 
P-HD-25 is also
appropriate for small group sizes.  
None of the priors displayed lead to inference with more than 0.1\% divergent
transitions.

\clearpage

\section{Gaussian responses: Latin square}
\setcounter{figure}{0}

We include additional background and all results from the latin square simulation study from Section 5.2 in the main article.

\subsection{Additional background}
\label{sec:supp:latin}

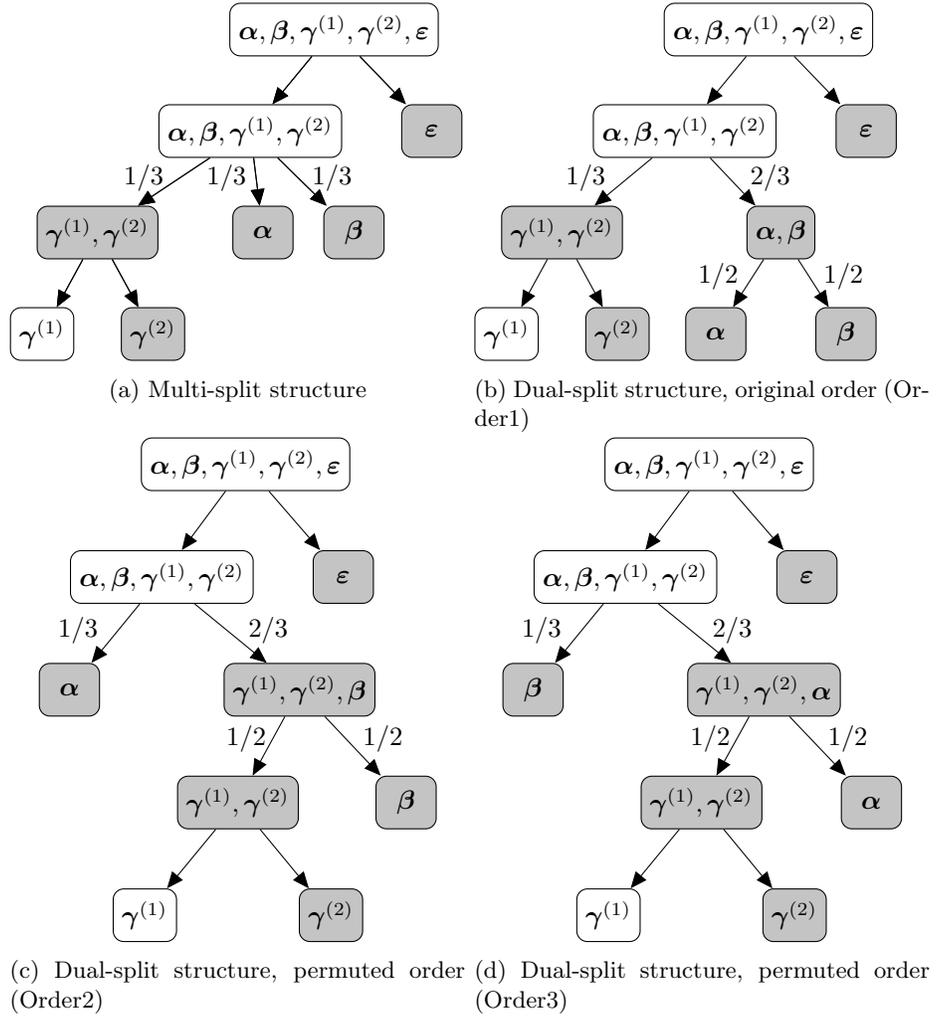
\begin{figure}
    \centering
        \begin{subfigure}[t]{0.45\textwidth}
	    \begin{tikzpicture} %
		    \node[draw, rounded corners, minimum height = 0.7cm, minimum width = 0.8cm, fill = white] (top) {$\boldsymbol{\alpha}, \boldsymbol{\beta}, \boldsymbol{\gamma}^{(1)}, \boldsymbol{\gamma}^{(2)}, \boldsymbol{\varepsilon}$} ; %
		    \node[draw, rounded corners, left=of top, yshift=-1.35cm, xshift = 2.5cm, minimum height = 0.7cm, minimum width = 0.8cm] (latent) {$\boldsymbol{\alpha}, \boldsymbol{\beta}, \boldsymbol{\gamma}^{(1)}, \boldsymbol{\gamma}^{(2)}$} ; %
		    \node[draw, rounded corners, right=of top, yshift=-1.35cm, xshift = -1.5cm, minimum height = 0.7cm, minimum width = 0.8cm, fill = basecolor] (residual) {$\boldsymbol{\varepsilon}$} ; %
		    \node[draw, rounded corners, left=of latent, yshift=-1.35cm, xshift = 1.0cm, minimum height = 0.7cm, minimum width = 0.8cm, fill = basecolor] (gamma) {$\boldsymbol{\gamma}^{(1)}, \boldsymbol{\gamma}^{(2)}$};
		    \node[draw, rounded corners, right=of latent, yshift = -1.35cm, xshift = -2.46cm, minimum height = 0.7cm, minimum width = 0.8cm, fill = basecolor] (alpha) {$\boldsymbol{\alpha}$};
		    \node[draw, rounded corners, right=of latent, yshift = -1.35cm, xshift = -1.25cm, minimum height = 0.7cm, minimum width = 0.8cm, fill = basecolor] (beta) {$\boldsymbol{\beta}$};
		    \node[draw, rounded corners, left=of gamma, yshift = -1.35cm, xshift = 1.5cm, minimum height = 0.7cm, minimum width = 0.8cm] (gamma1) {$\boldsymbol{\gamma}^{(1)}$};
		    \node[draw, rounded corners, right=of gamma, yshift = -1.35cm, xshift = -1.5cm, minimum height = 0.7cm, minimum width = 0.8cm, fill = basecolor] (gamma2) {$\boldsymbol{\gamma}^{(2)}$};
		    
		    \edge {top} {latent} ; %
		    \edge {top} {residual} ; %
		    \edge {latent} {alpha};
		    \edge {latent} {beta};
		    \edge {latent} {gamma};
		    \edge {gamma} {gamma1};
		    \edge {gamma} {gamma2};
		    \path[->, every node/.style={midway}]
		   		(top) edge[] node[xshift=-0.3cm, yshift = 0.1cm] {} (latent)
		   		(top) edge[] node[xshift= 0.3cm, yshift = 0.1cm] {} (residual)
		   		(latent) edge[] node[xshift = -0.4cm, yshift = 0.05cm] {1/3} (gamma)
		   		(latent) edge[] node[xshift = -0.4cm, yshift = 0.05cm] {1/3} (alpha)
		   		(latent) edge[] node[xshift = 0.4cm, yshift = 0.05cm] {1/3} (beta)
		   		(gamma) edge[] node[xshift = -0.3cm, yshift = 0.1cm] {} (gamma1)
		   		(gamma) edge[] node[xshift = 0.3cm, yshift = 0.1cm] {} (gamma2);
	    \end{tikzpicture}
	    \caption{Multi-split structure}
	    \label{fig:supp:latin:orig}
	\end{subfigure}
	\begin{subfigure}[t]{0.45\textwidth}
	    \begin{tikzpicture} %
		    \node[draw, rounded corners, minimum height = 0.7cm, fill = white] (top) {$\boldsymbol{\alpha}, \boldsymbol{\beta}, \boldsymbol{\gamma}^{(1)}, \boldsymbol{\gamma}^{(2)}, \boldsymbol{\varepsilon}$} ; %
		    \node[draw, rounded corners, left=of top, yshift=-1.35cm, xshift = 2.5cm, minimum height = 0.7cm] (latent) {$\boldsymbol{\alpha}, \boldsymbol{\beta}, \boldsymbol{\gamma}^{(1)}, \boldsymbol{\gamma}^{(2)}$} ; %
		    \node[draw, rounded corners, right=of top, yshift=-1.35cm, xshift = -1.5cm, minimum height = 0.7cm, minimum width = 0.8cm, fill = basecolor] (residual) {$\boldsymbol{\varepsilon}$} ; %
		    \node[draw, rounded corners, left=of latent, yshift=-1.35cm, xshift = 1.4cm, minimum height = 0.7cm, minimum width = 0.8cm, fill = basecolor] (gamma) {$\boldsymbol{\gamma}^{(1)}, \boldsymbol{\gamma}^{(2)}$};
		    \node[draw, rounded corners, right=of latent, yshift=-1.35cm, xshift = -1.4cm, minimum height = 0.7cm, minimum width = 0.8cm, fill = basecolor] (AB) {$\boldsymbol{\alpha}, \boldsymbol{\beta}$};
		    \node[draw, rounded corners, left=of AB, yshift = -1.35cm, xshift = 1.0cm, minimum height = 0.7cm, minimum width = 0.8cm, fill = basecolor] (alpha) {$\boldsymbol{\alpha}$};
		    \node[draw, rounded corners, right=of AB, yshift = -1.35cm, xshift = -1.0cm, minimum height = 0.7cm, minimum width = 0.8cm, fill = basecolor] (beta) {$\boldsymbol{\beta}$};
		    \node[draw, rounded corners, left=of gamma, yshift = -1.35cm, xshift = 1.5cm, minimum height = 0.7cm, minimum width = 0.8cm] (gamma1) {$\boldsymbol{\gamma}^{(1)}$};
		    \node[draw, rounded corners, right=of gamma, yshift = -1.35cm, xshift = -1.5cm, minimum height = 0.7cm, minimum width = 0.8cm, fill = basecolor] (gamma2) {$\boldsymbol{\gamma}^{(2)}$};
		    
		   	\path[->, every node/.style={midway}]
		   		(top) edge[] node[xshift=-0.3cm, yshift = 0.1cm] {} (latent)
		   		(top) edge[] node[xshift= 0.3cm, yshift = 0.1cm] {} (residual)
		   		(latent) edge[] node[xshift = -0.5cm, yshift = 0.05cm] {1/3} (gamma)
		   		(latent) edge[] node[xshift =  0.5cm, yshift = 0.05cm] {2/3} (AB)
		   		(gamma) edge[] node[xshift = -0.3cm, yshift = 0.1cm] {} (gamma1)
		   		(gamma) edge[] node[xshift = 0.3cm, yshift = 0.1cm] {} (gamma2)
		   		(AB) edge[] node[xshift = -0.4cm, yshift = 0.1cm] {1/2} (alpha)
		   		(AB) edge[] node[xshift = 0.4cm,  yshift = 0.1cm] {1/2} (beta);
	    \end{tikzpicture}
	    \caption{Dual-split structure, original order (Order1)}
	    \label{fig:supp:latin:perm1}
	\end{subfigure}
	\begin{subfigure}[t]{0.45\textwidth}
		\centering
	    \begin{tikzpicture} %
		    \node[draw, rounded corners, minimum height = 0.7cm, fill = white] (top) {$\boldsymbol{\alpha}, \boldsymbol{\beta}, \boldsymbol{\gamma}^{(1)}, \boldsymbol{\gamma}^{(2)}, \boldsymbol{\varepsilon}$} ; %
		    \node[draw, rounded corners, left=of top, yshift=-1.5cm, xshift = 2.5cm, minimum height = 0.7cm] (latent) {$\boldsymbol{\alpha}, \boldsymbol{\beta}, \boldsymbol{\gamma}^{(1)}, \boldsymbol{\gamma}^{(2)}$} ; %
		    \node[draw, rounded corners, right=of top, yshift=-1.5cm, xshift = -1.5cm, minimum height = 0.7cm, minimum width = 0.8cm, fill = basecolor] (residual) {$\boldsymbol{\varepsilon}$} ; %
		    \node[draw, rounded corners, left=of latent, yshift=-1.5cm, xshift = 1.4cm, minimum height = 0.7cm, minimum width = 0.8cm, fill = basecolor] (gamma) {$\boldsymbol{\alpha}$};
		    \node[draw, rounded corners, right=of latent, yshift=-1.5cm, xshift = -1.4cm, minimum height = 0.7cm, minimum width = 0.8cm, fill = basecolor] (AB) {$\boldsymbol{\gamma}^{(1)}, \boldsymbol{\gamma}^{(2)}, \boldsymbol{\beta}$};
		    \node[draw, rounded corners, left=of AB, yshift = -1.5cm, xshift = 2.0cm, minimum height = 0.7cm, minimum width = 0.8cm, fill = basecolor] (alpha) {$\boldsymbol{\gamma}^{(1)}, \boldsymbol{\gamma}^{(2)}$};
		    \node[draw, rounded corners, right=of alpha, yshift = -1.5cm, xshift = -1.0cm, minimum height = 0.7cm, minimum width = 0.8cm, fill = basecolor] (beta) {$\boldsymbol{\gamma}^{(2)}$};
		    \node[draw, rounded corners, left=of alpha, yshift = -1.5cm, xshift = 1.0cm, minimum height = 0.7cm, minimum width = 0.8cm] (beta2) {$\boldsymbol{\gamma}^{(1)}$};
		    \node[draw, rounded corners, right=of AB, yshift = -1.5cm, xshift = -1cm, minimum height = 0.7cm, minimum width = 0.8cm, fill = basecolor] (gamma2) {$\boldsymbol{\beta}$};
		    
		   	\path[->, every node/.style={midway}]
		   		(top) edge[] node[xshift=-0.3cm, yshift = 0.1cm] {} (latent)
		   		(top) edge[] node[xshift= 0.3cm, yshift = 0.1cm] {} (residual)
		   		(latent) edge[] node[xshift = -0.5cm, yshift = 0.05cm] {1/3} (gamma)
		   		(latent) edge[] node[xshift =  0.5cm, yshift = 0.05cm] {2/3} (AB)
		   		(alpha) edge[] node[xshift = 0.3cm, yshift = 0.1cm] {} (beta)
		   		(alpha) edge[] node[xshift = -0.3cm, yshift = 0.1cm] {} (beta2)
		   		(AB) edge[] node[xshift = -0.3cm, yshift = 0.1cm] {1/2} (alpha)
		   		(AB) edge[] node[xshift = 0.4cm, yshift = 0.1cm] {1/2} (gamma2);
	    \end{tikzpicture}
	    \caption{Dual-split structure, permuted order (Order2)}
	    \label{fig:supp:latin:perm2}
	\end{subfigure}
	\begin{subfigure}[t]{0.45\textwidth}
		\centering
	    \begin{tikzpicture} %
		    \node[draw, rounded corners, minimum height = 0.7cm, fill = white] (top) {$\boldsymbol{\alpha}, \boldsymbol{\beta}, \boldsymbol{\gamma}^{(1)}, \boldsymbol{\gamma}^{(2)}, \boldsymbol{\varepsilon}$} ; %
		    \node[draw, rounded corners, left=of top, yshift=-1.5cm, xshift = 2.5cm, minimum height = 0.7cm] (latent) {$\boldsymbol{\alpha}, \boldsymbol{\beta}, \boldsymbol{\gamma}^{(1)}, \boldsymbol{\gamma}^{(2)}$} ; %
		    \node[draw, rounded corners, right=of top, yshift=-1.5cm, xshift = -1.5cm, minimum height = 0.7cm, minimum width = 0.8cm, fill = basecolor] (residual) {$\boldsymbol{\varepsilon}$} ; %
		    \node[draw, rounded corners, left=of latent, yshift=-1.5cm, xshift = 1.4cm, minimum height = 0.7cm, minimum width = 0.8cm, fill = basecolor] (gamma) {$\boldsymbol{\beta}$};
		    \node[draw, rounded corners, right=of latent, yshift=-1.5cm, xshift = -1.4cm, minimum height = 0.7cm, minimum width = 0.8cm, fill = basecolor] (AB) {$\boldsymbol{\gamma}^{(1)}, \boldsymbol{\gamma}^{(2)}, \boldsymbol{\alpha}$};
		    \node[draw, rounded corners, left=of AB, yshift = -1.5cm, xshift = 2.0cm, minimum height = 0.7cm, minimum width = 0.8cm, fill = basecolor] (alpha) {$\boldsymbol{\gamma}^{(1)}, \boldsymbol{\gamma}^{(2)}$};
		    \node[draw, rounded corners, right=of alpha, yshift = -1.5cm, xshift = -1.0cm, minimum height = 0.7cm, minimum width = 0.8cm, fill = basecolor] (beta) {$\boldsymbol{\gamma}^{(2)}$};
		    \node[draw, rounded corners, left=of alpha, yshift = -1.5cm, xshift = 1.0cm, minimum height = 0.7cm, minimum width = 0.8cm] (beta2) {$\boldsymbol{\gamma}^{(1)}$};
		    \node[draw, rounded corners, right=of AB, yshift = -1.5cm, xshift = -1cm, minimum height = 0.7cm, minimum width = 0.8cm, fill = basecolor] (gamma2) {$\boldsymbol{\alpha}$};
		    
		   	\path[->, every node/.style={midway}]
		   		(top) edge[] node[xshift=-0.3cm, yshift = 0.1cm] {} (latent)
		   		(top) edge[] node[xshift= 0.3cm, yshift = 0.1cm] {} (residual)
		   		(latent) edge[] node[xshift = -0.5cm, yshift = 0.05cm] {1/3} (gamma)
		   		(latent) edge[] node[xshift =  0.5cm, yshift = 0.05cm] {2/3} (AB)
		   		(alpha) edge[] node[xshift = 0.3cm, yshift = 0.1cm] {} (beta)
		   		(alpha) edge[] node[xshift = -0.3cm, yshift = 0.1cm] {} (beta2)
		   		(AB) edge[] node[xshift = -0.3cm, yshift = 0.1cm] {1/2} (alpha)
		   		(AB) edge[] node[xshift = 0.4cm, yshift = 0.1cm] {1/2} (gamma2);
	    \end{tikzpicture}
	    \caption{Dual-split structure, permuted order (Order3)}
	    \label{fig:supp:latin:perm3}
	\end{subfigure}
  \caption{Two of the possible orderings for turning the triple split into a dual split. 
  		   \subref{fig:supp:latin:orig}) The multi-split structure of the HD prior, \subref{fig:supp:latin:perm1}) the original order used in simulation study in paper (Order1), \subref{fig:supp:latin:perm2}) one permuted order (Order2), and \subref{fig:supp:latin:perm3}) the other permuted order (Order2)}
  \label{fig:latin:diffOrder}
\end{figure}

\begin{figure}
	\centering
	\includegraphics[width = 1\textwidth]{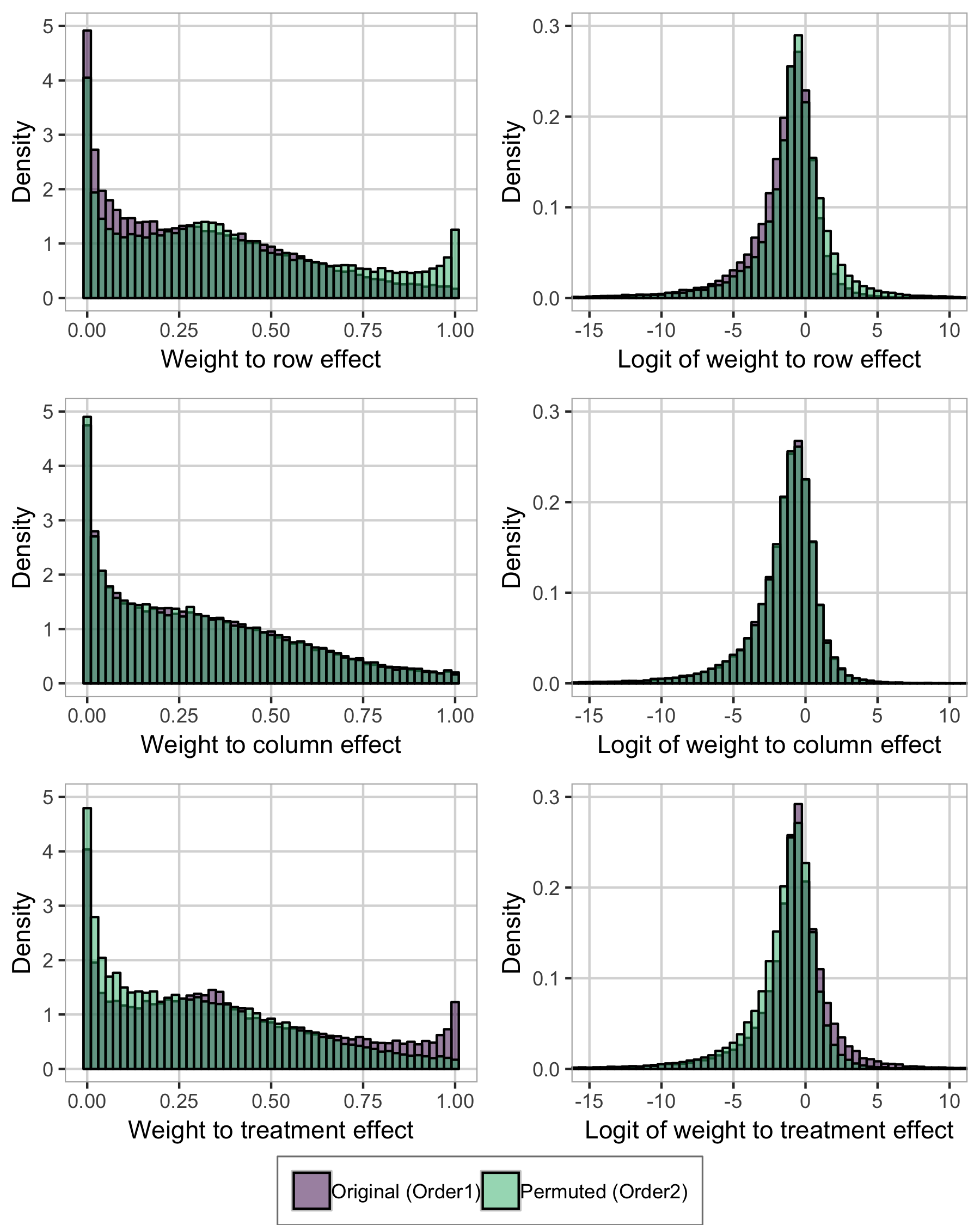}
	\caption{Comparison of priors on distribution of total latent variance to
			 row effect, column effect and treatement effect for the original order 
			 Order1 and the permuted order Order2. The distributions of the weights to the left, and of the logit weights on the right.}
	\label{fig:supp:latin:orders:weight}
\end{figure}

\begin{figure}
	\centering
	\includegraphics[width = 1\textwidth]{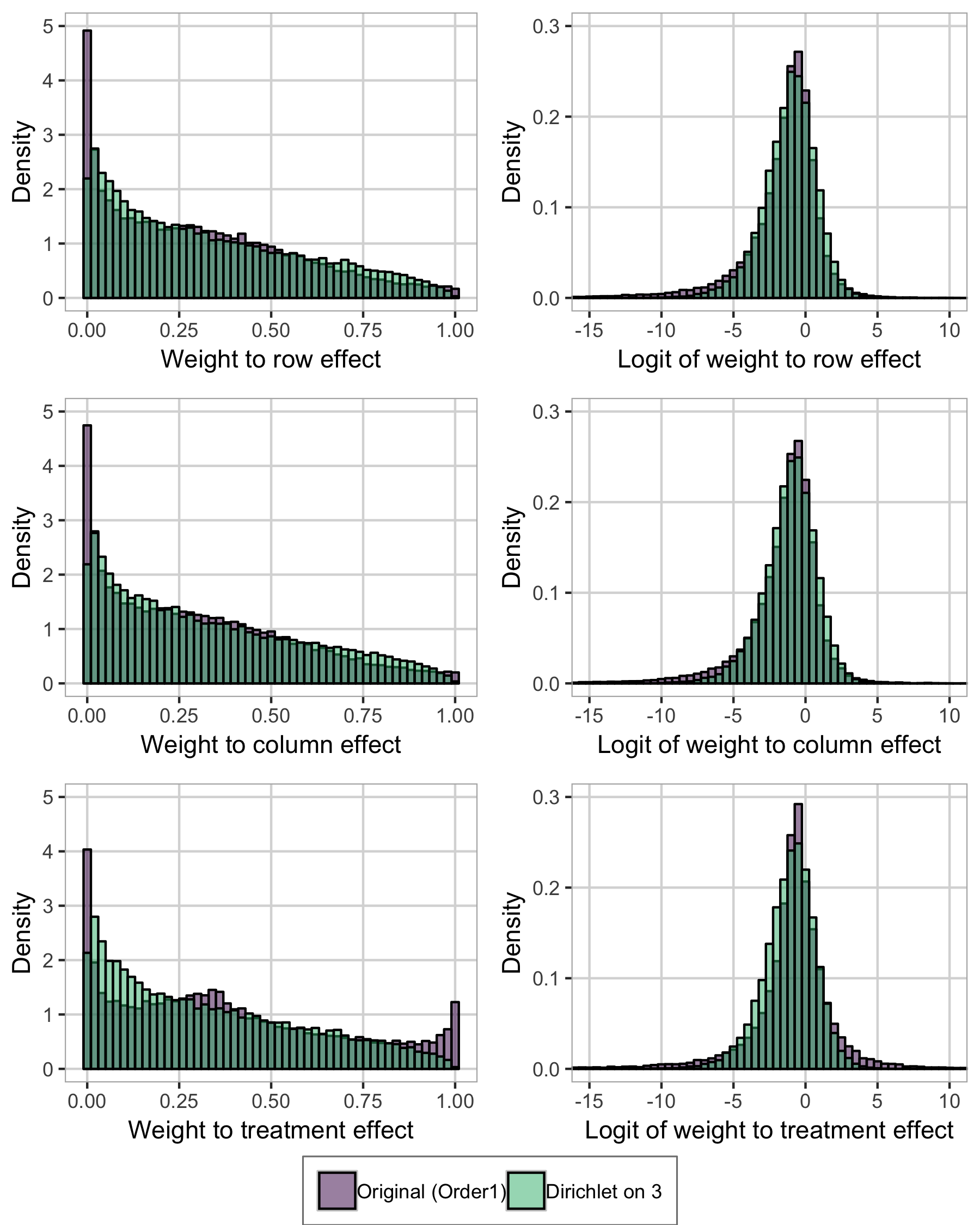}
	\caption{Comparison of priors on distribution of total latent variance to
			 row effect, column effect and treatement effect for the original order 
			 Order1 and a Dirichlet prior on the triple split. 
			 The distributions of the weights to the left, and of the logit weights on the right.}
	\label{fig:supp:latin:orders:weightdiri}
\end{figure}

The reasoning behind the tree structure for the prior in the latin square simulation study displayed in Figure \ref{fig:supp:latin:orig} is as follows: At the first level (top level)
the prior shrinks the latent part of the model, at the second level the total latent variance
is distributed with equal preference to the row effect, the column effect and the treatment effect,
and at the third level the treatment effect is shrunk towards the unstructured effect. 
We select an HD prior using the model structure in Figure \ref{fig:supp:latin:orig}. We also implement
the triple split as explained in
Section \ref{sec:supp:multPC}.
The original order chosen in the main article is denoted Order1 (\ref{fig:supp:latin:perm1}), and the permuted orders Order2 (\ref{fig:supp:latin:perm2}) and Order3 (\ref{fig:supp:latin:perm3}). 
The total variance of the 
latent model is split into $\omega^{(1)}$, $\omega^{(2)}$ and $\omega^{(3)}$, which are
the proportions of the latent variance going to the row effect, column effect and 
the treatment effect, respectively. Figure \ref{fig:supp:latin:orders:weight} shows the difference in marginal priors for
$\omega^{(1)}$, $\omega^{(2)}$ and $\omega^{(3)}$ for Order1 and Order2, on weight scale and on logit weight scale. Figure \ref{fig:supp:latin:orders:weightdiri}
shows the difference in the same marginal priors for Order1 and a Dirichlet prior on the triple split, 
where the latter is the default choice in the HD prior framework.

\begin{figure}
	\centering
	\includegraphics[width=0.5\textwidth]{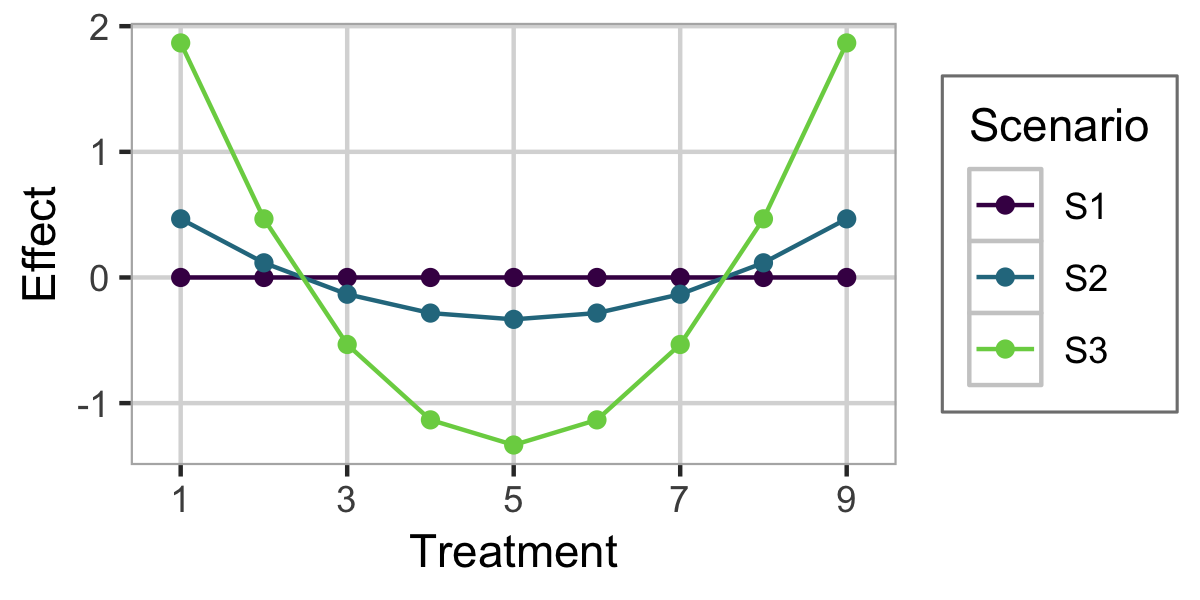}
	\caption{The true treatment effect for the simulated datasets in the latin square simulation study.}
	\label{fig:supp:latin:trueTreat}
\end{figure}

The true treatment effect $\boldsymbol{x} = (x_1, \ldots, x_9)$ we use in the latin square simulation study 
is given by $x_i = C\left((i-5)^2-20/3\right)$, $i = 1, \ldots, 9$ where $C = 0$ for scenario S1, 
$C = 0.05$ for scenario S2, and $C = 0.2$ for scenario S3.
These corresponds to signal to noise
ratios (SNRs) of $0\%$, $48\%$ and $94\%$ for S1, S2 and S3, respectively, as computed by
$\text{SNR} = S_{xx}/(S_{xx}+\sigma_\mathrm{t}^2)$, where $S_{xx} = \sum_{i = 1}^9 (x_i - \bar{x})^2$.
Figure \ref{fig:supp:latin:trueTreat} shows the true treatment effect for the three scenarios.

In the latin square experiment we use the following settings in the \texttt{R}-function \texttt{stan}: 
a burn-in of length 25 000, a total sample number (including burn-in) of 125 000, one chain which we thin to
every fifth sample, we initialize all parameters to zero, and 
use \texttt{adapt\_delta} equal to 0.95. We use default values for the rest of the settings. 
For the leave-one-out log predictive score (LOO-LPS), we use 1000 simulations for warm-up
and 2000 samples in total, which yields a low estimated variance of the LOO-LPS.
The simulation study ran on a computer cluster and takes no more than a couple of days, depending on the activity on the cluster.

\subsection{Results}
\label{sec:supp:latin:results}

We
have investigated the properties of 
the HD prior 
when the principles of the framework are tweaked.
What we investigate is varying values of the median $\omega_{\mathrm{m}}$ of the prior on the weight
indicating the proportion of treatment variance going to the structured effect, varying distributions on the distance in the original PC 
prior framework, varying the value of $\lambda$ for the multi-split, varying the type and ordering of the multi-split (see Figure \ref{fig:latin:diffOrder}), and we also study a joint prior where
we use a Dirichlet prior on all effects except the residuals, and on all five effects. 
We compare the HD prior to the following default priors, where all have Jeffreys' prior on the residual
variance and the following priors on the remaining variances or standard deviations: 
$\text{InvGamma}(1, 5 \times 10^{-5})$ (P-INLA), $\text{Half-Cauchy}(25)$ (P-HC), and $\text{PC}_{\mathrm{SD}}(3, 0.05)$ (P-PC).

\begin{figure}[h]
\centering
	\includegraphics[width = 0.833\textwidth]{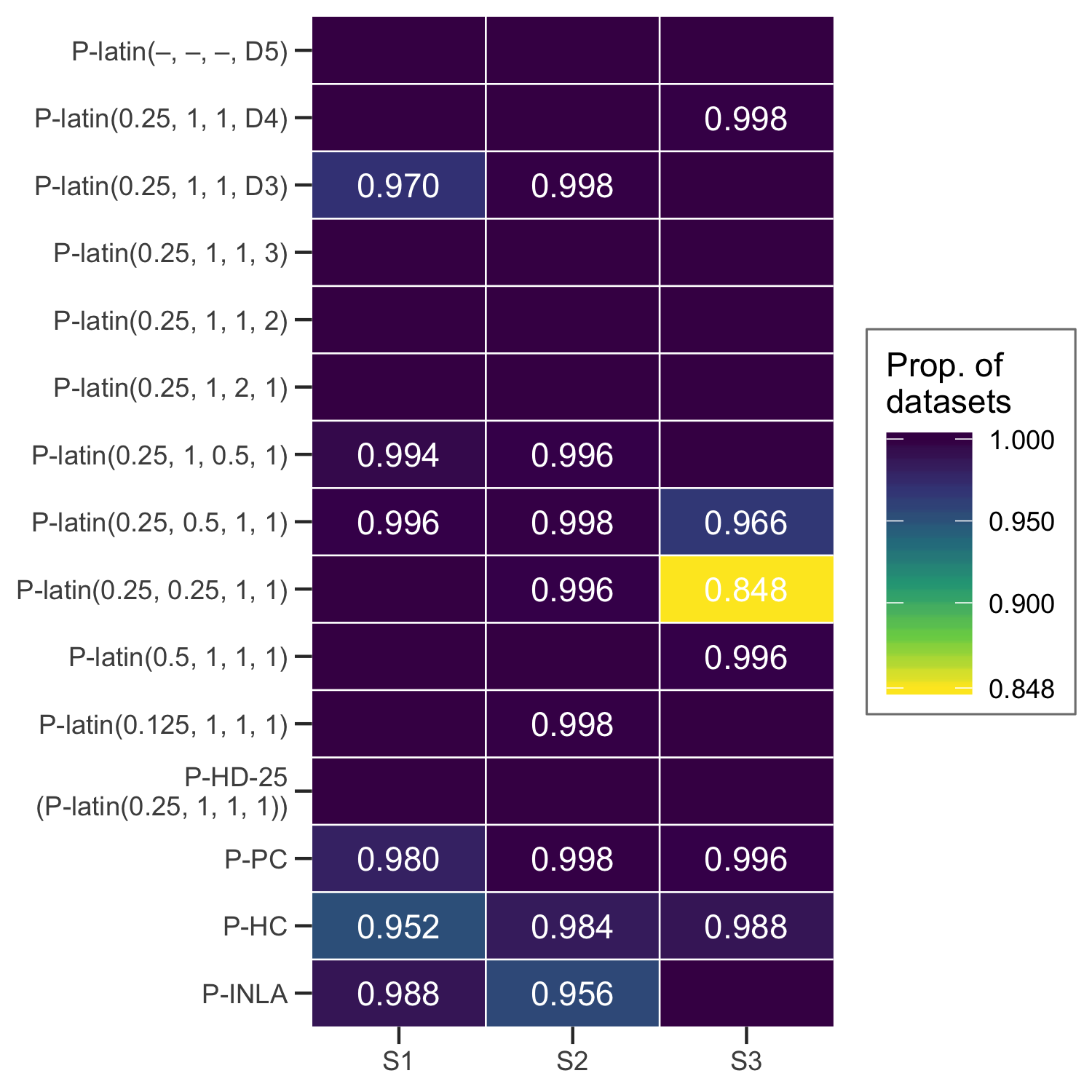}
	\caption{The proportion of datasets for each scenario and prior leading to at most 0.1\% divergent transitions during the inference in the latin square experiment simulation study. We say that the stability is 1.0 if all datasets for a given prior and scenario lead to no more than 0.1\% divergent transitions. No number means that the stability is 1.0. The bottom four priors are the main focus of the study, the top three are the Dirichlet priors, while the middle eight are the HD prior with varying values of $\omega_{\mathrm{m}}$, amount of shrinkage, varying values of $\lambda$, and varying ordering of the implementation of the triple split. The notation for the HD prior is P-latin($\omega_{\mathrm{m}}$, shape parameter, $\lambda$, order number/type).}
	\label{fig:supp:gaussian:acc_latin}
\end{figure}

For each scenario, we have removed the datasets that
lead to more than 0.1\% divergent transitions for at least one of the priors,
so all the results for a given scenario are based on the same datasets for all priors.
We use the proportion of datasets leading 
to at most 0.1\% divergent transitions during the inference as a measure of stability, for 
each prior and scenario. Figure \ref{fig:supp:gaussian:acc_latin} displays these
proportions for the latin square simulation study, and we see that it is not a big difference between P-INLA, 
P-HC, P-PC, and P-HD-25. However, when we lower the value of the shape parameter in the distribution we use on 
the distance (tweaking the third principle of the PC prior), the number of divergent transitions occurring during 
the inference increases, which indicates a more difficult posterior to draw samples from. 
When we change the 
values of $\omega_{\mathrm{m}}$, $\lambda$, or the way we implement the triple split (see 
Figure \ref{fig:latin:diffOrder}) the stability of the inference does not suffer.

\begin{figure}
\centering
	\includegraphics[width = 1\textwidth]{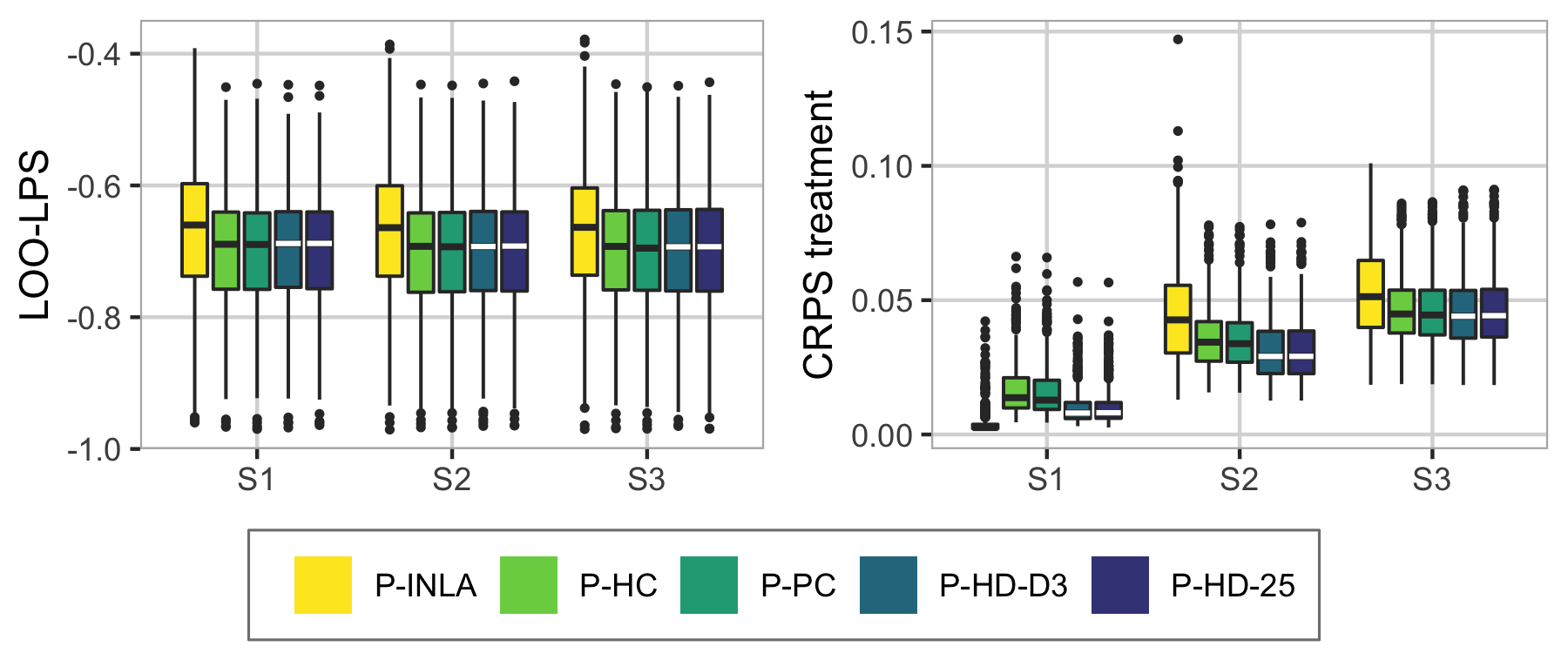}
	\caption{Results from the latin square simulation study.}
	\label{fig:supp:gaussian:latin_vanlig}
\end{figure}

\begin{figure}
\centering
	\includegraphics[width = 1\textwidth]{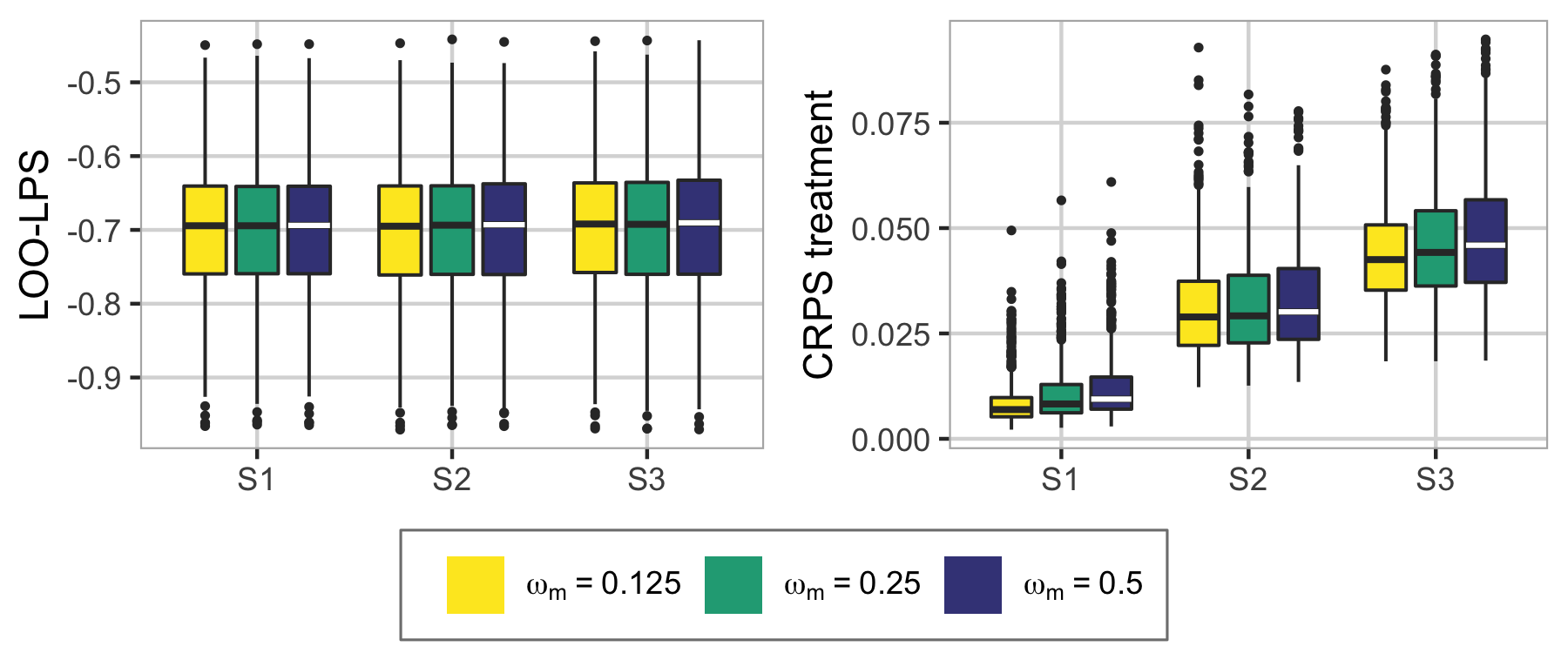}
	\caption{Results from the latin square simulation study when varying the position of the median $\omega_{\mathrm{m}}$ in the PC prior on the distance between $\bm{\gamma}^{(1)}$ and $\bm{\gamma}^{(2)}$. $\omega_{\mathrm{m}} = 0.25$ gives P-HD-25.}
	\label{fig:supp:gaussian:latin_median}
\end{figure}

\begin{figure}
\centering
	\includegraphics[width = 1\textwidth]{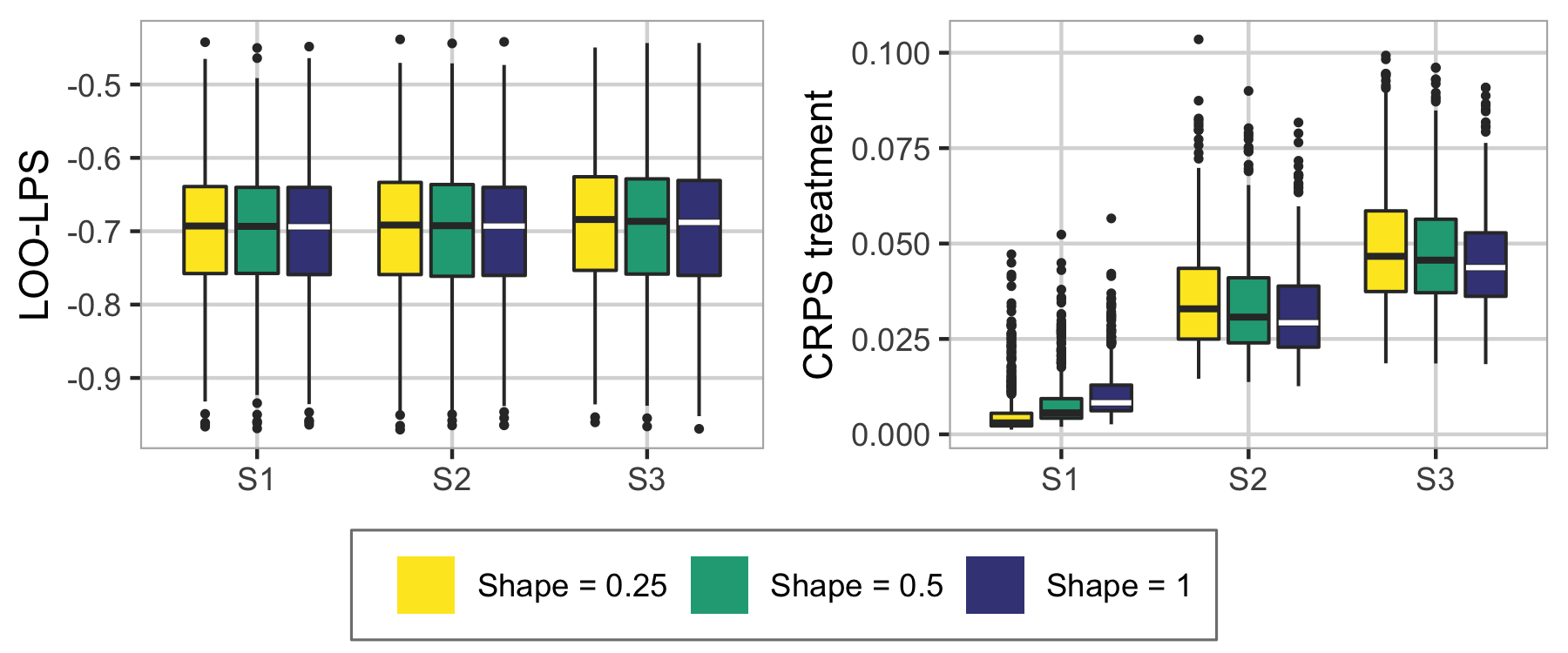}
	\caption{Results from the latin square simulation study when varying the shape parameter in the distribution on the distance in the PC prior for the split between unstructured and structured treatment effect. Shape parameter 1 gives the exponential distribution, which gives P-HD-25.}
	\label{fig:supp:gaussian:latin_shape}
\end{figure}

\begin{figure}
\centering
	\includegraphics[width = 1\textwidth]{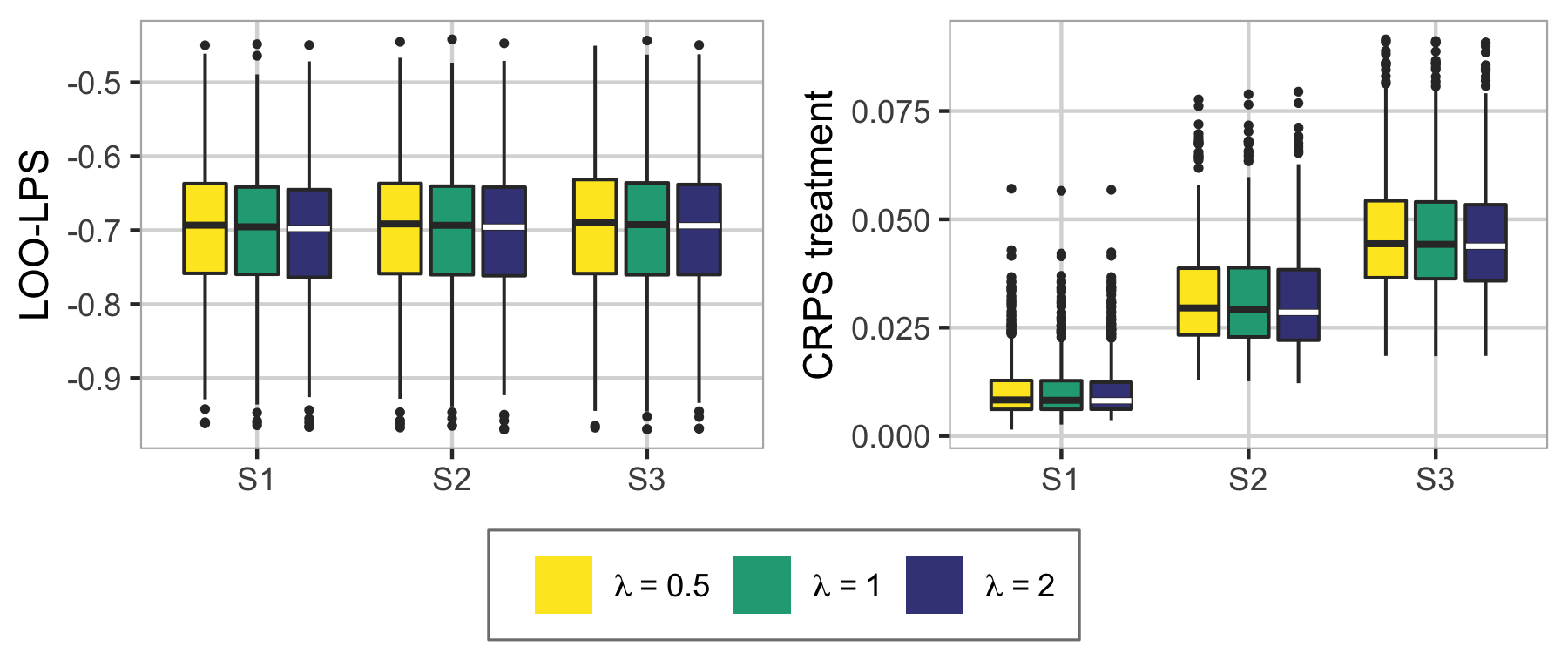}
	\caption{Results from the latin square simulation study when varying the value of $\lambda$ in the PC prior for the multi split. $\lambda = 1$ gives P-HD-25.}
	\label{fig:supp:gaussian:latin_lambda}
\end{figure}

\begin{figure}
\centering
	\includegraphics[width = 1\textwidth]{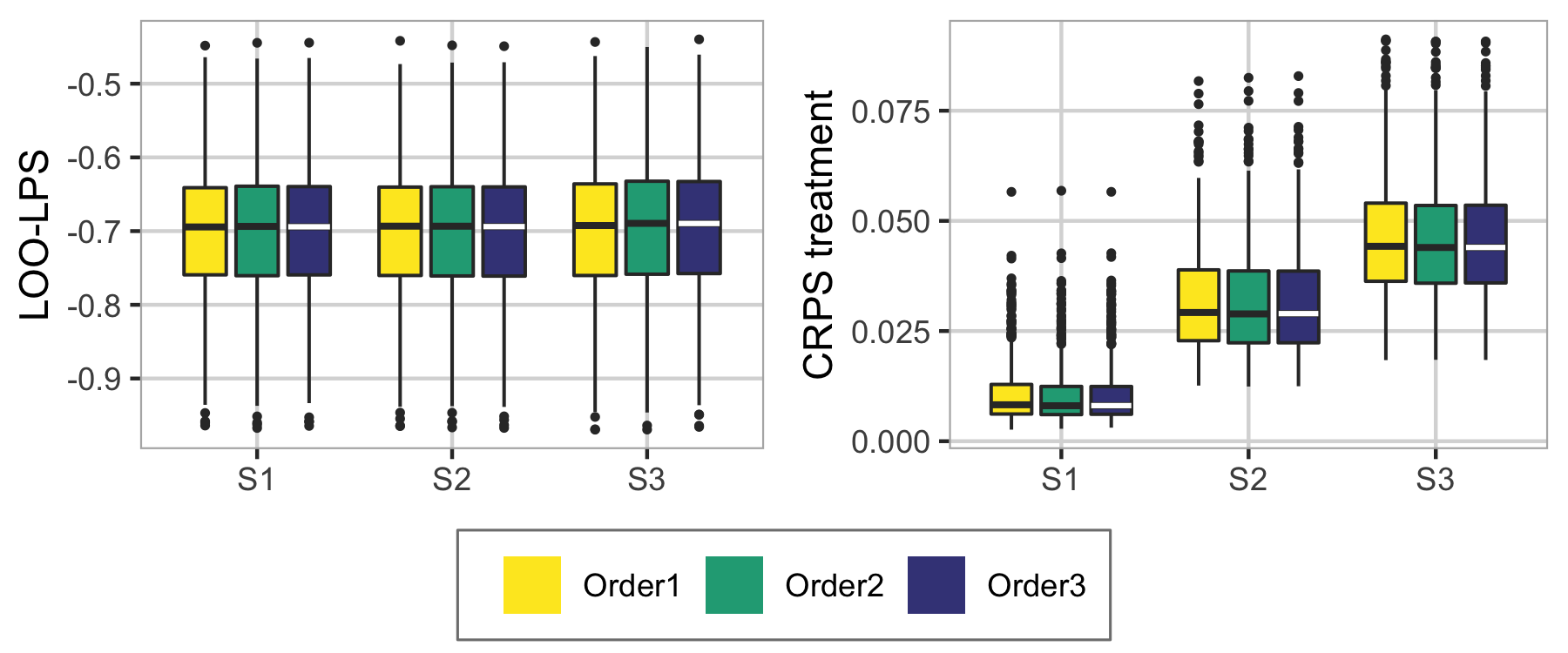}
	\caption{Results from the latin square simulation study when varying order of the implementation of the triple split in the PC prior. Order1 gives P-HD-25.}
	\label{fig:supp:gaussian:latin_order}
\end{figure}

\begin{figure}
\centering
	\includegraphics[width = 1\textwidth]{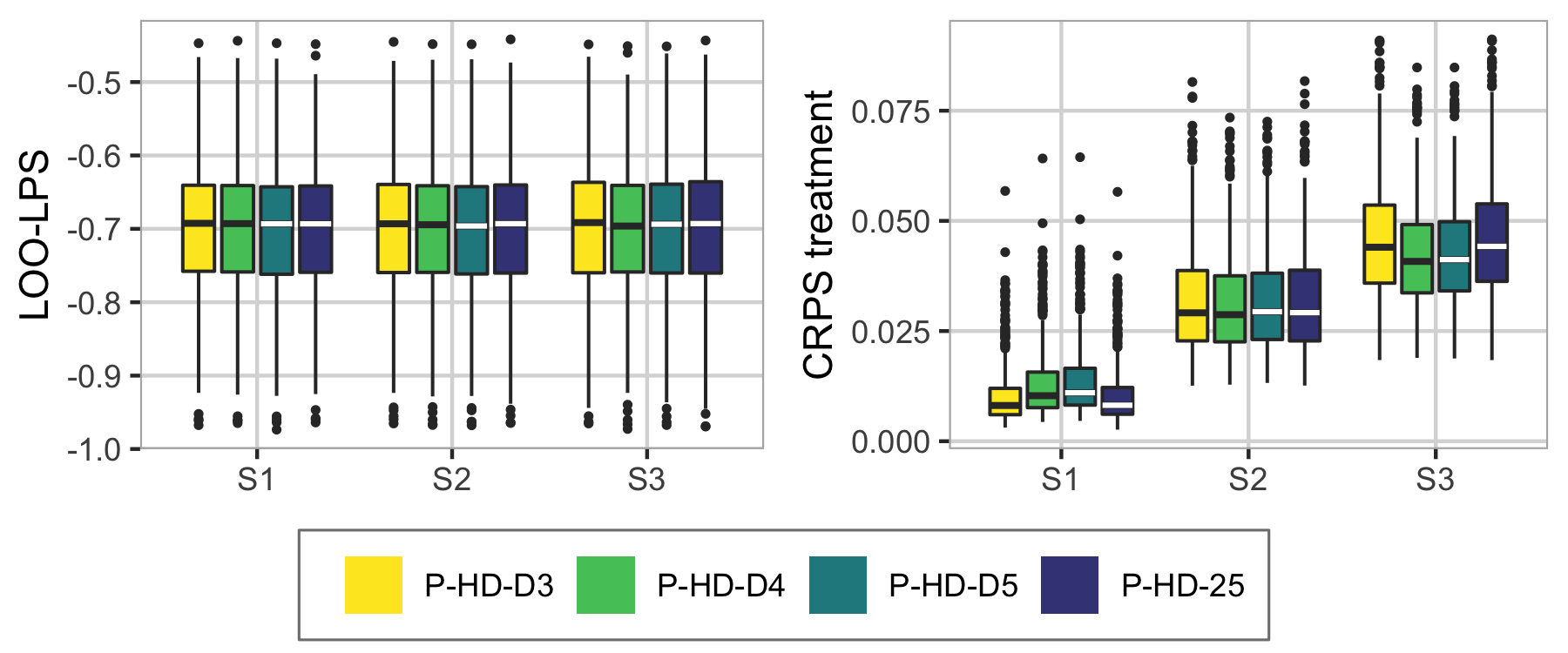}
	\caption{Results from the latin square simulation study for the Dirichlet prior. P-HD-D3 has a Dirichlet prior on the split between the row, column and treatment effects, P-HD-D4 has a Dirichlet prior between all effects except the residuals, and P-HD-D5 has a Dirichlet prior on all five random effects (including the residuals). The other weights has PC priors as in P-HD-25.}
	\label{fig:supp:gaussian:latin_diri}
\end{figure}

Figures \ref{fig:supp:gaussian:latin_vanlig}-\ref{fig:supp:gaussian:latin_diri} show
all results from the latin square simulation study. The box-plots include the 
median, the first and third quartile, 1.5 times the inter-quartile range (distance 
between first and third quartile), and outliers, if any. The six graphs all show the 
continuous rank probability score (CRPS) of the structured treatment effect 
$\bm{\gamma}^{(1)}$ and the leave-one-out log predictive score (LOO-LPS). In each plot, we have removed the datasets leading to 
too many (i.e., more than 0.1\%) divergent transitions in the inference for at least one of the three priors displayed. The order of the priors is the same in the legend and for each scenario in all plots, so P-INLA is the leftmost, so comes P-HC and so on.

Figure \ref{fig:supp:gaussian:latin_vanlig} shows the results that are also displayed in the main paper: P-INLA gives a lower LOO-LPS, i.e. a poorer model fit, than the other priors.
The CRPS is lowest for the HD prior with either triple
split implementation 
for scenarios S2 and S3.
Figure \ref{fig:supp:gaussian:latin_median} shows 
results for varying values of the median $\omega_{\mathrm{m}}$
for
the prior for selecting between $\bm{\gamma}^{(1)}$ and $\bm{\gamma}^{(2)}$ has little
effect on the results, 
and we see that a lower value of the median is 
slightly better when the true treatment effect is weak, and a higher value is slightly better when the true 
treatment effect is strong. The difference is however small. Figure 
\ref{fig:supp:gaussian:latin_shape} shows the results when we change the distribution we use on the 
distance between $\bm{\gamma}^{(1)}$ and $\bm{\gamma}^{(2)}$. 
Changing the exponential prior on the distance between $\bm{\gamma}^{(1)}$ and $\bm{\gamma}^{(2)}$
to a gamma prior with shape parameter $0.5$ or $0.25$, which has a stronger peak at 0, improves results for S1 (see Figure \ref{fig:supp:gaussian:latin_shape}), but
induces more instability in the inference (Figure \ref{fig:supp:gaussian:acc_latin}).
The results are also stable 
to changes in the hyperparameter for the two
dual-splits (Figure \ref{fig:supp:gaussian:latin_lambda})
and changes in the way
that the triple-split is implemented; either decomposed into dual-splits in different ways (Figure \ref{fig:supp:gaussian:latin_order}) or using a Dirichlet distribution (Figure \ref{fig:supp:gaussian:latin_diri}).


We have compared the HD prior with a Dirichlet prior on the triple split (P-HD-D3) to
HD priors with a
Dirichlet prior on
a quadruple split between $\bm{\alpha}$, $\bm{\beta}$, $\bm{\gamma}^{(1)}$ and $\bm{\gamma}^{(2)}$ (P-HD-D4) and between all five effects (P-HD-D5). The two latter perform worse than P-HD-D3 when the treatment effect has no structured contribution, 
scenario S1, in terms of CRPS (Figure \ref{fig:supp:gaussian:latin_diri}). Using P-HD-D4 and P-HD-D5 we lose the shrinkage properties between the 
unstructured and structured treatment effect, so we expect them to perform worse for S1. For S2 and S3 
they perform slightly better.
The LOO-LPS is not affected noticeably by the implementation of the triple split.

\clearpage

\clearpage

\clearpage

\subsection{Example}
\label{sec:supp:latinsquare:exCode}

We provide a script for \texttt{R} that can be used to simulate 
data and fit the latin square model. The 
script is available as part of the Supplementary Materials. This script can be used to look at differences between the priors and the resulting posteriors.

The following priors from the simulation study can be chosen:

\begin{enumerate}
	\item INLA default (P-INLA)
	\item Half-Cauchy (P-HC)
	\item Component-wise PC priors (P-PC)
	\item HD prior with PC priors on all splits (for example, P-HD-25). Here you can choose to change
		\begin{itemize}
			\item the median $\omega_\mathrm{m}$ for the proportion of treatment variance going to the structured effect [0.25 is default],
			\item the shape parameter for the gamma distribution on the distance between the unstructured and structured treatment effect [1 is default],
			\item A scaling factor for the value of $\lambda$ 
			used in the multi-splits [1 is default],
			\item the ordering of the triple split [1 is default, 2 and 3 are the other orderings].
		\end{itemize}
	\item HD prior with a combination of PC and Dirichlet priors (for example, P-HD-D3). Here you can choose to change
		\begin{itemize}
			\item the number of effects involved in the Dirichlet prior in the HD prior [3 is default, 4 and 5 are the other options].
		\end{itemize}
\end{enumerate}

After the prior has been chosen, the \textbf{scenario} can be selected: scenario S1 (no 
treatment effect), S2 (medium treatment effect) or S3 (strong treatment effect). See Section \ref{sec:supp:latin}
for details. A dataset of the same size as the ones in the simulation study is simulated, and the 
dataset can be reproduced using a seed value.

\texttt{Rstan} is used for the inference, and you can choose between the following
number of samples: \texttt{"low"} (250 (warmup) + 1000, 
only for testing, this will not give enough samples), \texttt{"medium"} (2500 (warmup) + 10000) or \texttt{"high"} (25000 (warmup) + 100000, this is used in the simulation study in the paper).

The sampler can be run without the likelihood to sample from the prior. 
A plot of the prior on total weight (the amount of the total variance) for each of the five effects in 
the model is available. The prior on total variance and the separate variances for the effects
are not shown as they do not have proper priors under the scale-invariant HD priors or
Jeffreys' prior on the residual variance.

For the posterior, the following scores and plots are provided:

\begin{itemize}
	\item The number of divergent transitions that occurred during the inference (see e.g. Section \ref{sec:supp:randintback}).
	\item The posterior total weights for the five model effects and the posterior total variance.
	\item The posterior standard deviations for the five model effects.
	\item The posterior mean of the structured treatment effect, with standard deviations, compared to the true effect.
	\item The average CRPS of the structured treatment effect, see Section 5.2 in the main article for details.
	\item The LOO-LPS (see Section 5.2 in the main article for details), with corresponding variance of the estimate, and the number of the 81 inferences with more than 1\% divergent transitions.
\end{itemize}

\clearpage

\section{Binomial responses}
\setcounter{figure}{0}

We include additional background and results from the Kenyan neonatal mortality simulation study and real application presented in Section 6 in the main article.

\subsection{Additional background}
\label{sec:supp:additional}

The DHS survey from 2014 is stratified by county and urban/rural 
and has two levels of clustering. Since the counties Nairobi and Mombasa are fully urban,
there are in total 92 strata. The households were selected within each stratum through
a two-stage clustered sampling design. Kenya was divided into 96251 enumeration areas (EAs) based
on the 2009 national census,
and the first stage of the sampling design consists of sampling
clusters 
from the list of EAs in the stratum
and the second stage consists of sampling households 
within the selected clusters. 
Within the selected households all women aged 15--49 who spent the last night in
the household are interviewed.

In Section 6.2 in the main paper, we simulate from a
model consisting of spatially structured and unstructured random effects and an i.i.d. effect of
cluster. Further, preliminary investigations showed that the design with 47 counties provides little
information about how the variance should be distributed between the structured and the unstructured
spatial effect. Therefore, we use the 290 constituencies of Kenya with 6 clusters per constituency to
replicate the size of the survey, but provide a spatial design where the data is more informative about
the relative sizes of the unstructured and structured spatial effects.
In Section 6.3 in the main paper we analyse the original data on the county-level
and include a random effect of household. 
The key focus of the application is to display how to use and select the new prior, 
and how the interpretability and transparency of the prior is helpful for
assessing and criticising the results.

\subsection{Simulation study}

We use the following input values to the function \texttt{stan} for the simulation study with neonatal 
mortality in Kenya: 25 000 samples for burn-in, in total 75 000 samples, 
one chain thinned to every fifth sample, all parameters initialized to zero,
\texttt{adapt\_delta} equal to 0.95, and default settings for all other input values.
The simulation study ran on a computer cluster and takes less than a week, depending on the activity on the cluster.

\begin{figure}[hb]
\centering
	\includegraphics[width = 0.66\textwidth]{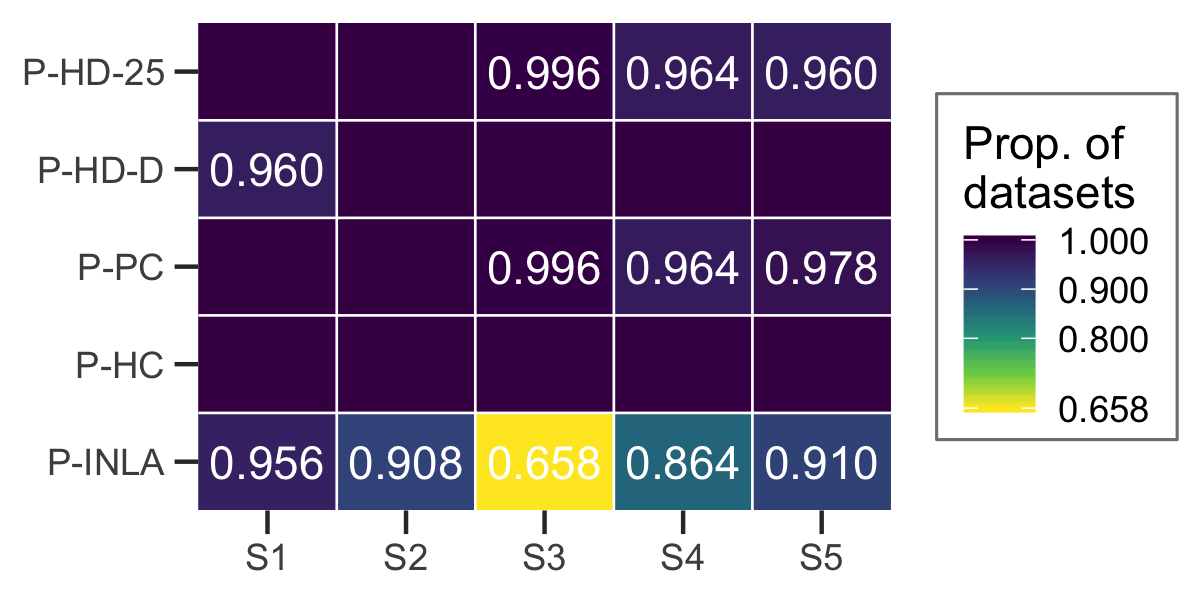}
	\caption{The proportion of datasets for each scenario and prior leading to at most 0.1\% divergent transitions during the inference in the neonatal mortality in Kenya simulation study. We say that the stability is 1.0 if all datasets for a given prior and scenario lead to no more than 0.1\% divergent transitions. No number means that the stability is 1.0.}
	\label{fig:supp:nonGauss:acc_kenya}
\end{figure}

\begin{figure}[t]
\centering
	\includegraphics[width = 1\textwidth]{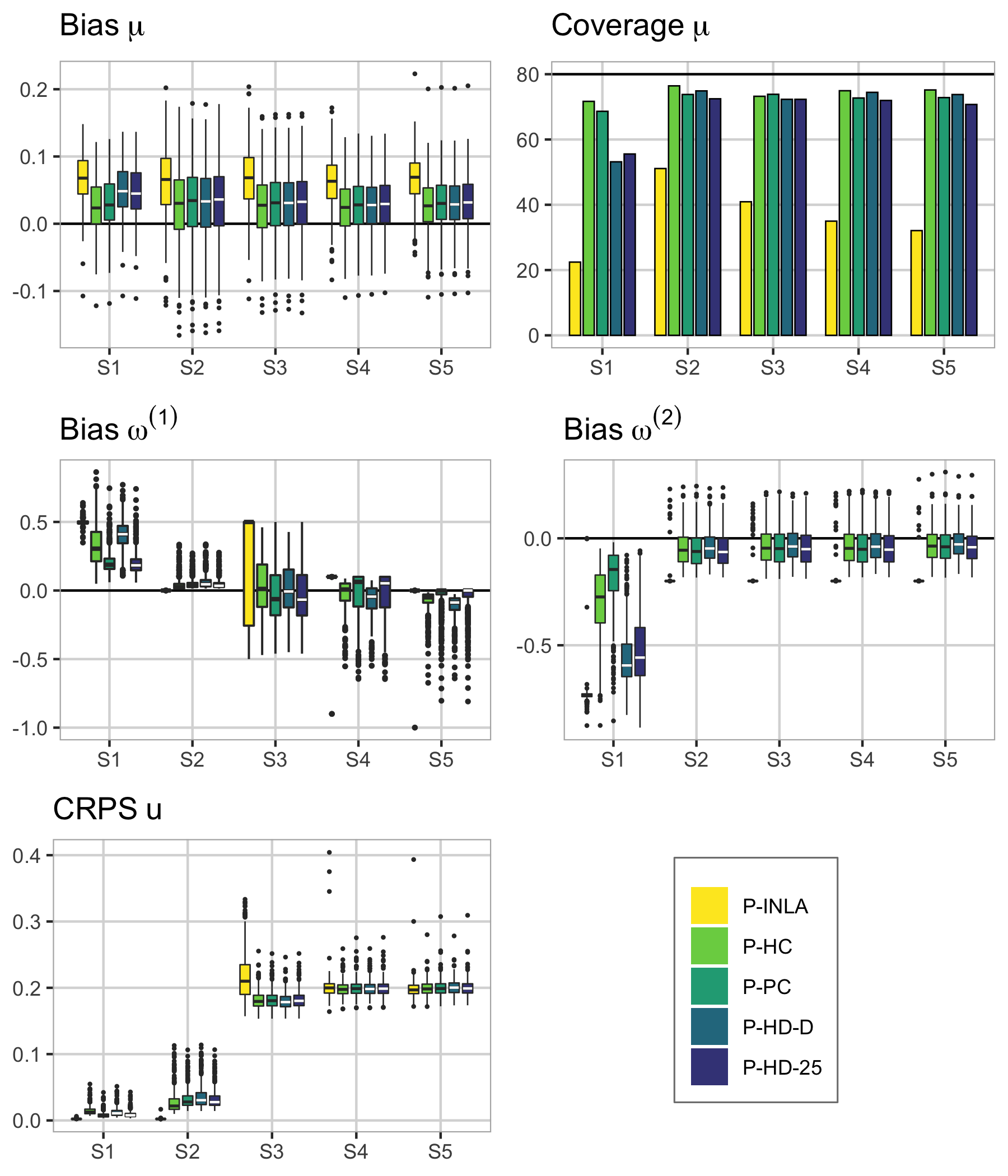}
	\caption{Upper left: bias of the intercept $\mu$, 
	upper right: the coverage of $\mu$,
	mid left: the bias of $\omega^{(1)}$,
	mid right: the bias of $\omega^{(2)}$, and
	lower left: CRPS of $\bm{u}$. 
	Scenario is indicated at the x-axes. The order of the priors is the same in the legend and for each scenario, so P-INLA is the leftmost, then comes P-HC and so on.
	The biases are calculated using the estimated median minus the true value, and the coverage is found by counting the number
	of times the true value lies in the 80\% credible interval.}
	\label{fig:supp:nonGauss:kenya_more_res}
\end{figure}

We include additional results from the Kenya neonatal mortality simulation study. Figure 
\ref{fig:supp:nonGauss:acc_kenya} shows the proportion of datasets leading to no more than 0.1\% 
divergent transitions during the inference. P-INLA is the only prior which 
leads to a large number of datasets giving divergent transitions, and mainly 
in scenario S3, the other three priors give stable inference for all scenarios. 
Figure \ref{fig:supp:nonGauss:kenya_more_res} shows the bias and coverage of $\mu$, the 
the bias of $\omega^{(1)}$ and $\omega^{(2)}$ and the 
CRPS of $\bm{u}$, for the five 
priors we have used in the simulation study. It is only for scenario S3, 
when the Dirichlet prior is closest to the truth, that P-HD-D is 
performing better than P-HD-25, in the other scenarios it is doing worse.

Figure \ref{fig:supp:nonGauss:kenya_more_res} shows that P-INLA gives
way too low coverage for $\mu$,
while the other priors leads to a better and similar coverage.
%
For scenarios S2-S5 the true value of the weight is 0.2, P-INLA is for most datasets estimating $\omega^{(1)}$ to be 0, giving a bias of -0.2. The other four priors are all slightly underestimating the weight in S2-S5. 
P-HD-D is as good as (only scenario S3) or worse than P-HD-25.
In scenario S1, the true weight is equal to 1 while the base model is 0, and all priors are underestimating the weight. P-INLA is doing worst with a bias around -0.75 for most datasets, while P-HD-25 is doing a bit better with a bias of around -0.5, and P-HC and P-PC are also underestimating the weight. This may be an indication that we get the prior back, and that the likelihood does not contribute much in the inference.

%

\clearpage

\subsection{Application}

\begin{figure}[h!]
\centering
	\includegraphics[width = 0.66\textwidth]{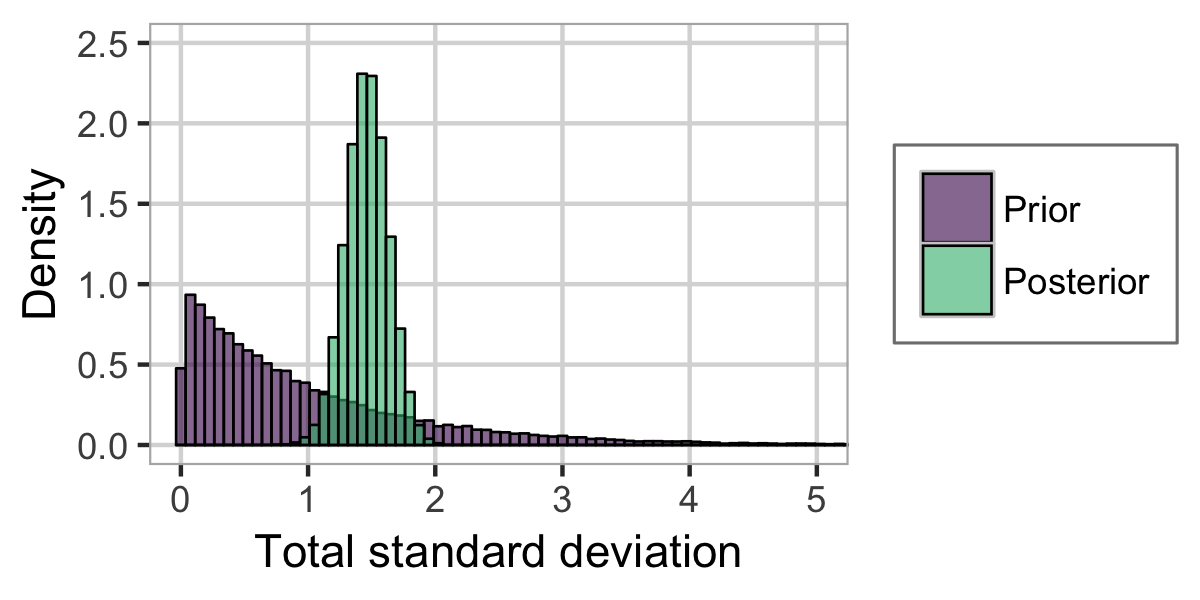}
	\caption{The prior and posterior of the total standard deviation $\sigma_{\mathrm{T}}$ from the analysis of the neonatal mortality in Kenya dataset.}
	\label{fig:supp:nonGauss:totstd}
\end{figure}

The prior and posterior of the total standard deviation from the Kenya neonatal mortality dataset analysis can be seen in Figure \ref{fig:supp:nonGauss:totstd}.

\begin{figure}[t]
\centering
	\includegraphics[width = 1\textwidth]{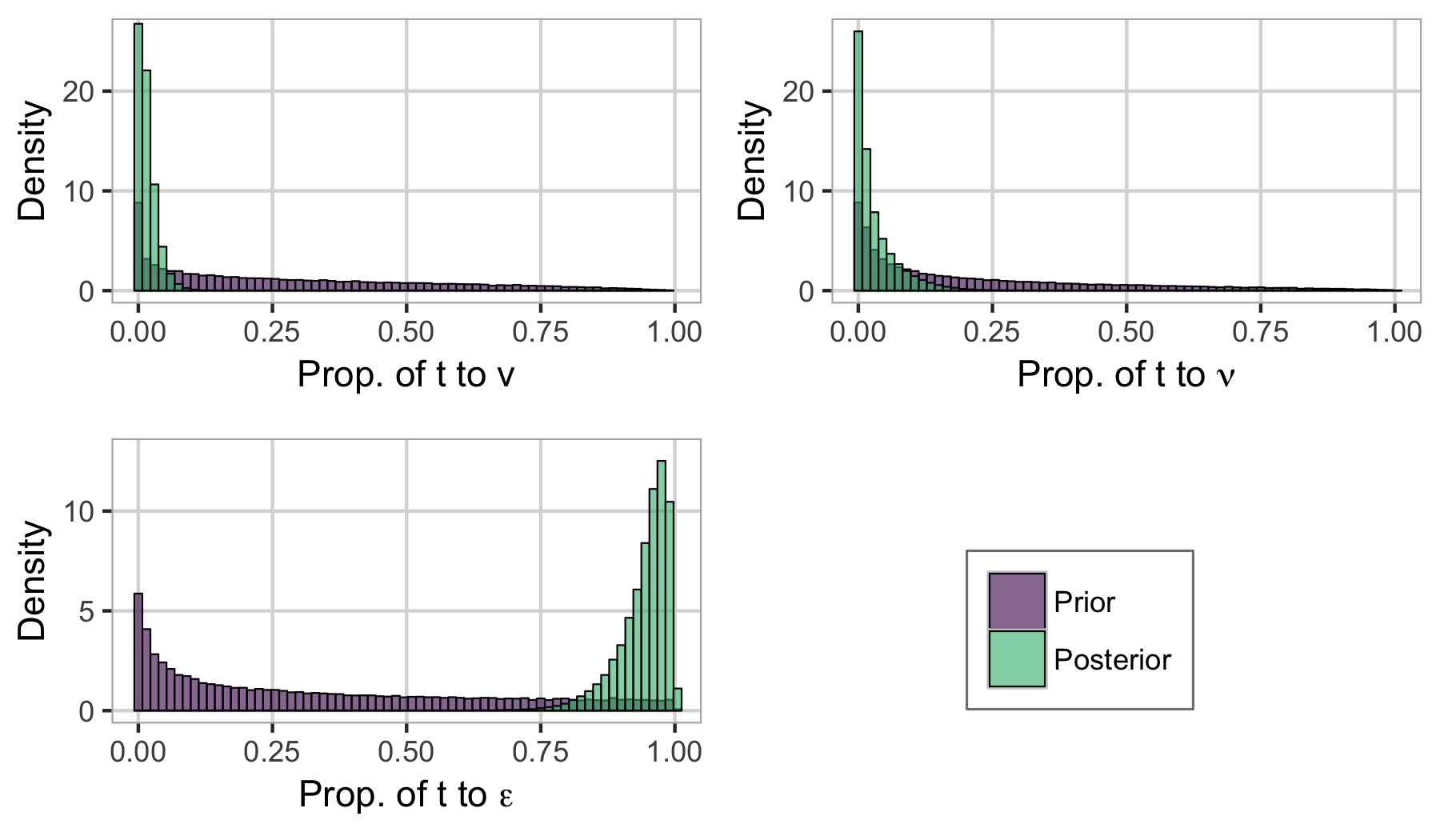}
	\caption{The priors and posteriors of the proportion of the total latent variance assigned to
			 the household effect, the cluster effect, and the unstructured spatial effect.}
	\label{fig:supp:nonGauss:kenya_totvar_vnueps}
\end{figure}

The prior and posterior distributions of the total weight of the unstructured random effects $v$ (unstructured county effect), $\nu$ (unstructured cluster effect) and 
$\varepsilon$ (unstructured household effect) can be seen in Figure \ref{fig:supp:nonGauss:kenya_totvar_vnueps}. The total weight is 
$\omega^{(1)}$ for $\varepsilon$, $\omega^{(2)}(1-\omega^{(1)})$ for $\nu$, and 
$(1-\omega^{(3)})(1-\omega^{(2)})(1-\omega^{(1)})$ for $v$. 
The medians of these three are $0.955$, $0.014$ and $0.011$, respectively.
It is clear that the household effect 
$\varepsilon$ explains most of the variance, the cluster effect $\nu$ explains some, and the 
unstructured county effect $v$ explains the least of the three.

\begin{figure}[h!]
\centering
	\includegraphics[width = 0.67\textwidth]{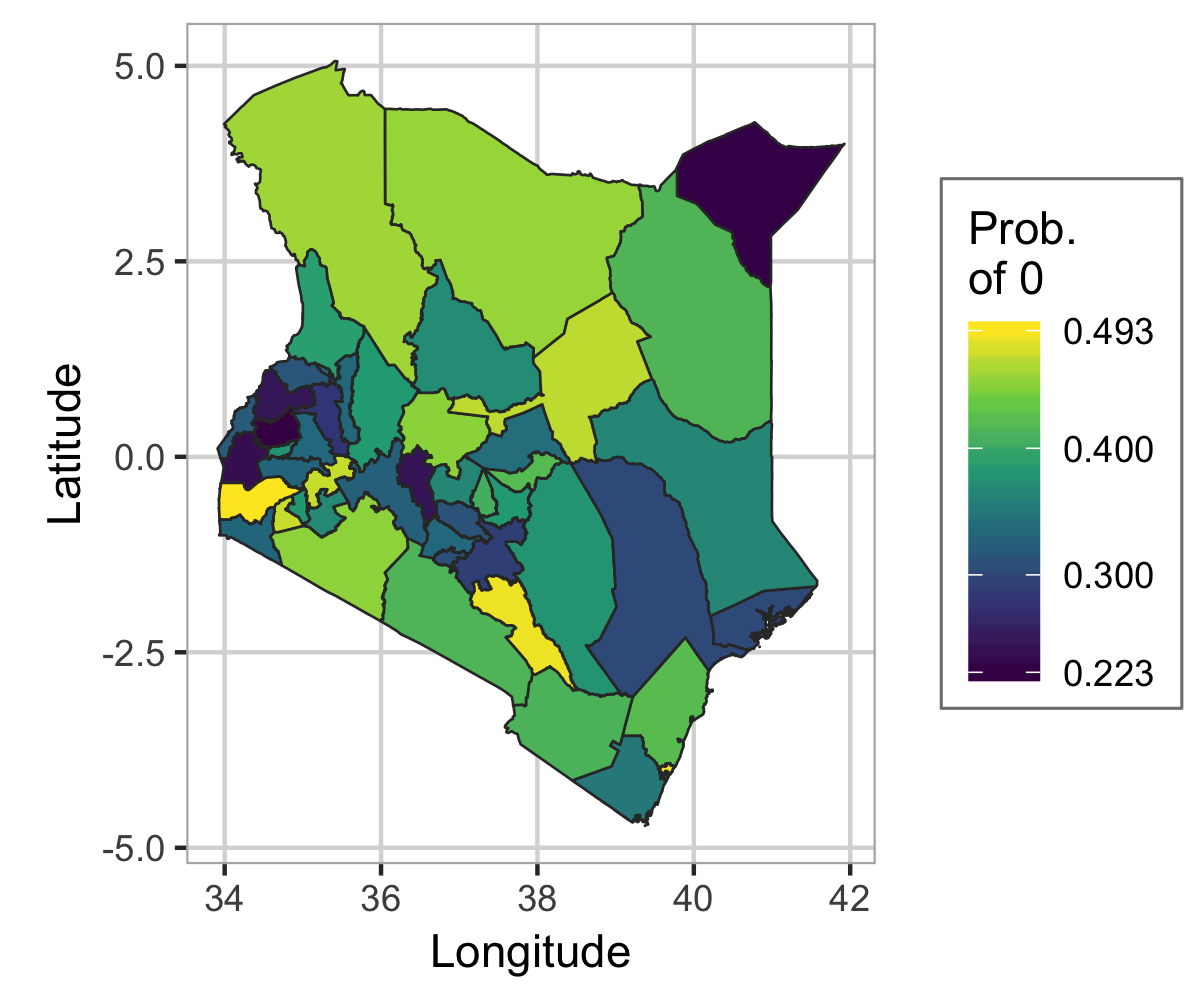}
	\caption{The significance of the spatial effect $\bm{u}$ visualized through the tail probabilities $\text{Prob}(u_i > 0)$ for the counties where the median of $\bm{u}$ is smaller than 0, and $\text{Prob}(u_i < 0)$ for the counties where the median of $\bm{u}$ is larger than 0.}
	\label{fig:supp:nonGauss:prob0}
\end{figure}

Figure \ref{fig:supp:nonGauss:prob0} shows how far a value of $0$ is from the posterior median of $\bm{u}$ expressed
by the posterior tail probability of getting 0 or further away from the median. 
We see that for many counties the posterior median of $u$ is close to 0 as expressed by the value 0.5 in the figure, and 0 is
at the most barely outside the interquartile range as expressed by a value of 0.25.

\bibliographystyle{humannat}
\bibliography{bib/references}

\end{document}